\documentclass[11pt,a4paper]{article}
\pdfoutput=1
\usepackage{jheppub}

\usepackage{color}
\usepackage{amsmath}
\usepackage{pifont}   
\usepackage{verbatim}       
\usepackage{acronym} 
    
\usepackage{amsfonts}
\usepackage{amssymb}
\usepackage{mathrsfs} 
\usepackage{graphicx}
\usepackage{multirow} 
\usepackage{slashed} 
\usepackage{array}

\usepackage{graphicx}
\usepackage{epsfig}
\usepackage{lscape}
\usepackage{rotating}
\usepackage{epsf}
\usepackage{bm}
\usepackage{booktabs}
\usepackage{array}
\usepackage{caption}
\usepackage{subcaption}
\usepackage[utf8]{inputenc}
\usepackage{float}

\usepackage{multirow}
\usepackage{array,booktabs,colortbl,multirow}
\usepackage{colortbl,xcolor,colordvi,color}
\usepackage{rotating}
\allowdisplaybreaks

\newcommand{\beq}{\begin{align}}
\newcommand{\eeq}{\end{align}}
\def\be{\begin{equation}}
\def\ee{\end{equation}}
\def\bea{\begin{eqnarray}}
\def\eea{\end{eqnarray}}
\def\bitem{\begin{itemize}}
\def\eitem{\end{itemize}}
\newcommand{\bec}{\begin{center}}
\newcommand{\eec}{\end{center}}
\newcommand{\ba}{\begin{array}}
\newcommand{\ea}{\end{array}}
\newcommand{\cmrule}{\midrule[0.25mm]}

\title{Classification of dark pion multiplets as dark matter candidates and collider phenomenology}

\author[1]{Hugues Beauchesne}
\author[2]{and Giovanni Grilli di Cortona}

\affiliation[1]{Department of Physics, Ben-Gurion University, \\Beer-Sheva 8410501, Israel}
\affiliation[2]{Institute of Theoretical Physics, Faculty of Physics, University of Warsaw, ul.~Pasteura 5, \\PL--02--093 Warsaw, Poland}

\emailAdd{beauches@post.bgu.ac.il, ggrillidc@fuw.edu.pl}

\abstract{New confining sectors can contain a set of pseudo-Goldstone mesons that exhibit a complicated structure in terms of stability and relative masses. Stable ones can act as dark matter candidates, while their interactions with the unstable ones determine their relic abundances. The overall structure, by specifying which channels are kinematically forbidden or not, affects the cosmology, constraints and collider phenomenology. In this paper, we present a classification of these pseudo-Goldstone meson structures. We find that the structures can be classified into three categories, corresponding to strong, suppressed and essentially non-existent constraints from indirect detection. Limits on decay lengths of the unstable mesons and dark jet properties are presented for several benchmark models.}

\begin{document}

\maketitle

\section{Introduction}\label{Sec:Introduction}
There is little doubt about the existence of Dark Matter (DM) at this point. A plethora of observations such as galactic rotational curves, large and small scale structures and anisotropies in the cosmic microwave background all point toward its existence. Beside some basic facts such that it must be long-lived, non-baryonic and have an abundance roughly five times that of baryonic matter~\cite{Aghanim:2018eyx}, the exact nature of dark matter is however still unknown.

One fairly understudied possibility is that dark matter consists of the stable pseudo-Goldstone mesons of a new confining sector. In such a scenario, a set of fermions charged under a new confining group is introduced. Depending on their quantum numbers, masses and other interactions, some approximate chiral symmetry can exist between them. If the fermions are light enough compared to the confinement scale of the new group, this approximate chiral symmetry will however be broken spontaneously to a smaller subgroup by a condensate of the dark fermions. To each of the spontaneously broken symmetry corresponds a pseudo-Goldstone boson or so-called dark pion. If enough symmetries are left unbroken by the condensate and explicit symmetry breaking, some of the dark pions can be stable and act as dark matter candidates. The other dark pions eventually decay away, but their interactions with the stable ones can potentially explain the dark matter abundance we observe today.

Such a scenario is interesting to study for several reasons. First, dark pions are present in many existing models of beyond the Standard Model physics and are theoretically well-motivated. They are ubiquitous in models of Neutral Naturalness, of which Twin Higgs \cite{Chacko:2005pe, Barbieri:2005ri} is arguably the archetype. They also have been proposed as a potential explanation for dark matter \cite{Hochberg:2014kqa, Cline:2013zca} and certain excesses \cite{Freytsis:2016dgf}, as well as in more esoteric settings \cite{Strassler:2006qa}. Second, multiplets of dark pions allow for explanations of the dark matter abundance that fall outside the standard paradigm of thermal freeze out of Weakly Interacting Massive Particles (WIMP) \cite{Lee:1977ua, Vysotsky:1977pe, Kolb:1990vq}. Strongly Interacting Massive Particles (SIMP) \cite{Hochberg:2014dra}, in which $3\to 2$ processes determine the DM relic abundance, is a well-known example of this, but dark pions also naturally accommodate more exotic mechanisms such as Codecaying Dark Matter \cite{Dror:2016rxc, Farina:2016llk, Okawa:2016wrr} and ELDER Dark Matter \cite{Kuflik:2015isi}. Such scenarios are typically far less constrained by direct detection than WIMPs. Third, there has recently been an increasing amount of work dedicated to discovering jets of dark hadrons, or so called dark jets. From a theoretical point of view, this includes emerging jets \cite{Schwaller:2015gea}, semivisible jets \cite{Cohen:2015toa} and soft-bombs \cite{Knapen:2016hky}. From an experimental point of view, collider searches have been performed in Ref.~\cite{Sirunyan:2018njd}. Of course, one of the practical problems of looking for dark jets is that they can be realized at colliders in many different ways, depending on their multiplicity, thrust and the stability of their constituent dark hadrons. Being able to narrow down what dark jets should look like would certainly simplify experimental search strategies.

With this context in mind, the goal of this article is to classify multiplets of dark pions in terms of their symmetry breaking structure, analyze their resulting viability as dark matter candidates and map this to the corresponding dark jet properties at colliders. More precisely, the symmetry breaking takes place in two steps: Spontaneous Chiral Symmetry Breaking (CSB) of a flavor group to a smaller subgroup and Explicit Breaking (EB) of the latter by interactions or masses. The combination of these two breakings results in a set of dark pions with a possibly complicated structure of mass versus stability. This structure crucially affects the evolution of the dark matter abundance in the early Universe. Making sure that the resulting dark matter is compatible with the Universe as we observe it today imposes constraints on the structure of the dark pion sector. This in turn is reflected in how dark jets should look like at colliders. What we do in this article is classify these structures, apply cosmological constraints and make predictions as to the characteristics of dark jets. More specifically, we will concentrate on multiplets that contain both stable and unstable dark mesons at the same time. If all dark pions were stable, this would reduce to SIMP or possibly standard WIMP thermal freeze-out, while, if they were all unstable, they obviously could not serve as dark matter candidates. In a lot of ways, this paper is a generalization of Ref.~\cite{Beauchesne:2018myj}. Previous work on combinations of stable and unstable dark mesons also include Refs.~\cite{Buckley:2012ky, Freytsis:2016dgf, Okawa:2016wrr, Heikinheimo:2014xza, Bernreuther:2019pfb}.

The final result of this paper will be that dark meson multiplets can be separated into three distinct categories. In the first category, all stable pions can annihilate with another to produce unstable ones, even at low temperature. This results in strong bounds from indirect detection and jets that are similar to QCD jets with a higher confinement scale. In the second category, some stable pions cannot annihilate with other ones to produce unstable pions, but some others can. This considerably alleviates the bounds from indirect detection, opening the way to less QCD-like jets. When the mass splitting between the dark pions is small, the dark jets can take many forms. However, a larger splitting between the dark pions forces the jets to be increasingly narrow, with a larger proportion of stable mesons, larger multiplicity and with short decay lengths for their unstable dark pions. In the third category, all annihilations of two stable pions to a final state containing an unstable pion are forbidden. This means that bounds from indirect detection are essentially inexistent. The resulting dark jets are similar to those of the second category, but with a larger upper limit on the decay length of the unstable mesons.

The article is organized as follows. First, we classify the different patterns of symmetry breaking. The mechanism through which they can lead to the correct dark matter relic abundance is then described. The process through which we obtain current cosmological bounds is explained afterward. Limits on the mass splitting and decay lengths are then presented for a set of benchmark models. Collider properties such as the fraction of the dark jets that consists of stable mesons and the thrust and multiplicity of the dark jets are discussed considering these constraints. We finish with some concluding remarks. The first appendix elaborates on the technical details of the procedure through which we obtain the correct dark matter abundance. The second appendix discusses more carefully our benchmark symmetry breaking patterns.

\section{Concepts and classification of symmetry breaking patterns}\label{Sec:CSBP}
We begin this article by introducing the concepts and notation that we will use throughout the paper. We will then classify the different combinations of chiral and explicit symmetry breaking.

\subsection{Notation}\label{sSec:Notation}
First, we define more clearly what we mean by a new confining sector. We begin by assuming a new confining gauge group $\mathcal{G}$. Considering the goals of this paper, we will generally assume its confinement scale is at a value that can reasonably be probed at current colliders, but this is not crucial from a cosmological point of view. We then introduce a set of $N$ massive Dirac dark quarks $q_i$ of mass $m_i$. By definition, they are charged under $\mathcal{G}$ but neutral under SM gauge groups. We assume there is an at least approximate flavor global symmetry between the dark quarks characterized by the group $G$. We also assume that the dark quarks communicate with the SM sector, but that the interactions between the two sectors are rather weak.

Next, we consider three patterns of chiral symmetry breaking. They are:
\begin{equation}\label{eq:CSBP}
  \begin{tabular}{cccc}
  $G$                  & $\to$ & $H$      & $\#$Pions     \\\hline
  $SU(N)\times SU(N)$  & $\to$ & $SU(N)$  & $N^2 - 1$     \\
  $SU(2N)$             & $\to$ & $Sp(2N)$ & $2N^2 - N - 1$\\
  $SU(2N)$             & $\to$ & $SO(2N)$ & $2N^2 + N - 1$
  \end{tabular}
\end{equation}
The groups on the left $G$ represent the original flavor symmetries and the groups on the right $H$ what they are broken to by the condensate. The first pattern is typical of dark quarks charged under complex representations of $\mathcal{G}$, the second to pseudoreal representations and the third to real representations \cite{Peskin:1980gc}. To each broken symmetry corresponds a pseudo-Goldstone meson or dark pion. Together, they form a representation of $H$ whose size is included in Eq.~(\ref{eq:CSBP}). To alleviate the text, we will in general refer to dark pions as simply pions, as there will be no risks of confusing them with the usual SM pions. They are referred to as pseudo-Goldstone bosons as we do assume the presence of mass terms that render the spontaneously broken symmetries approximate and give mass to the pions. The pion decay constant associated to the pions is labelled as $f$. The advantage of including real and pseudoreal representations for dark quarks is that it opens many possibilities for the structure of the dark sector. Perhaps most interestingly is that dark color singlets can then be constructed with two quarks or two antiquarks, not necessarily a quark and an antiquark. This is specially useful for real dark quarks, where a dark color singlet can be constructed with twice the same quark, resulting in a meson with a mass independent of the masses of the other dark quarks.

Then, we assume that the flavor group $H$ that was left unbroken by the condensate is explicitly broken by masses or interactions to some smaller subgroup $h$. The pions will then decompose themselves into a set of representations of $h$. This symmetry will generally be sufficient to maintain some of them stable, but not all of them. We will assume that no particles beside the pions exist that are charged under this group. Otherwise, the stability of the pions would not be insured anymore and this would effectively be equivalent to the symmetry being broken. The breaking of $H$ will also in general result in the different multiplets of $h$ having different masses.

To summarize, the pattern of symmetry breaking is:
\begin{equation}\label{eq:SBPsummary}
  G \xrightarrow[]{\text{CSB}} H \xrightarrow[]{\text{EB}} h.
\end{equation}
The first breaking determines the number of pions and the second breaking their masses and stability. The combination of these breakings determines a structure of pions with given masses and stability. This structure in turn controls which processes are kinematically forbidden or not. This determines the cosmology and the collider phenomenology.

\subsection{Classification of symmetry breaking patterns}\label{sSec:Classification}
With the notation established, the structures of pions resulting from a given combination of chiral and explicit symmetry breaking can be classified into three categories from both a theoretical and phenomenological point of view. In the hope of making the discussion more concrete, we will provide an example for each case.

\subsubsection*{Category I: Unforbidden}
A structure of pions is said to belong to category I if, for each stable pion, there exist another stable pion with which it can annihilate to produce at least one unstable pion via a kinematically allowed $2\to 2$ pion scattering process.\footnote{To avoid any possible confusion, a kinematically allowed process is defined as one in which the total mass of the particles coming in is larger than the total mass of the particles going out.} In more casual terms, it simply means that every stable pion can annihilate with another to produce at least an unstable one, even at very low temperature.

\begin{figure}[t!]
  \centering
  \begin{subfigure}{0.31\textwidth}
    \centering
    \includegraphics[width=\textwidth]{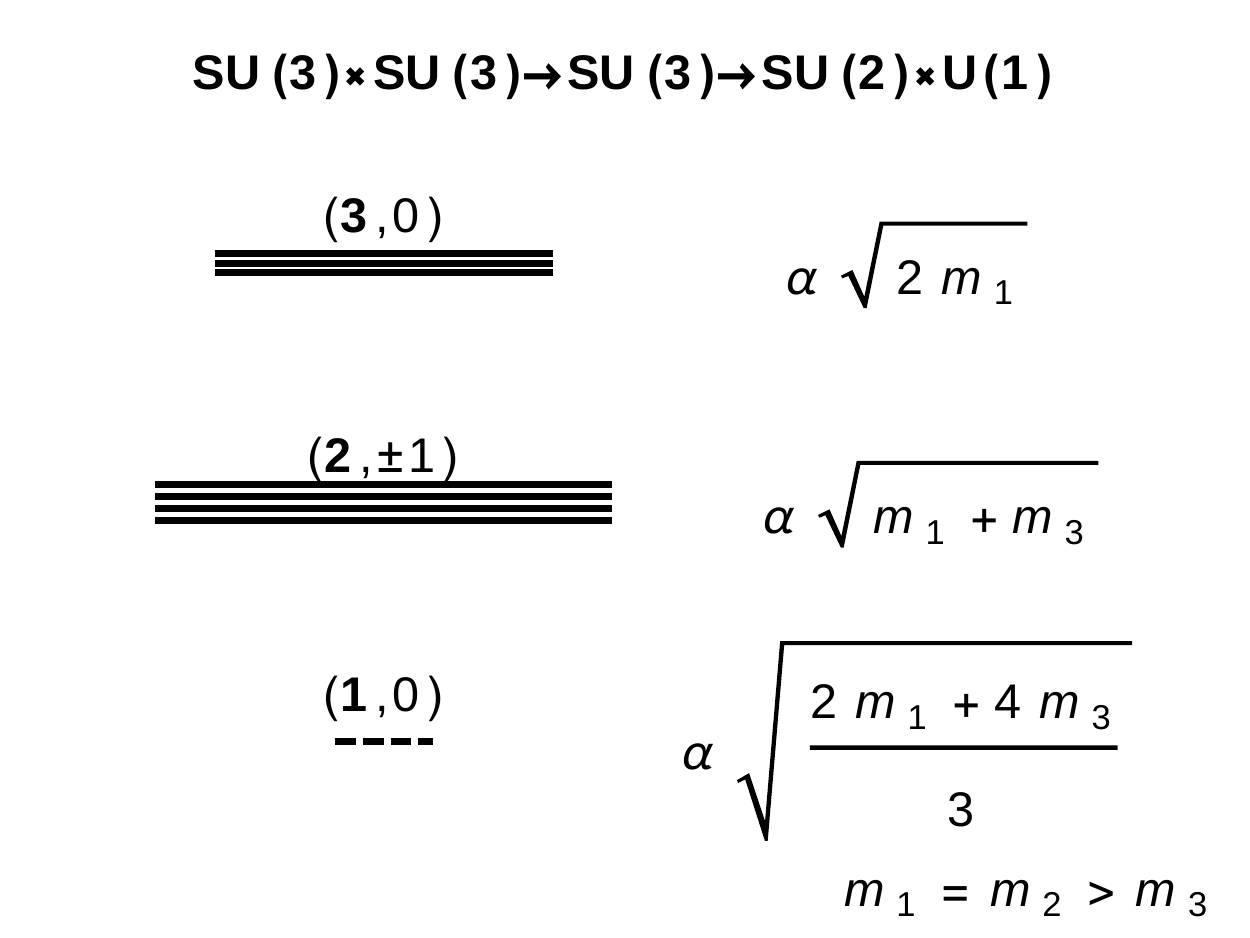}
    \caption{Category I}
    \label{fig:ExampleCategoryI}
  \end{subfigure}
  ~
    \begin{subfigure}{0.31\textwidth}
    \centering
    \includegraphics[width=\textwidth]{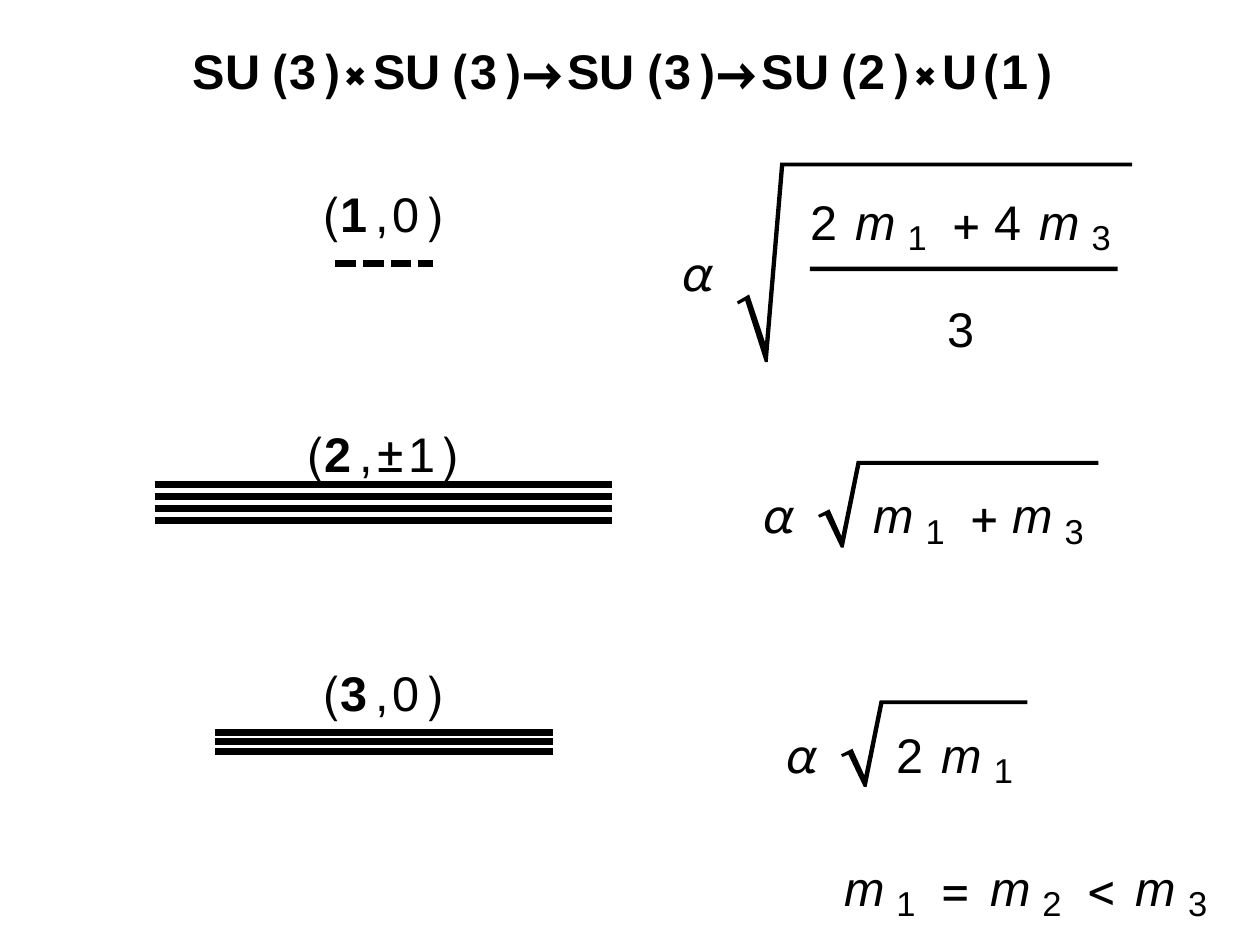}
    \caption{Category II}
    \label{fig:ExampleCategoryII}
  \end{subfigure}
  ~
    \begin{subfigure}{0.31\textwidth}
    \centering
    \includegraphics[width=\textwidth]{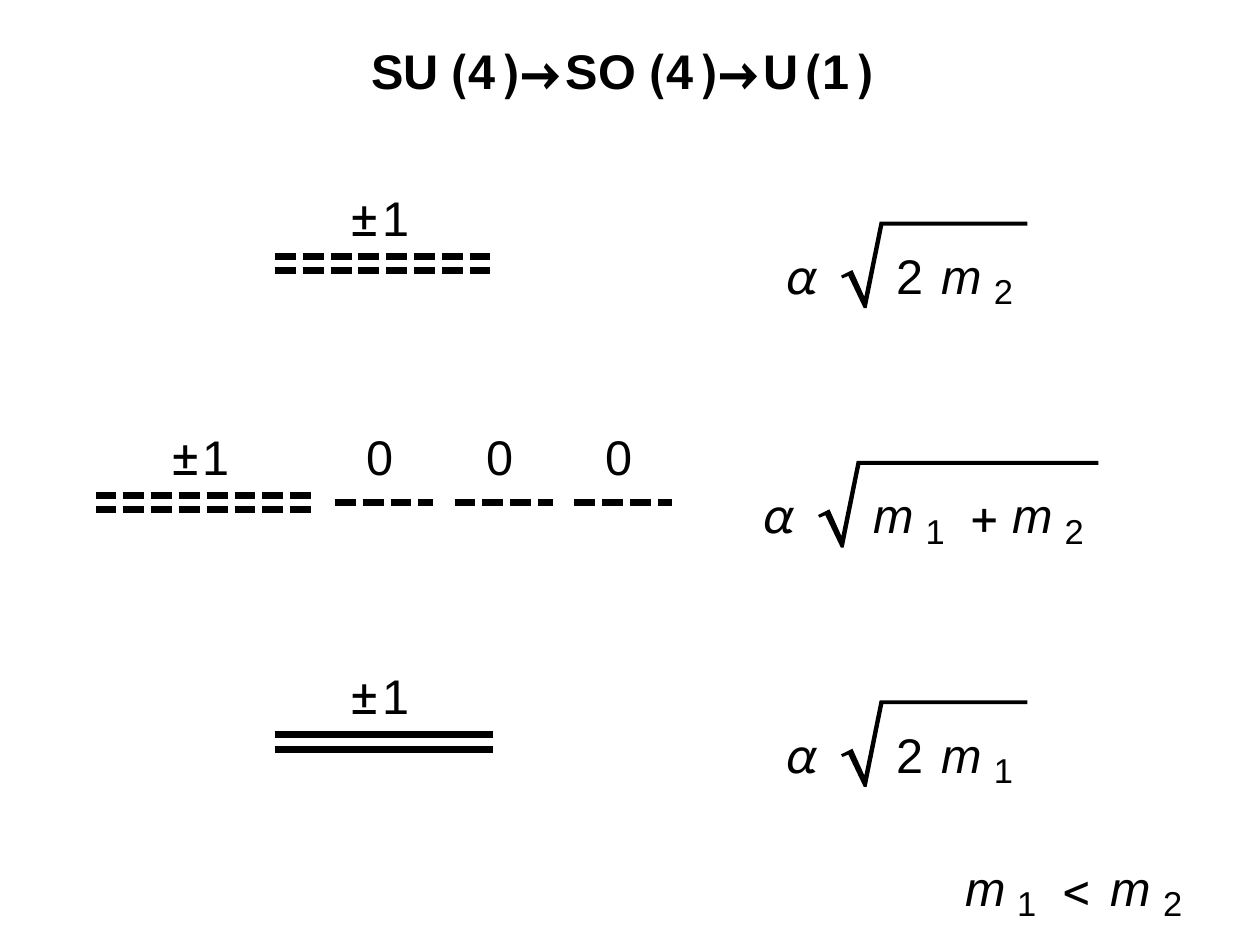}
    \caption{Category III}
    \label{fig:ExampleCategoryIII}
  \end{subfigure}
\caption{Examples of the three categories of symmetry breaking. Solid lines represent stable particles and dashed lines unstable ones. Double/triple/quadruple lines represent particles degenerate in mass. The numbers above the lines are the quantum numbers of the pions under $h$. The masses of the pion multiplets (up to a global multiplicative constant) are shown on the right.}\label{fig:ExampleCategories}
\end{figure}

An example of such a spectrum is shown in Fig.~\ref{fig:ExampleCategoryI}. In it, three dark quarks transform under a complex representation of $\mathcal{G}$. This results in an $SU(3) \times SU(3) \to SU(3)$ pattern of chiral symmetry breaking, leading to eight pions. The resulting $SU(3)$ is then explicitly broken to $SU(2)\times U(1)$. This splits the pions into multiplets $(\mathbf{3}, 0)$, $(\mathbf{2}, \pm 1)$ and $(\mathbf{1}, 0)$. Assuming the dark quark not charged under $SU(2)$ is $q_3$ and that it is lighter than the two others, the lightest state is then the singlet. Since it does not carry any charge under $SU(2)\times U(1)$, this group is insufficient to maintain the singlet stable and it will in general decay, barring any additional symmetry. All other pions can be proven to be stable because of conservation of group charges and energy. By the same principles, it is easy to verify that every stable pion can annihilate with its conjugate to produce a pair of singlets, thereby satisfying the condition of category~I.

\subsubsection*{Category II: Partially forbidden}
A structure of pions is said to belong to category II if, for some but not all of its stable pions, there exist another stable pion with which they can annihilate to produce at least one unstable pion via a kinematically allowed $2\to 2$ pion scattering process. Simply put, there are some stable pions that can collide with some other stable pions to produce unstable pions and some that can't.

An example of such a spectrum is shown in Fig.~\ref{fig:ExampleCategoryII}. It is similar to the example of category I, but with $m_3 > m_1$ instead. It is simple to use group charges and energy conservation to prove that the triplet and doublets are once again stable. It is also easy to verify that the $(\mathbf{3}, 0)$ cannot produce unstable pions via a kinematically allowed annihilation with another stable pion. However, the annihilation of $(\mathbf{2}, +1)$ with $(\mathbf{2}, -1)$ to $(\mathbf{3}, 0)$ and $(\mathbf{1}, 0)$ is kinematically allowed and has a non-zero amplitude even at leading order. All in all, this means that only a fraction of the stable pions can produce unstable pions at very low velocity.

\subsubsection*{Category III: Forbidden}
A structure of pions is said to belong to category III if, for every stable pion, there does not exist another stable pion with which it can annihilate to produce at least one unstable pion via a kinematically allowed $2\to 2$ pion scattering process. In simple terms, it means that it is impossible for two stable pions to produce an unstable one via a collision at low temperature.

An example of such a spectrum is shown in Fig.~\ref{fig:ExampleCategoryIII}. It consists of two quarks transforming under a real representation of $\mathcal{G}$. The pattern of spontaneous symmetry breaking is then $SU(4)\to SO(4)$, resulting in nine pions. The group $SO(4)$ is then broken explicitly to its $U(1)$ subgroup in which dark quarks have charge $+1/2$ and dark antiquarks $-1/2$, i.e. the equivalent of the baryon number up to a normalization chosen for convenience. Pions made of a quark and an antiquark therefore have a charge of 0, while those made of two quarks or two antiquarks have respectively a charge of $+1$ or $-1$. Only the pion made of twice the lightest quark and its conjugate are therefore stable. They also cannot annihilate between each other to produce unstable pions via kinematically allowed processes.

\subsubsection*{Additional comments}
This classification is very pragmatic. Indeed, the annihilation of stable pions to unstable ones provides constraints from indirect detection. For category I, all stable pions can contribute to the indirect detection signal, which results in strong constraints. For category II, only a subset of stable pions can contribute significantly to the indirect detection signal, which significantly reduces the bounds from indirect detection. For category III, none of the stable pions can contribute significantly to the indirect detection signal, which results in essentially non-existent constraints. Of course, these categories also affect the cosmological evolution.

Radiative corrections can potentially determine in which category a set of pions belongs to. For example, radiative corrections make the charged pion heavier than the neutral one in the Standard Model. Ultimately, the only factor that matters is the masses of the pions, not whether they come from the quark masses or radiative corrections. To avoid any potential confusion, we will only use benchmarks for which the category is not determined by radiative corrections given sufficient mass splitting between the dark quarks. As such, we will also neglect all radiative corrections. Doing so would anyhow require knowledge beyond the simple dark quark content of the new confining sector.

In some cases, different multiplets can be degenerate in mass. Unless there is some symmetry involved, radiative corrections will generally lift this degeneracy. Scattering between such particles are typically impeded by a small phase-space.

A series of examples that will also serve as benchmarks is shown in Table~\ref{tab:Benchmarks}. We refer to Appendix~\ref{app:SBB} for a complete description of every benchmark model. Note that discrete subgroups could of course also be used, even though we did not include any examples.

\begin{table}[t]
{\footnotesize
\setlength\tabcolsep{5pt}
\begin{center}
\scriptsize
\begin{tabular}{cccccccc}
\toprule
Cat.     & Label & $G$                  & $\to$ & $H$     & $\to$ & $h$                              & Condition(s)                          \\
\cmrule
         & CIa   & $SU(3) \times SU(3)$ & $\to$ & $SU(3)$ & $\to$ & $SU(2) \times U(1)$              & $m_1 = m_2 > m_3$                     \\
I        & CIb   & $SU(6)$              & $\to$ & $Sp(6)$ & $\to$ & $Sp(4) \times U(1)$              & $m_1 = m_2 > m_3$                     \\
         & CIc   & $SU(3) \times SU(3)$ & $\to$ & $SU(3)$ & $\to$ & $U(1)  \times U(1)$              & $m_1 > m_2 > m_3$                     \\
\cmrule 
         & CIIa  & $SU(4)$              & $\to$ & $SO(4)$ & $\to$ & $U(1)  \times U(1)$              & $m_1 < m_2$                           \\
         & CIIb  & $SU(3) \times SU(3)$ & $\to$ & $SU(3)$ & $\to$ & $SU(2) \times U(1)$              & $m_1 = m_2 < m_3$                     \\
II       & CIIc  & $SU(6)$              & $\to$ & $Sp(6)$ & $\to$ & $Sp(4) \times U(1)$              & $m_1 = m_2 < m_3$                     \\
         & CIId  & $SU(5) \times SU(5)$ & $\to$ & $SU(5)$ & $\to$ & $SU(3) \times SU(2) \times U(1)$ & $m_1 = m_2 < m_3 = m_4 = m_5$         \\
         & CIIe  & $SU(4) \times SU(4)$ & $\to$ & $SU(4)$ & $\to$ & $SU(2) \times U(1) \times U(1)$  & $m_1 = m_2 < m_3 \leq m_4$            \\
\cmrule 
         & CIIIa & $SU(4)$              & $\to$ & $SO(4)$ & $\to$ & $U(1)$                           & $m_1 < m_2$                           \\
III      & CIIIb & $SU(6)$              & $\to$ & $SO(6)$ & $\to$ & $U(1)$                           & $m_1 > m_2 > m_3$, $2m_2 > m_1 + m_3$ \\
         & CIIIc & $SU(4) \times SU(4)$ & $\to$ & $SU(4)$ & $\to$ & $SU(2)$                          & $m_1 = m_2 > m_3 = m_4$               \\
\bottomrule
\end{tabular}
\end{center}
}
\caption{Benchmark scenarios for the two categories of pion structures. The condition represents a relation that is assumed between the dark quark masses. It is sometimes crucial for the structure to be in a given category but not always. See Appendix~\ref{app:SBB} for more details.} 
\label{tab:Benchmarks}
\end{table}

\section{Overview of the cosmological evolution}\label{Sec:CosmoEvolution}
Having established the notation, we now proceed to discuss the cosmological evolution of the pion densities. This section will include a discussion of our assumptions, the different regimes of density evolution and how the pion structures affect the abundances.

\subsection{Assumptions}\label{sSec:Assumptions}
We begin this section by stating the assumptions we will make during the computation of the dark matter relic abundance. None of them are crucial to our discussion and are simply meant to simplify the presentation. First, we assume that annihilation of pions directly to SM particles is negligible compared to annihilation to other pions. Otherwise, bounds from direct detection would be difficult to satisfy (see Ref.~\cite{Beauchesne:2018myj}). Dark pions would also act mostly as standard WIMPs, which are already well studied. Second, we assume that the dark sector maintains the same temperature as the SM sector. This simplifies the presentation tremendously as there is a very large number of ways for the two sectors to exchange energy. In practice, requesting that the mesons decay fast enough to SM particles implies a minimum amount of interaction between the two sectors. If the pions interact with light particles, it should be sufficient for the two sectors to maintain kinematic equilibrium (see Ref.~\cite{Beauchesne:2018myj}), but could prove to be insufficient if the pions interact exclusively with heavy SM particles. One way or the other, this should not affect qualitatively any results. Third, we will neglect the presence of other composite particles. As they should be heavier than the pseudo-Goldstone mesons, they should be negligible in most regions of parameter space and not change the conclusions in the small regions where their effects become important.\footnote{We also neglect the presence of any asymmetry that could lead to asymmetric dark matter \cite{Petraki:2013wwa, Kaplan:2009ag, Zurek:2013wia}, as this is beyond the scope of this paper.}

\subsection{Different evolution regimes}\label{sSec:Regimes}
\begin{figure}[t!]
  \centering
  \begin{subfigure}{0.47\textwidth}
    \centering
    \caption{CIa: Codecaying}
    \includegraphics[width=\textwidth]{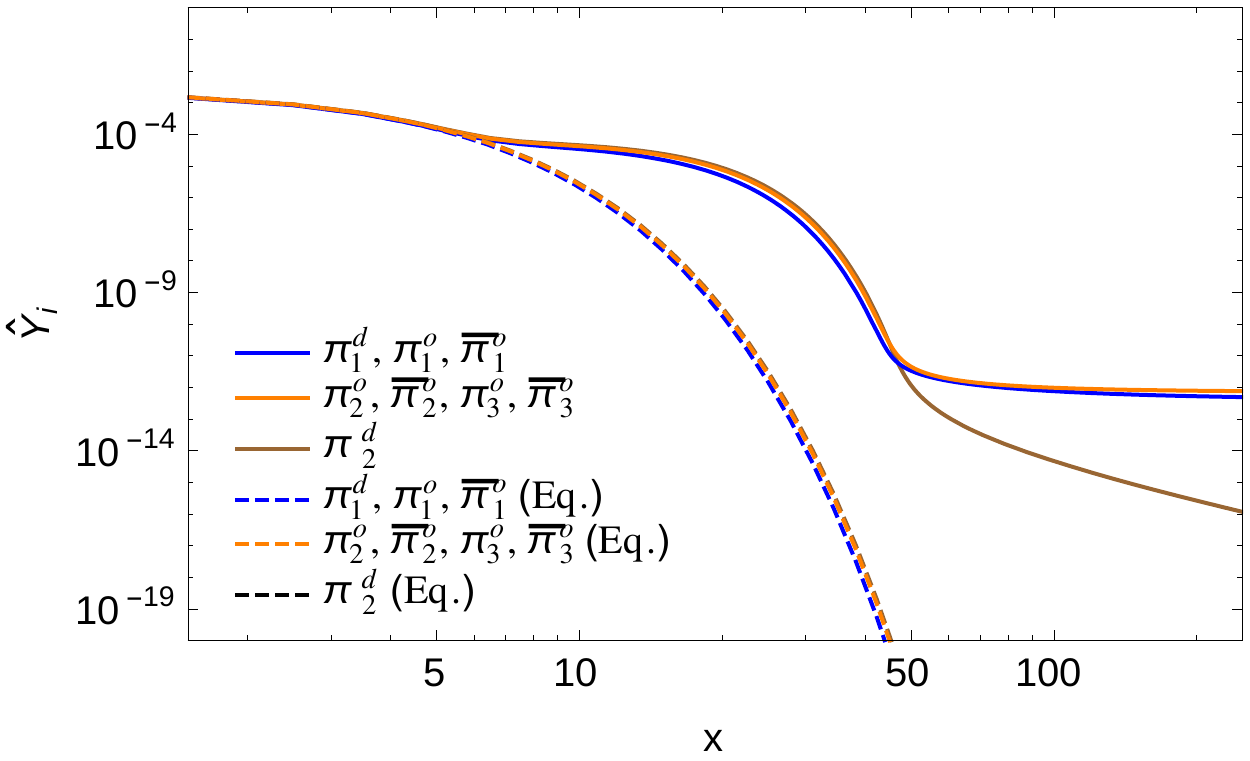}
    \label{fig:DMEvolution1}
  \end{subfigure}
  ~
  \begin{subfigure}{0.47\textwidth}
    \centering
    \caption{CIa: Coupling-independent}
    \includegraphics[width=\textwidth]{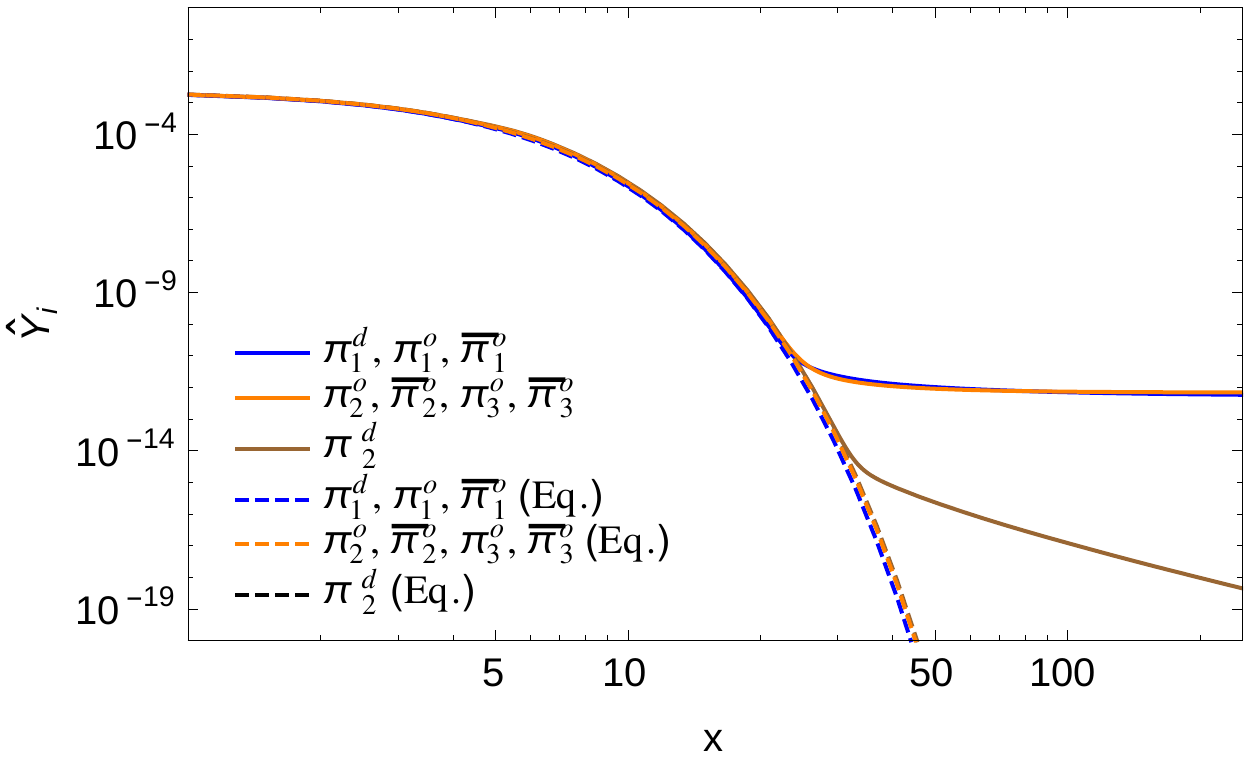}
    \label{fig:DMEvolution2}
  \end{subfigure}
  ~
  \begin{subfigure}{0.47\textwidth}
    \centering
    \caption{CIIb: Codecaying}
    \includegraphics[width=\textwidth]{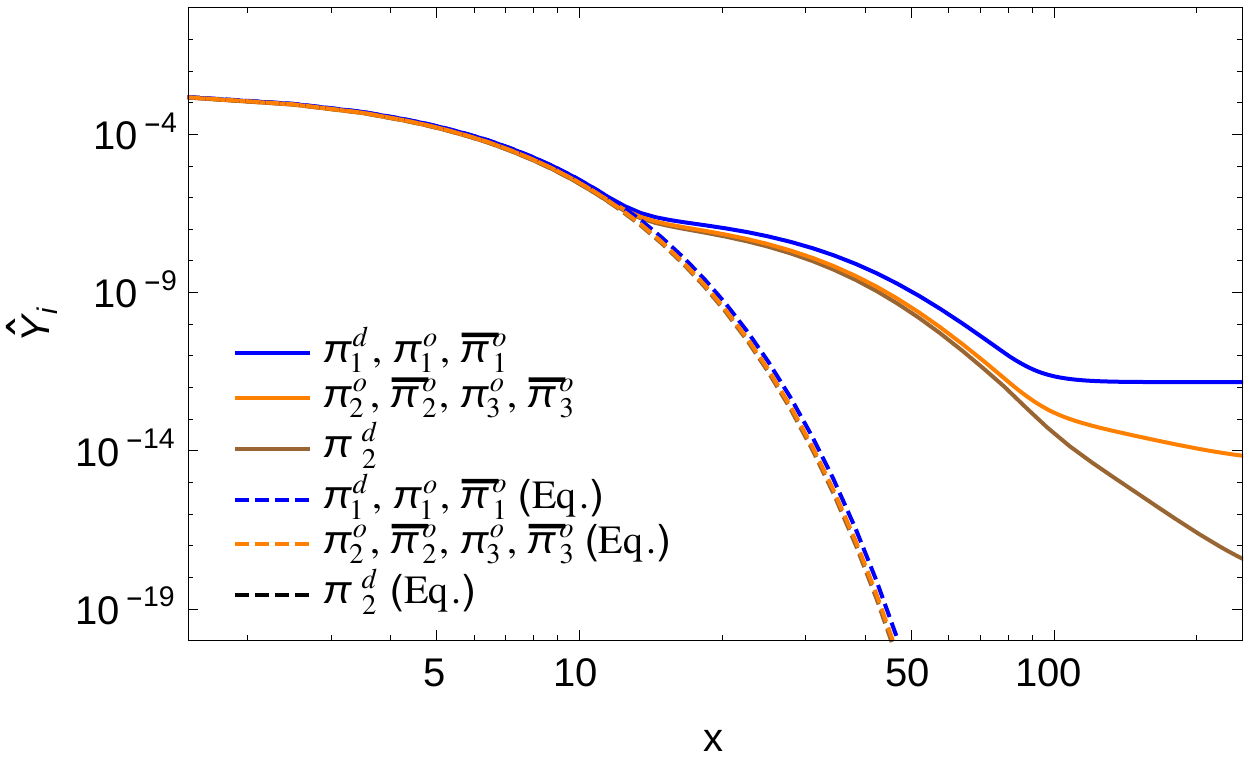}
    \label{fig:DMEvolution3}
  \end{subfigure}
  ~
  \begin{subfigure}{0.47\textwidth}
    \centering
    \caption{CIIb: Coupling-independent}
    \includegraphics[width=\textwidth]{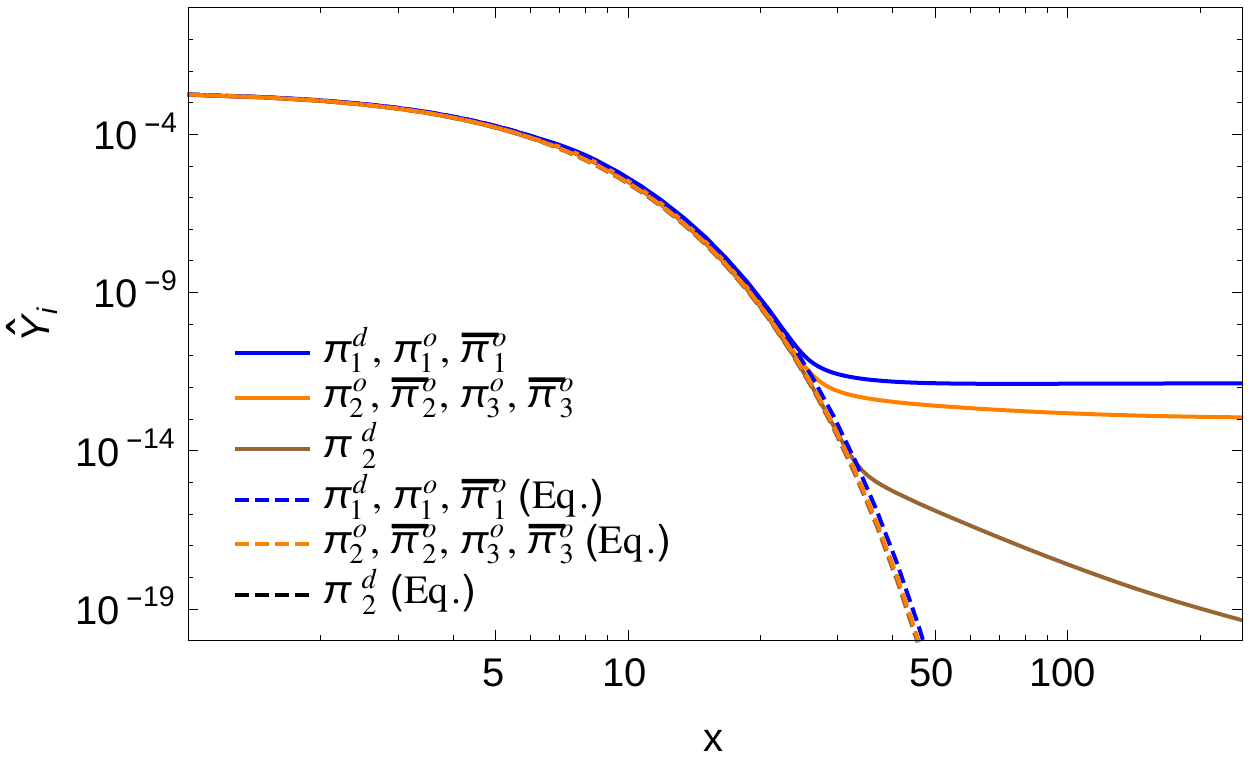}
    \label{fig:DMEvolution4}
  \end{subfigure}
  ~
  \begin{subfigure}{0.47\textwidth}
    \centering
    \caption{CIIIa: Codecaying}
    \includegraphics[width=\textwidth]{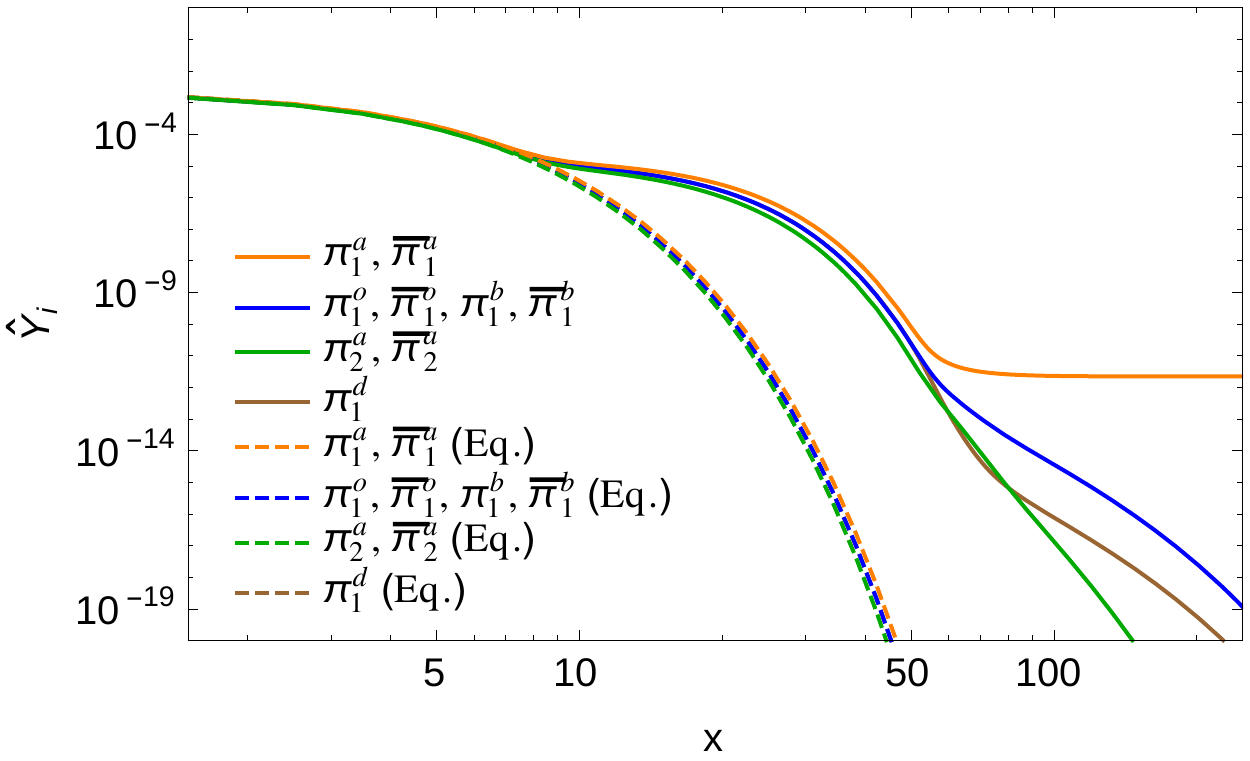}
    \label{fig:DMEvolution5}
  \end{subfigure}
  ~
  \begin{subfigure}{0.47\textwidth}
    \centering
    \caption{CIIIa: Coupling-independent}
    \includegraphics[width=\textwidth]{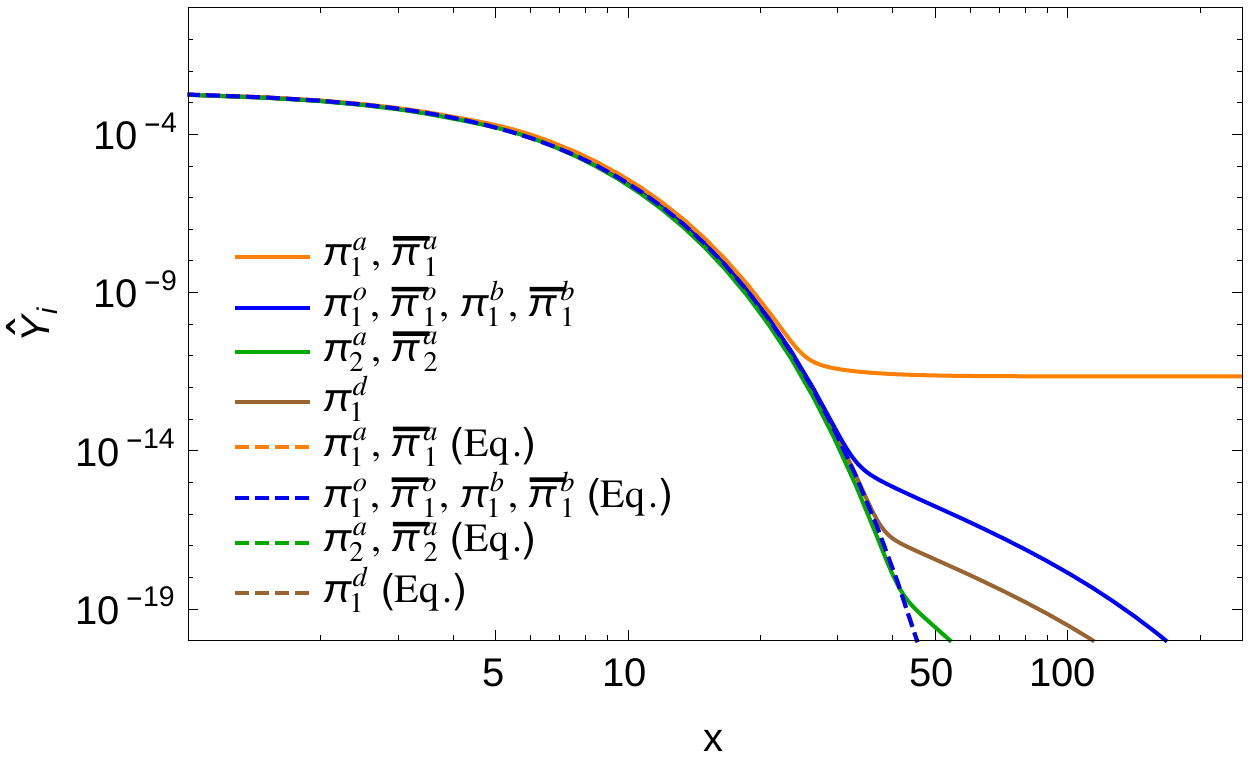}
    \label{fig:DMEvolution6}
  \end{subfigure}
\caption{Examples of the cosmic evolution of the pion abundances for the different categories and regimes. For category I, the parameters are set to: $m_1/m_3=1.1$, $m_{\pi_{\mathbf{3}}^s}= 100$~GeV and (a) $\Gamma_{\pi_1^d}=10^{-16}$~GeV, (b) $\Gamma_{\pi_1^d}=10^{-14}$~GeV. For category II, the parameters are set to: $m_3/m_1=1.1$, $m_{\pi_{\mathbf{3}}^s}= 100$~GeV and (c) $\Gamma_{\pi_1^d}=10^{-15}$~GeV, (d) $\Gamma_{\pi_1^d}=10^{-13}$~GeV. For category III, the parameters are set to: $m_2/m_1=1.1$, $m_{\pi_1^a}= 100$~GeV and (c) $\Gamma_{\pi_1^d}=10^{-16}$~GeV, (d) $\Gamma_{\pi_1^d}=10^{-14}$~GeV. In the last case, the decay widths of the other unstable pions are set to $\Gamma_{\pi_1^d}/10$. In all cases, the pion decay constant is adjusted to reproduce the correct DM abundance. The number of colors $N_c$ is set to 3 for CIa and CIIb and to 4 for CIIIa. See Secs.~\ref{sec:DecayLengths} and \ref{sapp:Notation} for the notation.}\label{fig:DMEvolution}
\end{figure}

With these assumptions established, the cosmology can fall into two distinct regimes which we present here. This discussion mirrors the previous work of Ref.~\cite{Beauchesne:2018myj}. Examples of the thermal evolution are shown in Fig.~\ref{fig:DMEvolution} for a benchmark of each category. The $\hat{Y}_i$'s represent the number density of a given multiplet divided by the entropy density and the size of the multiplet. We refer to Appendix~\ref{app:RDMAC} for the mathematical details.

\subsubsection*{Codecaying Dark Matter regime}
The first regime is characterized by the unstable pions having small decay widths, where what is meant by small will soon be clear. The evolution of the dark matter goes through (up-to) three steps as it passes from small to large $x\equiv m_{\pi_0}/T$, where $\pi_0$ is some arbitrary reference pion. This is illustrated in Figs.~\ref{fig:DMEvolution1}, \ref{fig:DMEvolution3} and \ref{fig:DMEvolution5}.

First, the Wess-Zumino-Witten (WZW) term allows for the presence of $3\to 2$ processes \cite{Wess:1971yu, Witten:1983tw, Witten:1983tx}. This term is present as long as $\pi_5(G/H) = \mathbb{Z}$. This is satisfied for all patterns of chiral symmetry breaking of Eq.~(\ref{eq:CSBP}) with at least two dark quarks, except $SU(2)\times SU(2) \to SU(2)$ which we will not be using anyhow. At small enough $x$, $3\to 2$ processes are sufficient to maintain the pions at their thermal equilibrium densities. This is the basic mechanism behind SIMP dark matter. If the WZW term is not allowed, this step is simply absent.

Second, as $x$ increases and the pion densities decrease, $3\to 2$ processes eventually become inefficient and pions become unable to maintain their thermal equilibrium densities. The $2\to 2$ processes are still efficient however and the pions therefore maintain chemical equilibrium with each other. The unstable ones decay and are replaced by the annihilation of stable pions to unstable ones. This leads to an overall decrease of the pion abundance.

Finally, the pions decouple from each other. The end result is a net density of stable pions, while the unstable ones simply decay away.

For a fixed mass structure, two sets of parameters affect the relic abundance. First, decreasing the pion decay constant allows the pions to maintain chemical equilibrium for a longer time, allowing more stable pions to be converted to unstable ones which then decay. In addition, it renders the $3\to 2$ processes more efficient. Both of these factors lead to a smaller DM relic abundance. Second, larger pion decay widths mean that the unstable pions will decay faster and will again reduce the DM relic abundance.

\subsubsection*{Coupling-independent regime}
As one increases the decay widths of the unstable pions, their densities approach their equilibrium values. Eventually, a point is reached where unstable pions are still in thermal equilibrium when the stable pions decouple from them. Past this point is the so-called coupling-independent regime of Ref.~\cite{Beauchesne:2018myj}, dubbed this way because further increasing the decay widths of the unstable pions does not affect the relic densities anymore. Effectively, the unstable pions become just another component of the plasma. This regime is illustrated in Figs.~\ref{fig:DMEvolution2}, \ref{fig:DMEvolution4} and \ref{fig:DMEvolution6}. For a fixed mass structure, the only factor that affects the DM relic abundance is how much the pions interact with each other, i.e. the pion decay constant $f$.

\subsection{Cosmological evolution for the different pion structures}\label{sSec:CosmoEvolCat}
The evolution of the pion densities follows the order of the previous subsection for all pion mass structures. The final relative abundances of the pions are however strongly dependent on the category.

In category I , the annihilation of two stable pions to at least an unstable one is always kinematically allowed. This means that all stable pions have similar cross sections for production of unstable pions, up to phase-space factors and numerical coefficients of $\mathcal{O}(1)$. This signifies that all the stable pions will decouple from the unstable ones at a similar temperature. This results in all the stable pions having similar densities, as can be seen in Figs.~\ref{fig:DMEvolution1} and \ref{fig:DMEvolution2}.

In category II, only some of the stable pions can annihilate to at least an unstable pion via a kinematically allowed process. Those that are not allowed to do so will have cross sections for the annihilation to unstable ones that are Boltzmann suppressed. This results in them decoupling when their densities are much larger than those that can annihilate to unstable ones. The end result is that the stable pions that cannot annihilate to unstable ones dominate the relic density, as can be seen in Figs.~\ref{fig:DMEvolution3} and \ref{fig:DMEvolution4}.

In category III, none of the stable pions can annihilate to at least an unstable pion via a kinematically allowed process. The lightest pions typically dominate the relic density, as can be seen in Figs.~\ref{fig:DMEvolution5} and \ref{fig:DMEvolution6}.

\section{Constraints}\label{Sec:Constraints}
In this section, we discuss the procedures through which constraints are applied and we refer to Sec.~\ref{sec:DecayLengths} for limits on different benchmark models. In this work, we use the experimental value for the relic abundance $\Omega_\mathrm{obs} h^2 = 0.1200\pm0.0012$ at $68\%$ confidence \cite{Aghanim:2018eyx}.

\subsection{Indirect detection}\label{sSec:ConstraintsID}
In models where the dark matter populates rich hidden sectors, annihilations may proceed via sequential decays, giving rise to multibody final states. In fact, depending on the spectrum of the dark sector, stable dark particles colliding with each other may produce unstable dark states. These subsequently decay back into Standard Model particles, which may in turn decay until only stable particles like positrons, antiprotons and photons are left. These particles can potentially be detected and/or affect their environment. Their indirect detection signals depend essentially on the annihilation rate of particles within the hidden sector.

The injection of photons and other high energy secondary particles is constrained by the measurement of the CMB by Planck \cite{Ade:2015xua}, the bounds set by the Fermi-LAT collaboration from DM searches in the Dwarf Spheroidal Galaxies of the Milky Way \cite{Drlica-Wagner:2015xua} and by measurements of the positron flux by AMS-02 \cite{Aguilar:2014mma, Accardo:2014lma}. As a consequence, constraints from the above experiments may provide an upper bound on the self-interaction strength of the pions. These cascade decays have been studied in Ref. \cite{Elor:2015bho}. We compute our limits using their bounds on $b\bar{b}$ final states, which should be rather accurate for all hadronic decays and correspond to $95\%$ confidence. Our constraints will be computed using an effective cross section defined by
\begin{equation}\label{eq:EffCS}
  \langle \sigma v \rangle_{\text{eff}} = \frac{1}{Y_\text{tot}^2}\left(\frac{\Omega_\text{tot}}{\Omega_\text{obs}}\right)^2 \sum_{\substack{P\in \mathcal{P}^{2\to 2} \\ \pi_{P_1}, \pi_{P_2} \text{ stable}}} \#\pi_{\text{unstable out}}^P \hat{Y}_{P_1}\hat{Y}_{P_2} \langle \sigma v\rangle^{T\to 0}_{P:\pi_{P_1}\pi_{P_2}\to \pi_{P_3} \pi_{P_4}},
\end{equation}
where $Y_{\mathrm{tot}}$ is the total number density of pions per entropy density, $\Omega_\text{tot}$ the total relic density, $\Omega_\text{obs}$ its observed value, $\mathcal{P}^{2\to 2}$ the set of all distinct $2\to2$ scattering between representations, $\hat{Y}_A$ the number density of pions from a representation $A$ per entropy density divided by the size of the representation and $\#\pi_{\text{unstable\,\,out}}^P$ the number of unstable pions in the out state of a process $P$. See Appendix~\ref{app:RDMAC} for more details. It is obtained by simply combining the rates of unstable pion production from all channels. This expression assumes the proportionality of local and global relative abundances, or so-called proportionality ansatz \cite{Bertone:2010rv, Bertone:2012fua, Anderhalden:2012qt, Bhattacharya:2013hva, Blennow:2015gta, Herrero-Garcia:2017vrl}, which should work to great approximation in the mass range we study. It is also possible to set limits studying the indirect detection signals from the annihilation of two dark pions directly to SM particles, although they are typically weaker than bounds from direct detection if we assume rather weak interactions between the SM and the dark sector~\cite{Beauchesne:2018myj}.  

Finally, it is expected that the indirect detection bounds will change in the near future. The lack of a signal from Planck, Fermi-LAT and AMS-02 will lead to tighter limits, but not change the results qualitatively \cite{Elor:2015bho}. An improvement of roughly one order of magnitude for the Fermi-LAT bounds will be possible depending on new dwarf spheroidal galaxy discovery \cite{Charles:2016pgz}. Finally, future more powerful instruments, such as GAMMA-400 \cite{Galper:2013sfa,Cumani:2015ava}, HERD (High Energy cosmic Radiation Detection) \cite{Zhang:2014qga,Huang:2015fca} or CTA (Cherenkov Telescope Array) \cite{Consortium:2010bc, Doro:2012xx, Silverwood:2014yza} may set even stronger limits at different mass scales.

\subsection{Other constraints}\label{sSec:OtherConstraints}
In our convention, an approximate perturbative unitarity limit of $m_\pi/f\lesssim 4 \pi$ can be set. As a consequence, as we will see in Sec.~\ref{sec:DecayLengths}, there will be an upper bound on the decay lengths of unstable pions. This bound will be the stronger limit in benchmarks where the effective thermal averaged cross section is suppressed or vanishing (category II and III). However, this bound should be taken more as a rough estimate, as chiral perturbation theory breaks down around this limit and additional resonances must be taken into account.

Furthermore, the decay of the unstable pions may disturb Big Bang Nucleosynthesis (BBN). In fact, if the unstable pions decay hadronically, they will inject relatively long-lived hadrons that can modify the ratio of protons to neutrons. As a consequence we must require an upper bound on the decay time of about $0.1$ second for a $95\%$ confidence level \cite{Kawasaki:2004qu,Jedamzik:2006xz,Jedamzik:2009uy}. It turns out, however, that the constraints from BBN are always weaker than the constraints from indirect detection or unitarity. Also, bounds from dark matter self-interaction \cite{Clowe:2003tk, Markevitch:2003at, Randall:2007ph, Rocha:2012jg, Peter:2012jh} are overshadowed by the other constraints.

Finally, the assumption that the annihilation of pions directly to SM particles is negligible compared to the annihilations between pions makes direct detection constraints irrelevant. 

\section{Limits on mass splitting and decay lengths}\label{sec:DecayLengths}
In this section, we present limits on decay lengths as a function of the mass splitting for all three possible structure categories. Of course, it is impossible to consider every pattern of symmetry breaking for every category. We will therefore present a series of benchmarks that illustrate general features. These features should apply roughly to all models within a given category.

\subsection{Limits on decay lengths for category I}\label{sSec:DLcatI}
Category I is characterized by all stable pions being able to contribute to the indirect detection signal. This simplifies the analysis, but also means that bounds from indirect detection are very strong.

\subsubsection*{One unstable pion}
We begin by considering patterns with only one unstable pion. The pattern CIa is a good example of this. This benchmark is characterized by five parameters:
\begin{equation}\label{eq:ParametersCIa}
  \frac{m_3}{m_1}, m_{\pi_{\mathbf{3}}^s}, \Gamma_{\pi_1^d}, f, N_c,
\end{equation}
where $m_{\pi_{\mathbf{3}}^s}$ is the mass of the triplet and $\Gamma_{\pi_1^d}$ the decay width of the unstable pion. We will fix $f$ by requiring that the abundance of the stable dark pions corresponds to the currently observed dark matter abundance. We set $N_c$ to 3, as we will do for all $SU(N)$ cases in this paper. The number of colors $N_c$ will be set to 4 for all $Sp(2N)$ and $SO(2N)$ cases.\footnote{The dark quarks are assumed to be in the fundamental representation of an $SU(N_c)$ confining group for complex representations, the fundamental of $Sp(N_c)$ with $N_c$ even for pseudoreal representations and the vectorial representation of $O(N_c)$ for real representations. See Appendix~\ref{app:RDMAC} for more details.} This parameter is only relevant for the $3\to 2$ processes, which generally do not change qualitatively the final abundances. We will also swap the decay width for the decay length for practicality. This conveniently leaves us with only three parameters.

\begin{figure}[t!]
  \centering
   \captionsetup{justification=centering}
  \begin{subfigure}{0.48\textwidth}
    \centering
    \caption{$m_{\pi_{\mathbf{3}}^s}=25$ GeV}
    \includegraphics[width=\textwidth]{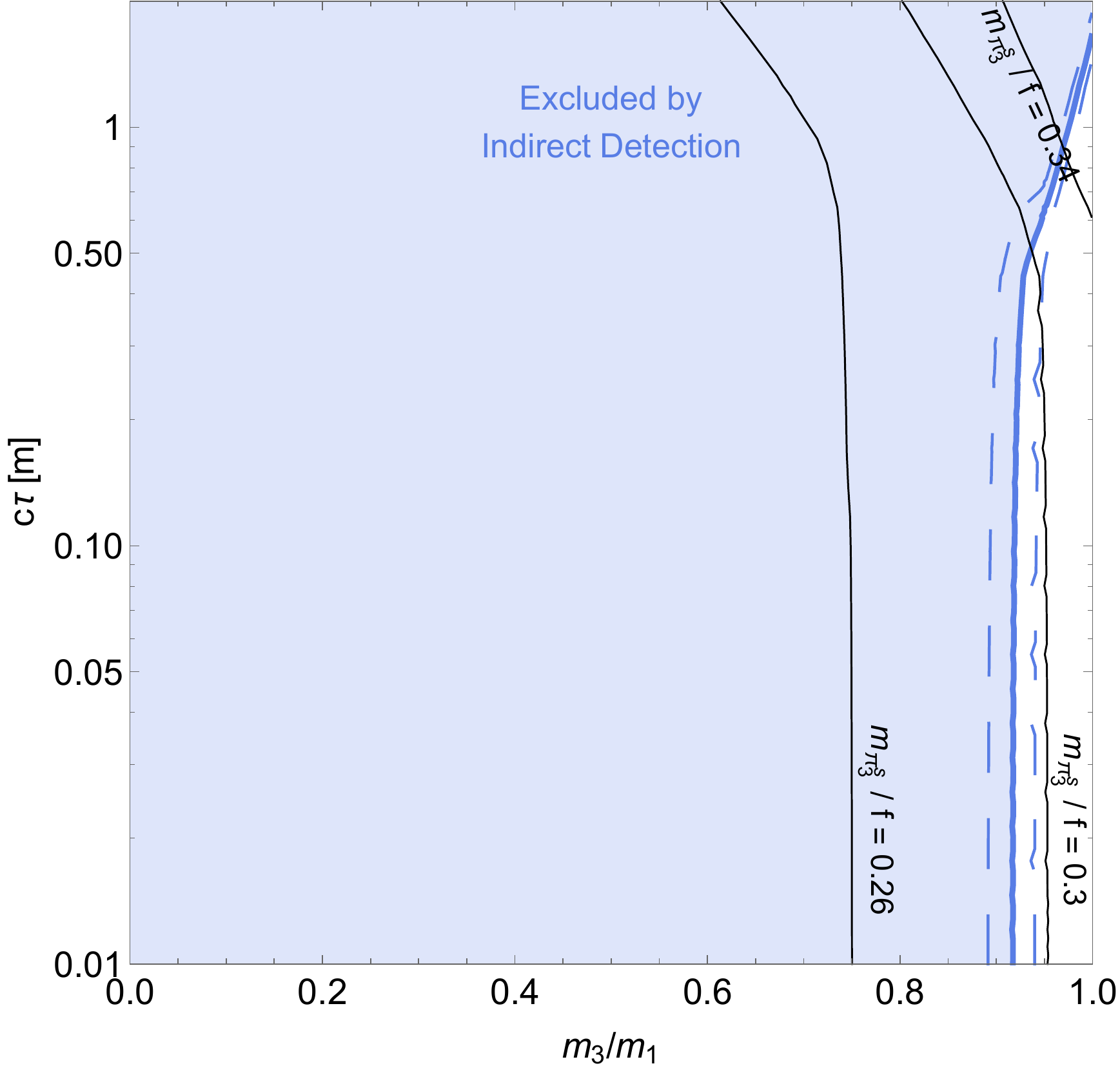}
    \label{fig:CIa25}
  \end{subfigure}
  ~
  \begin{subfigure}{0.48\textwidth}
    \centering
    \caption{$m_{\pi_{\mathbf{3}}^s}=50$ GeV}
    \includegraphics[width=\textwidth]{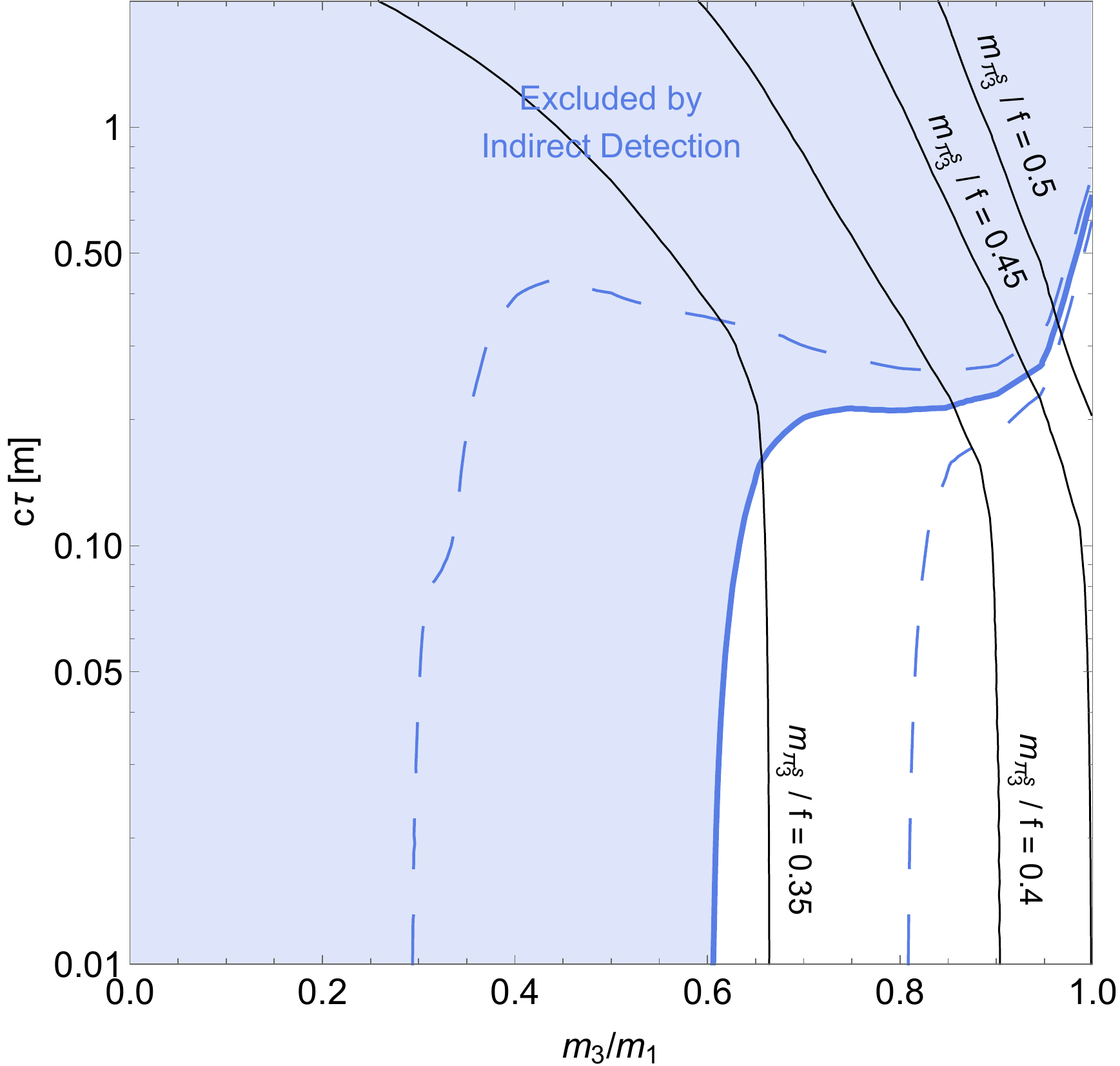}
    \label{fig:CIa50}
  \end{subfigure}
  ~
  \begin{subfigure}{0.48\textwidth}
    \centering
    \caption{$m_{\pi_{\mathbf{3}}^s}=75$ GeV}
    \includegraphics[width=\textwidth]{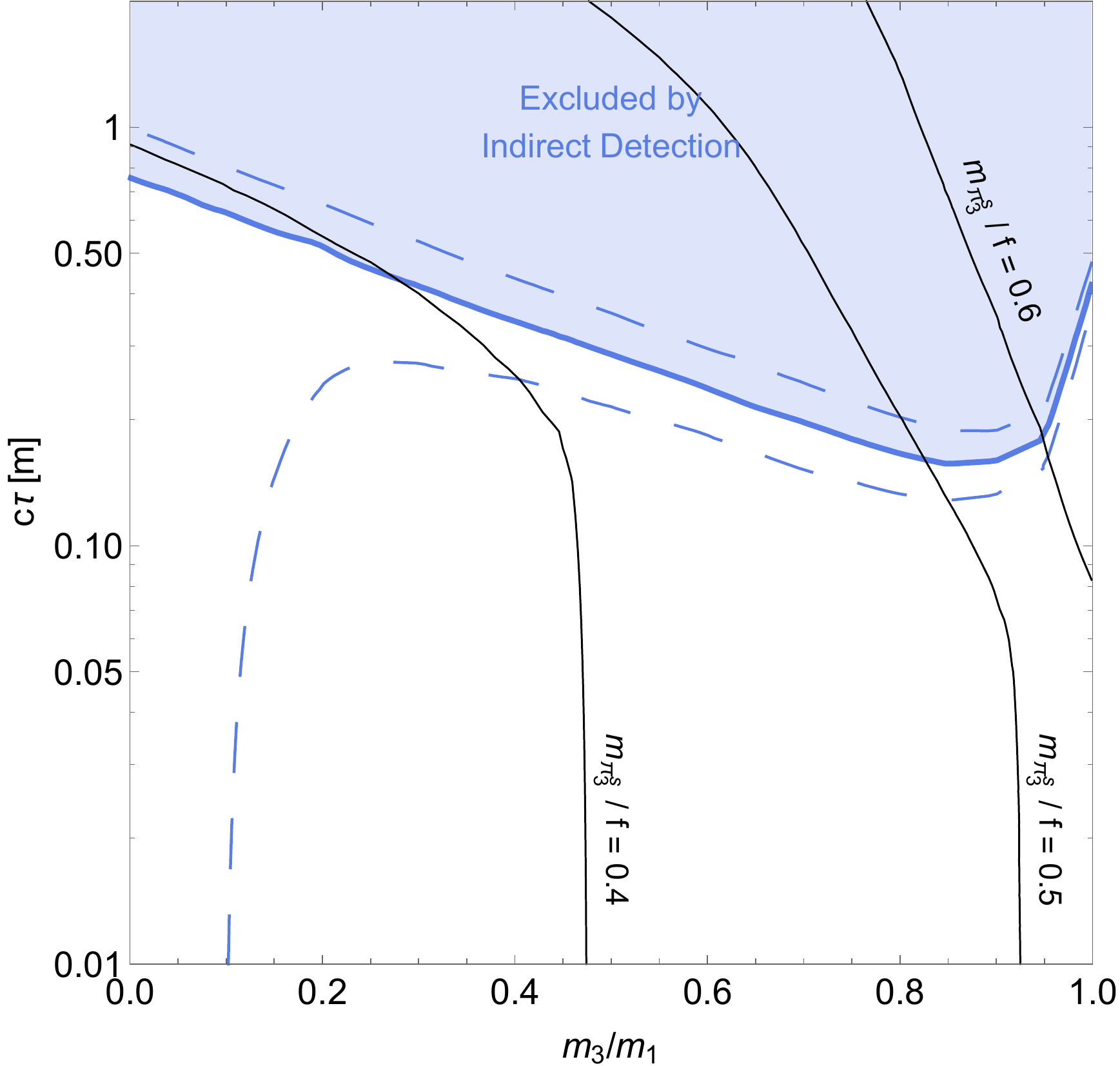}
    \label{fig:CIa75}
  \end{subfigure}
  \begin{subfigure}{0.48\textwidth}
    \centering
    \caption{$m_{\pi_{\mathbf{3}}^s}=100$ GeV}
    \includegraphics[width=\textwidth]{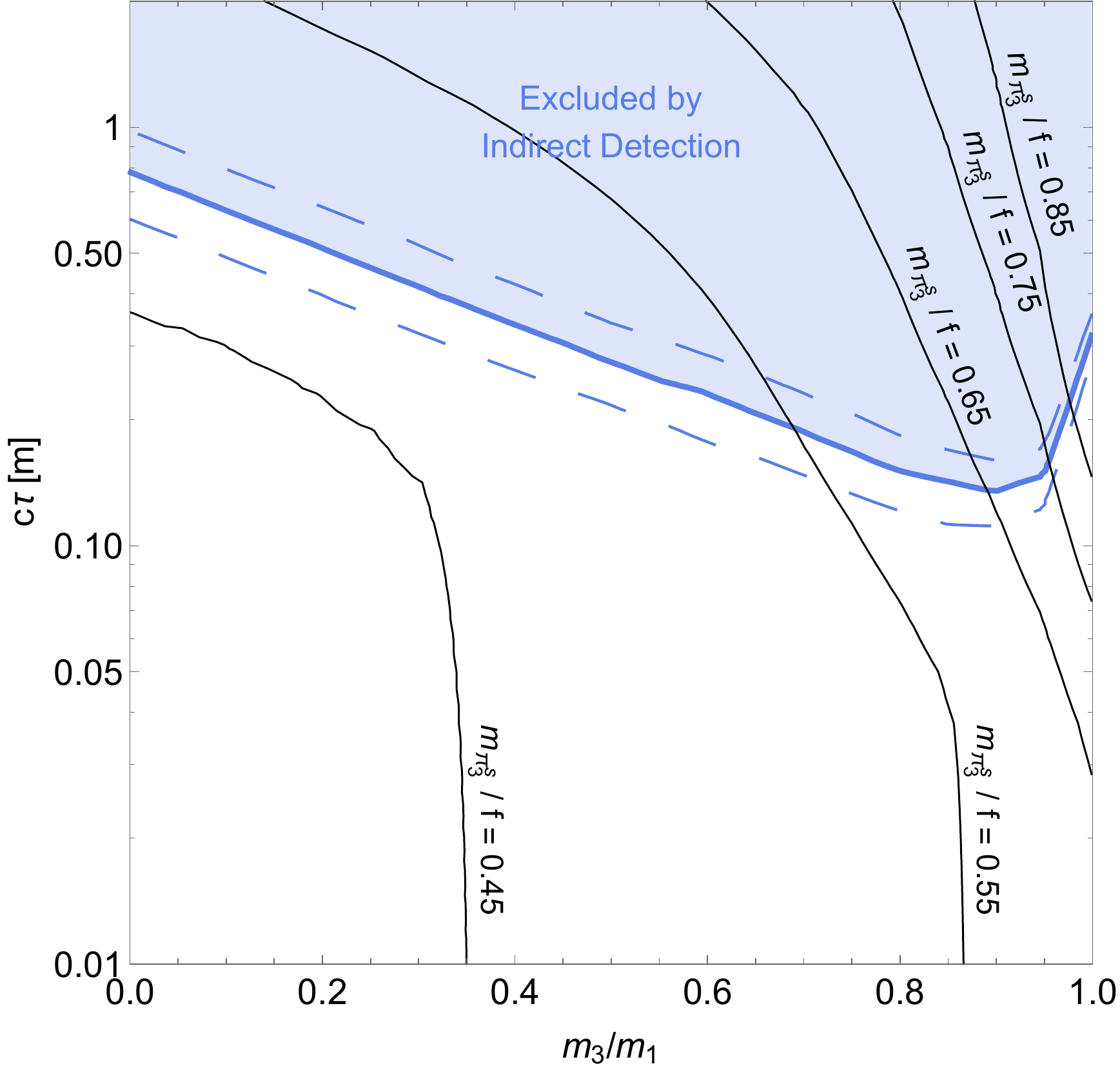}
    \label{fig:CIa100}
  \end{subfigure}
  \captionsetup{justification=justified}
\caption{Allowed region of parameter space for the benchmark CIa for different masses of the triplet. The black contour lines represent constant values of $m_{\pi_{\mathbf{3}}^s}/f$. The blue region is excluded by indirect detection dark matter searches. The dashed blue lines correspond to what the exclusion limit would be if the limit on the indirect detection effective cross section were to vary by $\pm 10\%$.}\label{fig:CIa}
\end{figure}

\begin{figure}[t!]
  \centering
   \captionsetup{justification=centering}
  \begin{subfigure}{0.48\textwidth}
    \centering
    \caption{$SU(4)\times SU(4) \to SU(4) \to SU(3) \times U(1)$}
    \includegraphics[width=\textwidth]{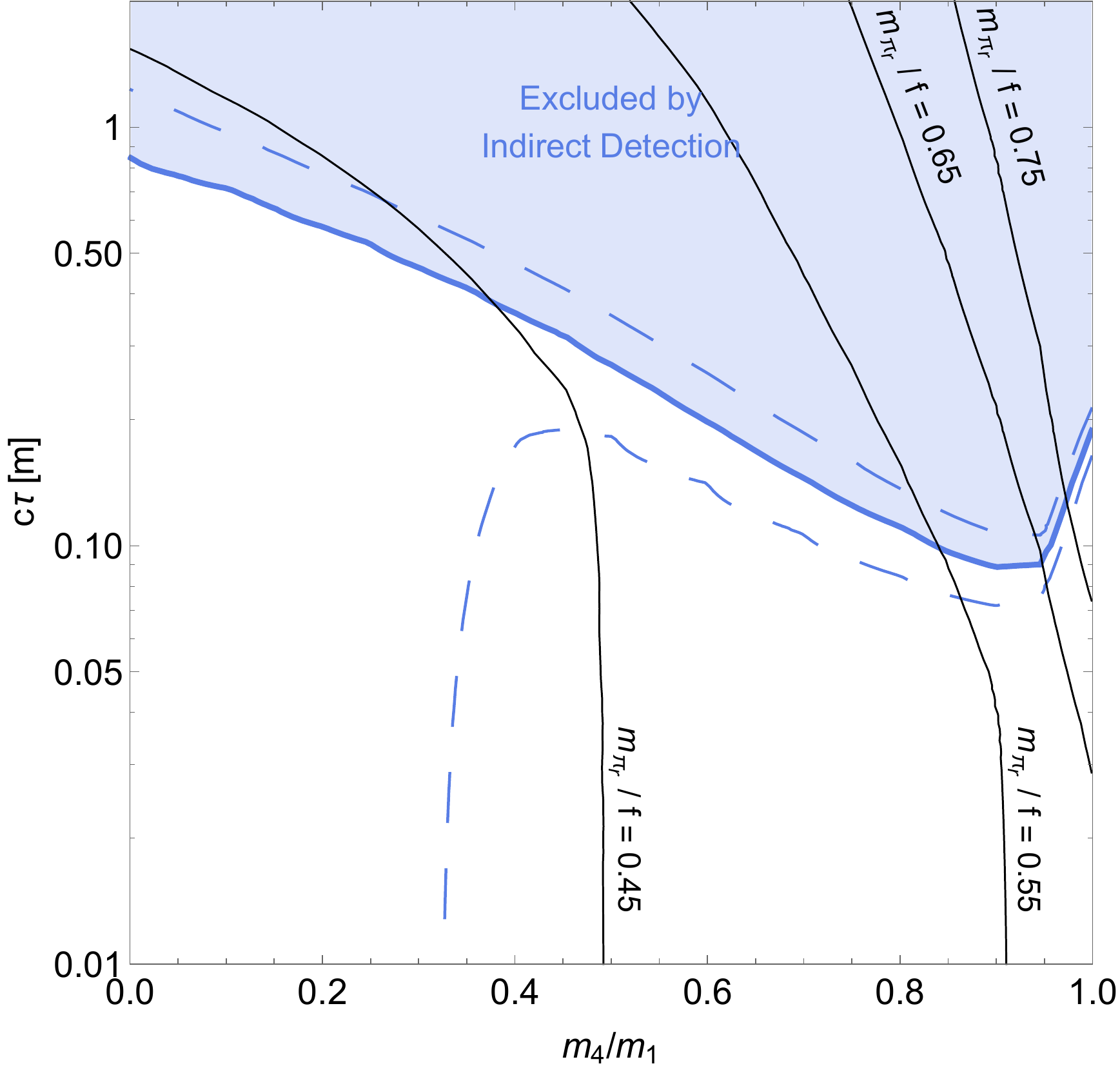}
    \label{fig:CatIa}
  \end{subfigure}
  ~
  \begin{subfigure}{0.48\textwidth}
    \centering
    \caption{$SU(6)\to Sp(6) \to Sp(4) \times U(1)$}
    \includegraphics[width=\textwidth]{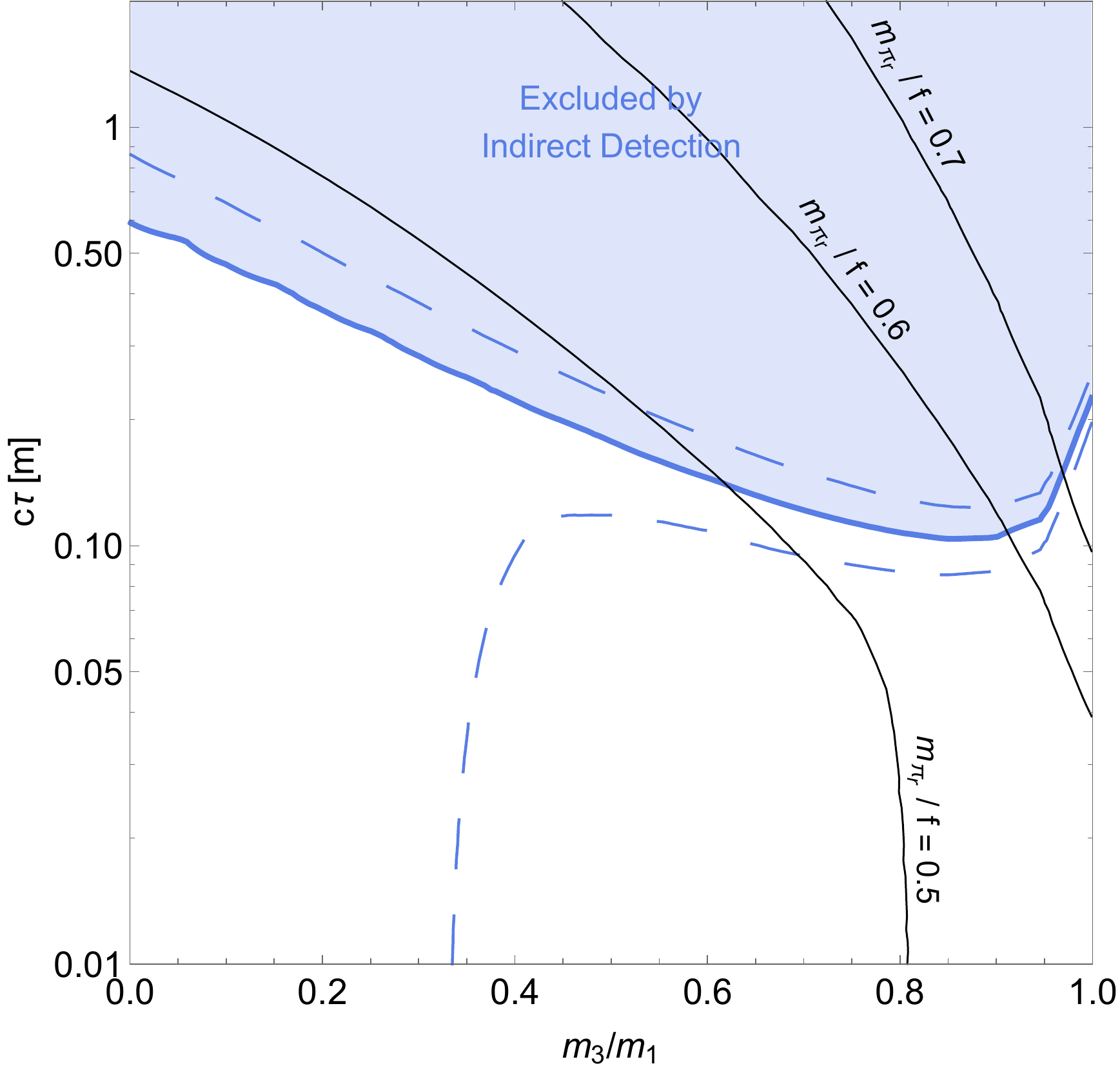}
    \label{fig:CatIb}
  \end{subfigure}
  ~
  \begin{subfigure}{0.48\textwidth}
    \centering
    \caption{$SU(5)\times SU(5) \to SU(5) \to SU(4) \times U(1)$}
    \includegraphics[width=\textwidth]{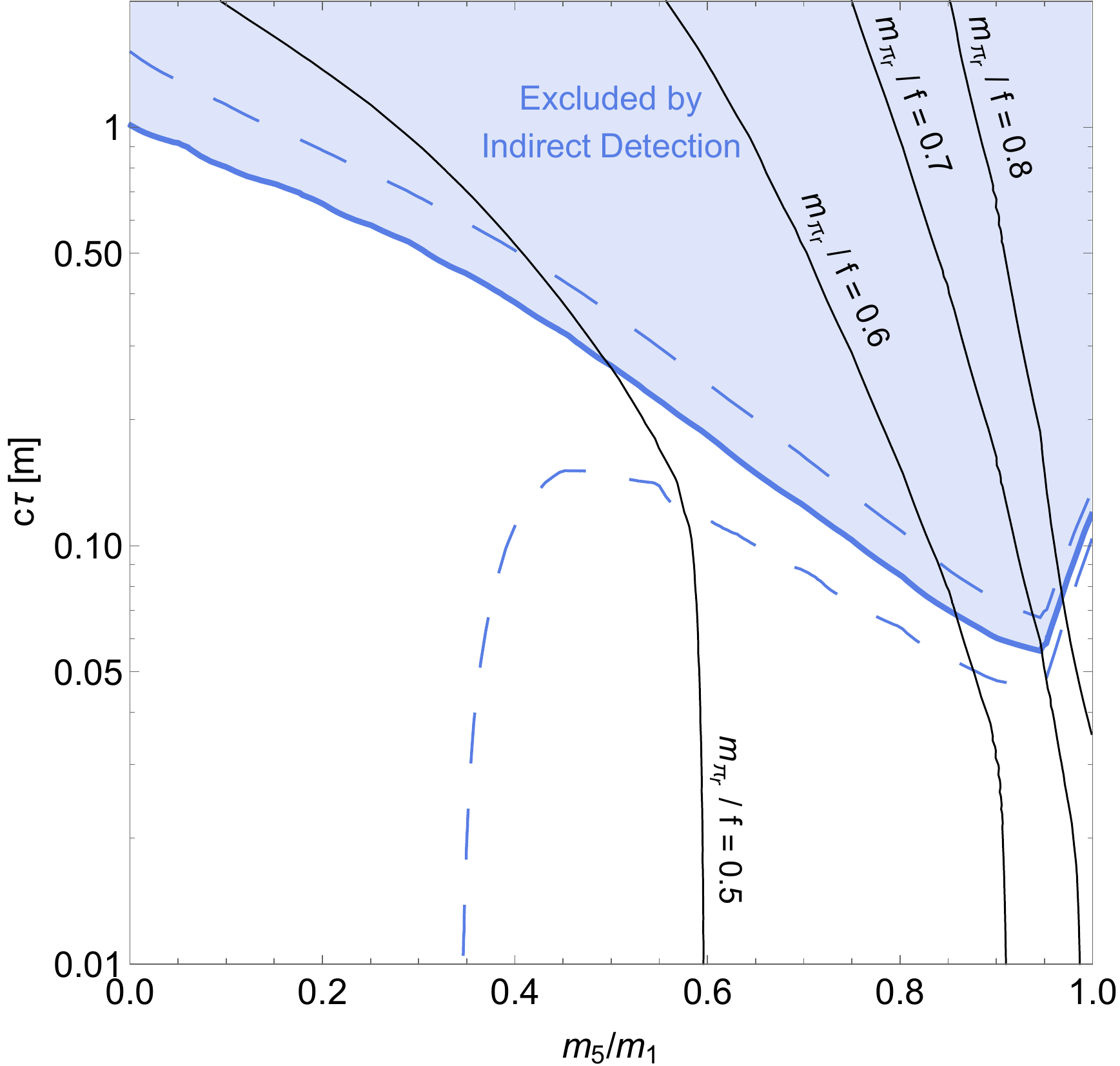}
    \label{fig:CatIc}
  \end{subfigure}
  \begin{subfigure}{0.48\textwidth}
    \centering
    \caption{$SU(8)\to Sp(8) \to Sp(6) \times U(1)$}
    \includegraphics[width=\textwidth]{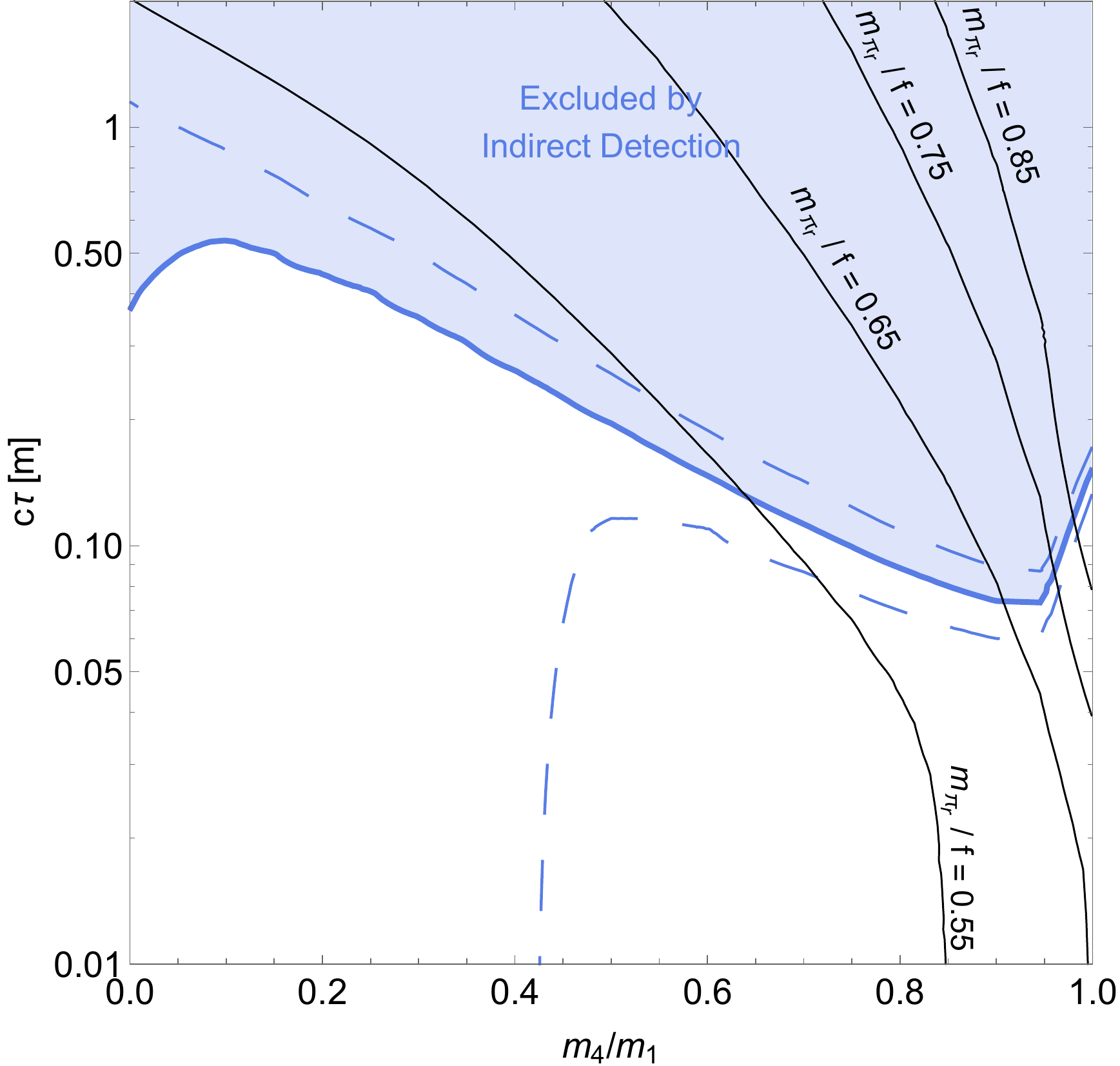}
    \label{fig:CatId}
  \end{subfigure}
  \captionsetup{justification=justified}
\caption{Allowed region of parameter space for different benchmarks of category I. For (a), (b), (c) and (d), the masses of the $(\mathbf{8}, 0)$ of $SU(3) \times U(1)$, the $(\mathbf{5}, 0)$ of $Sp(4) \times U(1)$, the $(\mathbf{15}, 0)$ of $SU(4) \times U(1)$ and the $(\mathbf{14}, 0)$ of $Sp(6) \times U(1)$ are respectively set to 75~GeV. The black contour lines represent constant values of $m_{\pi_r}/f$, where $m_{\pi_r}$ is simply the representation mass kept constant, i.e. 75~GeV. The blue region is excluded by indirect detection dark matter searches. The dashed blue lines correspond to what the exclusion limit would be if the limit on the indirect detection effective cross section were to vary by~$\pm 10\%$.}\label{fig:CatIOther}
\end{figure}

We show in Fig.~\ref{fig:CIa} the allowed parameter space for this benchmark as a function of $m_3/m_1$ and the decay length of $\pi_1^d$ for different fixed $m_{\pi_{\mathbf{3}}^s}$. The figure also contains contours of constant $m_{\pi_{\mathbf{3}}^s}/f$. At small decay length, these are vertical as the system is in the coupling-independent regime. This ratio starts to increase abruptly once the system enters the coupling independent regime, as $f$ must be made smaller to compensate for a too small decay width. Eventually, the pions are required to interact too much with each other and the system becomes incompatible with the limits from indirect detection. As can be seen, the region of $m_3/m_1$ close to 1 is always unexcluded. This is simply because in this limit the $s$-component of the effective cross section goes to zero. As the mass of the triplet increases, the allowed region of parameter space increases. Do note that a large part of the coupling-independent region of parameter space is very close to the current experimental limit from indirect detection. To illustrate this, we include blue dashed curves that indicate what the limits from indirect detection would be if the limit on the effective cross section were to change by $\pm 10\%$. Obviously, improving the bounds from indirect detection would massively reduce the amount of parameter space available. One can see that the ratio $m_{\pi_{\mathbf{3}}^s}/f$ is forced to be of $\mathcal{O}(0.1)$ to $\mathcal{O}(1.0)$ for pions masses between 25 and 100 GeV. In addition, unless the quarks are degenerate in mass, there is an upper limit on the decay length of the pions of the $\mathcal{O}(10)$ cm to $\mathcal{O}(1)$ m. Additional examples are shown in Fig.~\ref{fig:CatIOther} for different breaking patterns, keeping one of the pions at 75 GeV.

\subsubsection*{Two unstable pions}
\begin{figure}[t!]
  \centering
   \captionsetup{justification=centering}
  \begin{subfigure}{0.48\textwidth}
    \centering
    \caption{$\Gamma_{\pi_1^m}=\Gamma_{\pi_2^m}$}
    \includegraphics[width=\textwidth]{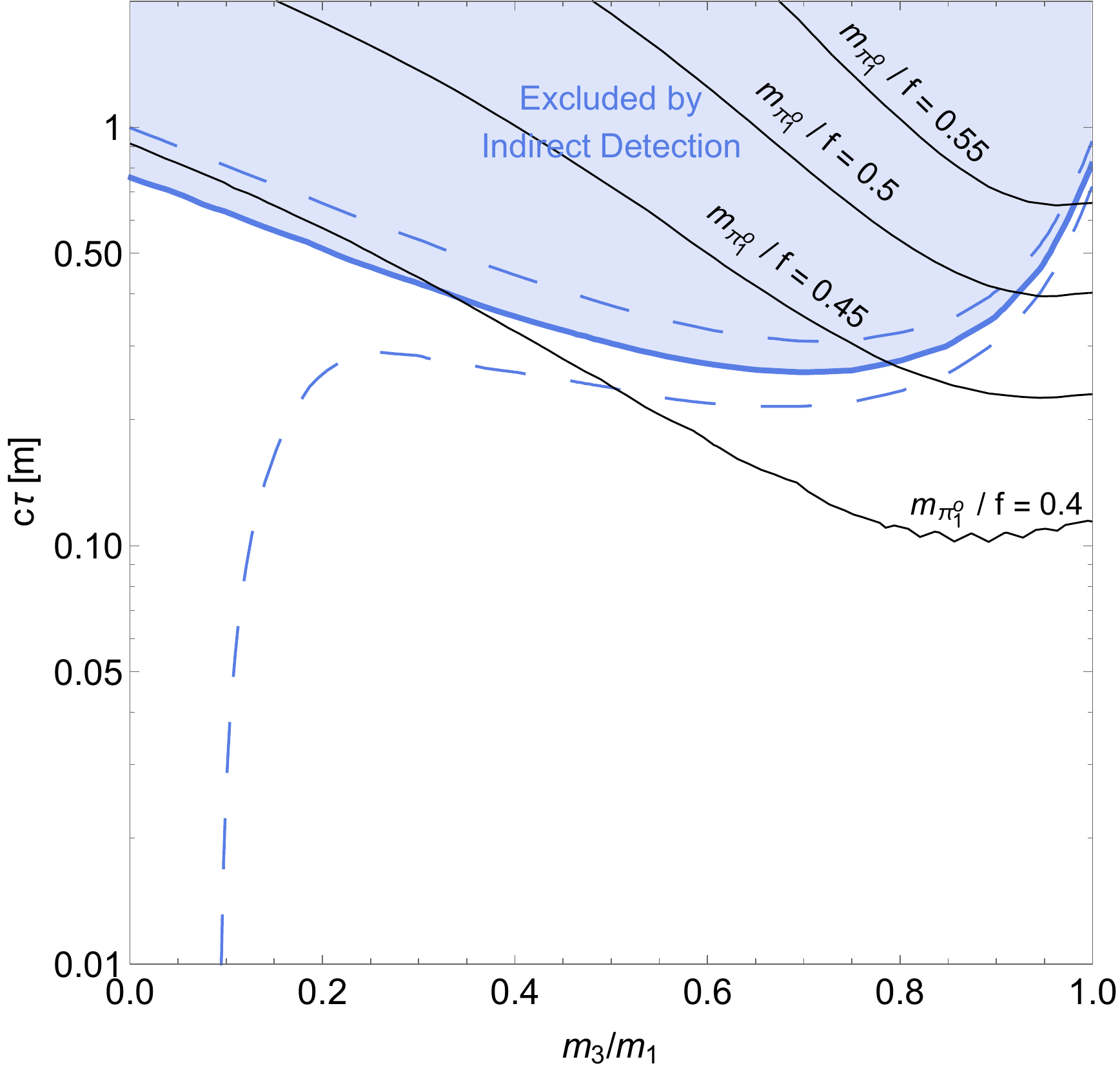}
    \label{fig:CatIe}
  \end{subfigure}
  ~
  \begin{subfigure}{0.48\textwidth}
    \centering
    \caption{$\Gamma_{\pi_1^m}=2\Gamma_{\pi_2^m}$}
    \includegraphics[width=\textwidth]{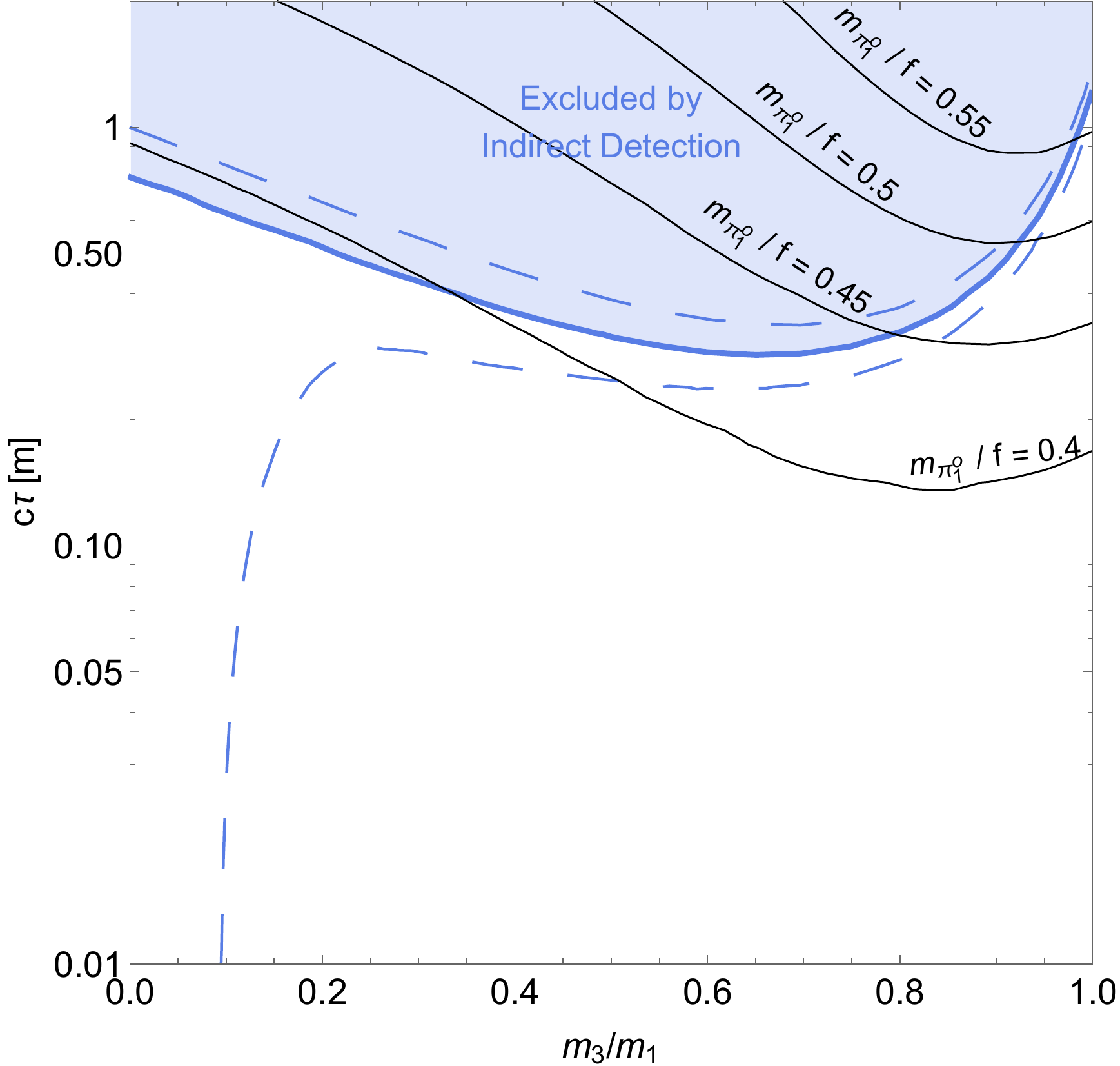}
    \label{fig:CatIf}
  \end{subfigure}
  \captionsetup{justification=justified}
\caption{Allowed region of parameter space for the benchmark CIc for different ratios of the decay widths. The mass of $\pi^o_1$ is kept at a constant value of 75~GeV. The contour lines represent constant values of $m_{\pi^o_1}/f$. The blue region is excluded by indirect detection dark matter searches. The dashed blue lines correspond to what the exclusion limit would be if the limit on the indirect detection effective cross section were to vary by~$\pm 10\%$. The decay length is that of the lightest unstable pion.}\label{fig:CatI2Unstable}
\end{figure}

We then consider patterns with two unstable pions. The pattern CIc is a good example of this. This benchmark is characterized by seven parameters (see Appendix~\ref{app:SBB} for notation):
\begin{equation}\label{eq:ParametersCIb}
  \frac{m_3}{m_1}, \frac{m_2}{m_1}, m_{\pi_1^o}, \Gamma_{\pi_1^m}, \Gamma_{\pi_2^m}, f, N_c.
\end{equation}
The parameters $f$ and $N_c$ are set as before, which leaves us with five parameters. For illustration purposes, we set $m_2/m_1$ to 1 and assume that $\Gamma_{\pi_1^m}$ and $\Gamma_{\pi_2^m}$ are proportional, leaving three parameters. The results are shown in Fig.~\ref{fig:CatI2Unstable}. At $m_3/m_1$ close to 1, the upper limit on the decay length of the lightest unstable pion is raised in comparison to benchmark CIa. This is simply because the presence of another unstable pion means that the lightest one doesn't need to decay as fast. As $m_3/m_1$ approaches 0, the difference between these two cases vanishes as the heaviest unstable pion is simply not present in sufficient quantity to affect the relic abundances.

\subsection{Limits on decay lengths for category II}\label{sSec:DLcatII}
Category II is characterized by only a subset of the stable pions being able to contribute to indirect detection. This generally results in suppressed bounds from indirect detection.

\subsubsection*{One unstable pion}
\begin{figure}[t!]
  \centering
   \captionsetup{justification=centering}
  \begin{subfigure}{0.48\textwidth}
    \centering
    \caption{$m_{\pi_{\mathbf{3}}^s}=25$ GeV}
    \includegraphics[width=\textwidth]{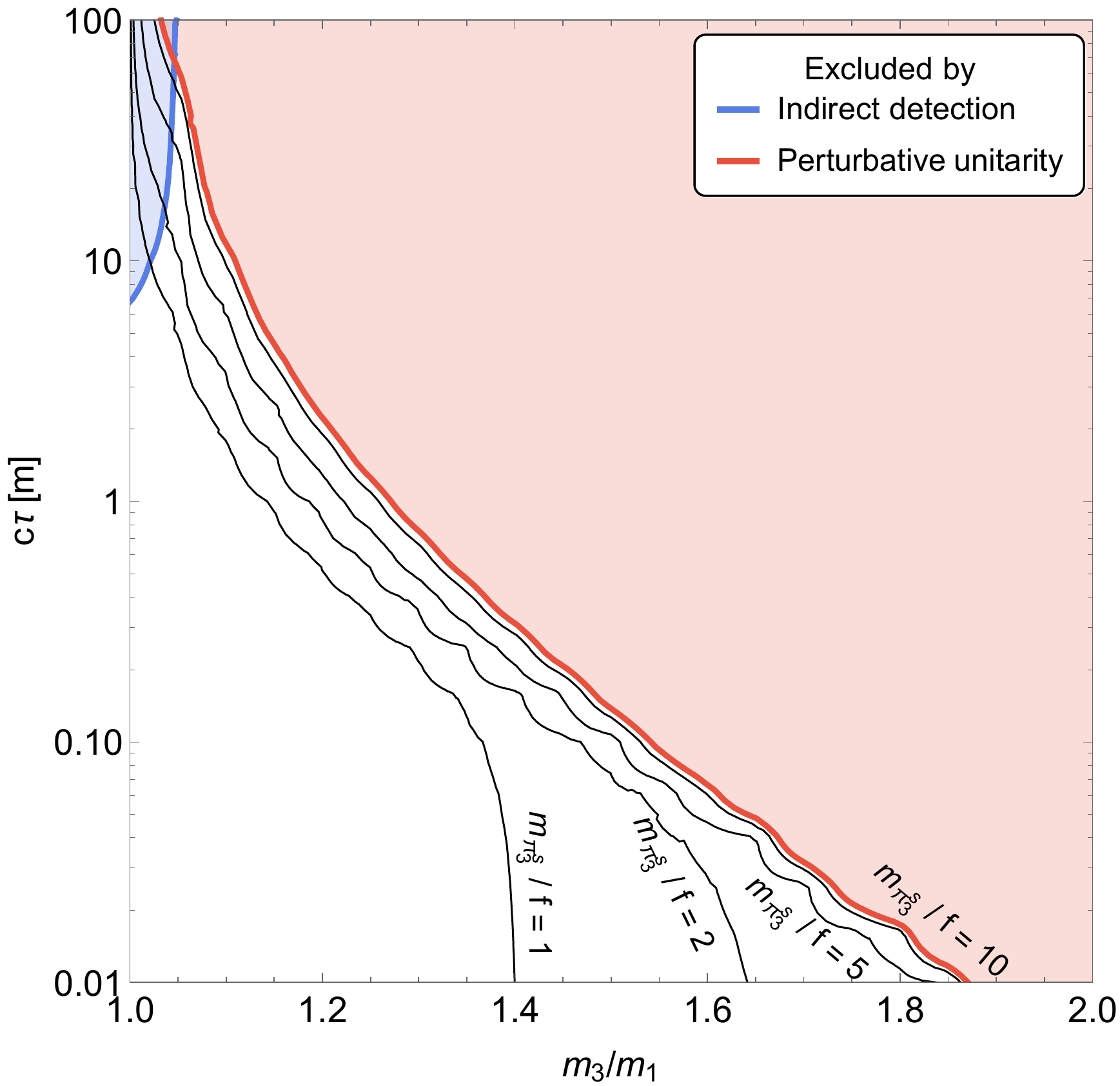}
    \label{fig:CIIb25}
  \end{subfigure}
  ~
  \begin{subfigure}{0.48\textwidth}
    \centering
    \caption{$m_{\pi_{\mathbf{3}}^s}=50$ GeV}
    \includegraphics[width=\textwidth]{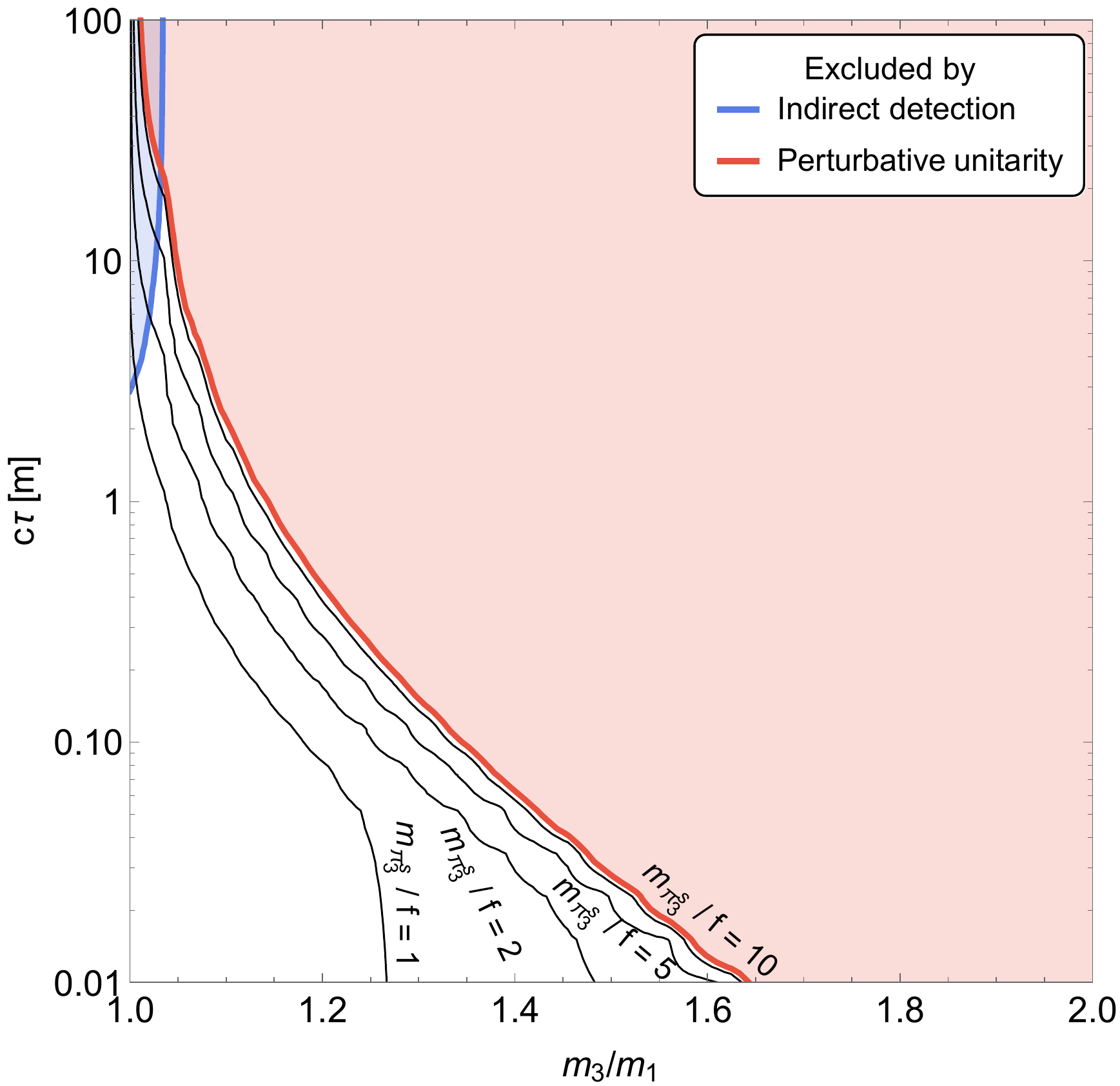}
    \label{fig:CIIb50}
  \end{subfigure}
  ~
  \begin{subfigure}{0.48\textwidth}
    \centering
    \caption{$m_{\pi_{\mathbf{3}}^s}=75$ GeV}
    \includegraphics[width=\textwidth]{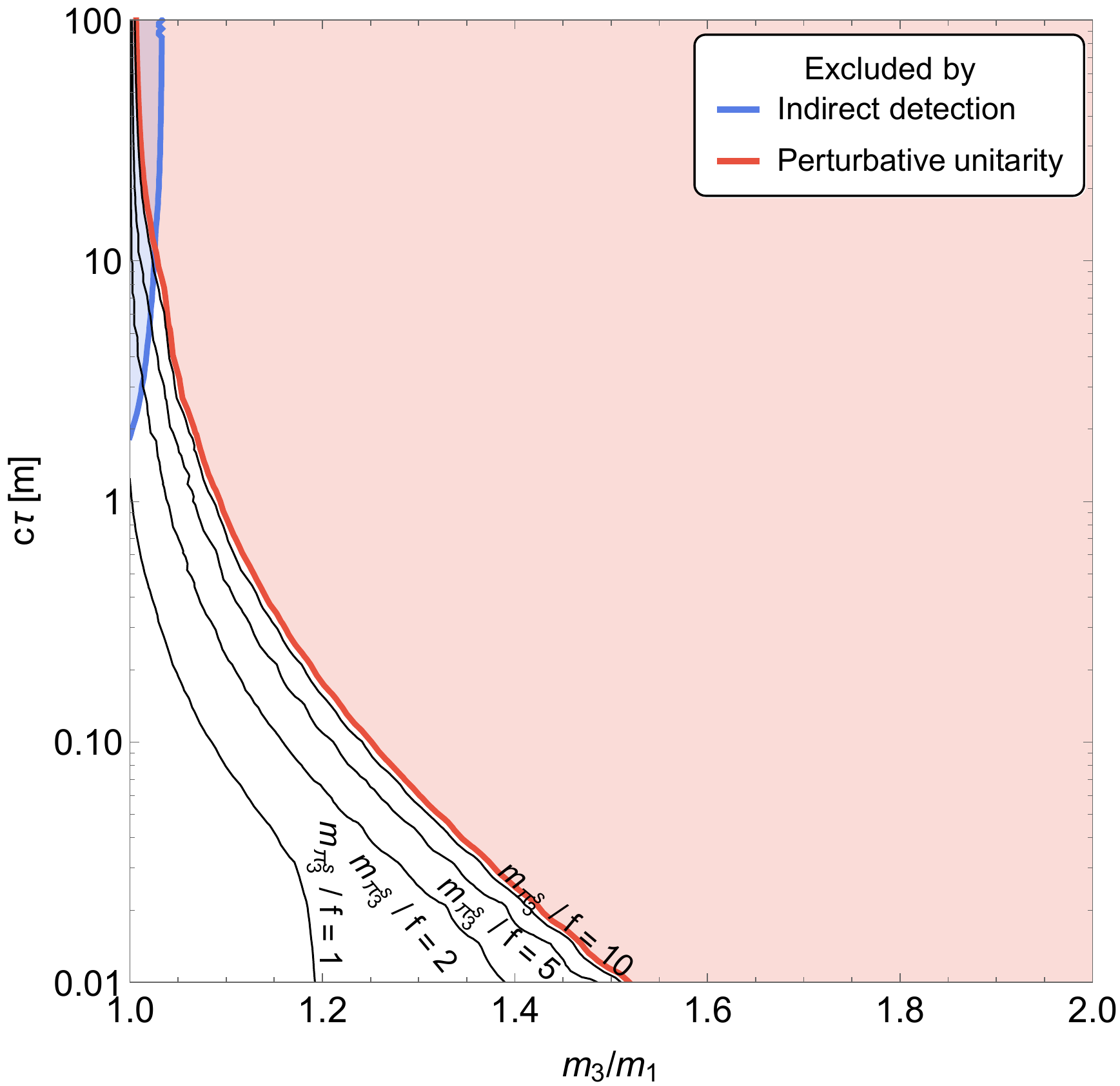}
    \label{fig:CIIb75}
  \end{subfigure}
  \begin{subfigure}{0.48\textwidth}
    \centering
    \caption{$m_{\pi_{\mathbf{3}}^s}=100$ GeV}
    \includegraphics[width=\textwidth]{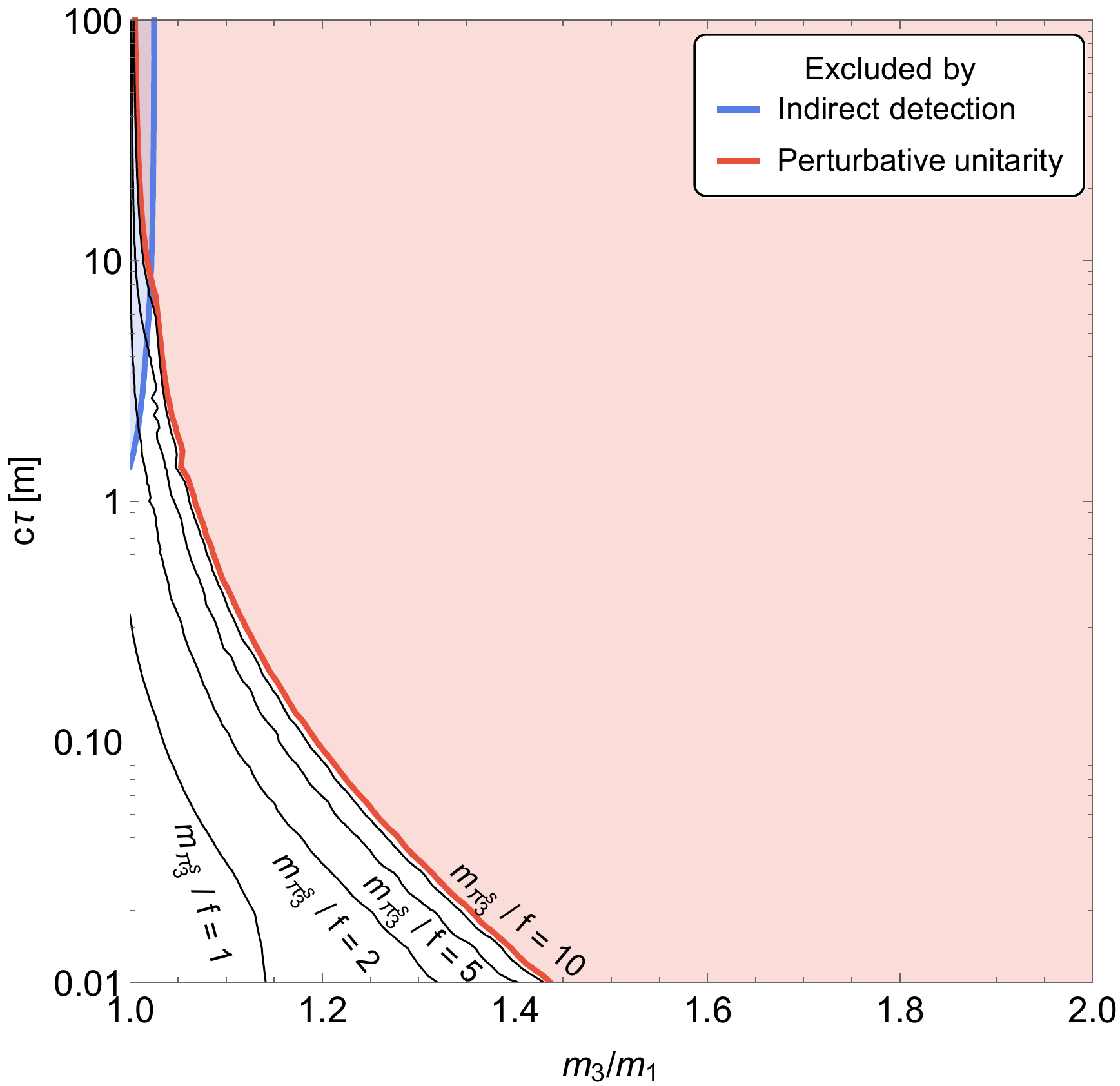}
    \label{fig:CIIb100}
  \end{subfigure}
  \captionsetup{justification=justified}
\caption{Allowed region of parameter space for the benchmark CIIb for different masses of the triplet. The black contour lines represent constant values of $m_{\pi_{\mathbf{3}}^s}/f$. The blue region is excluded by indirect detection DM searches, while the red one by perturbative unitarity constraints.}\label{fig:CIIb}
\end{figure}

\begin{figure}[t!]
  \centering
   \captionsetup{justification=centering}
  \begin{subfigure}{0.48\textwidth}
    \centering
    \caption{$SU(4)\times SU(4) \to SU(4) \to SU(3) \times U(1)$}
    \includegraphics[width=\textwidth]{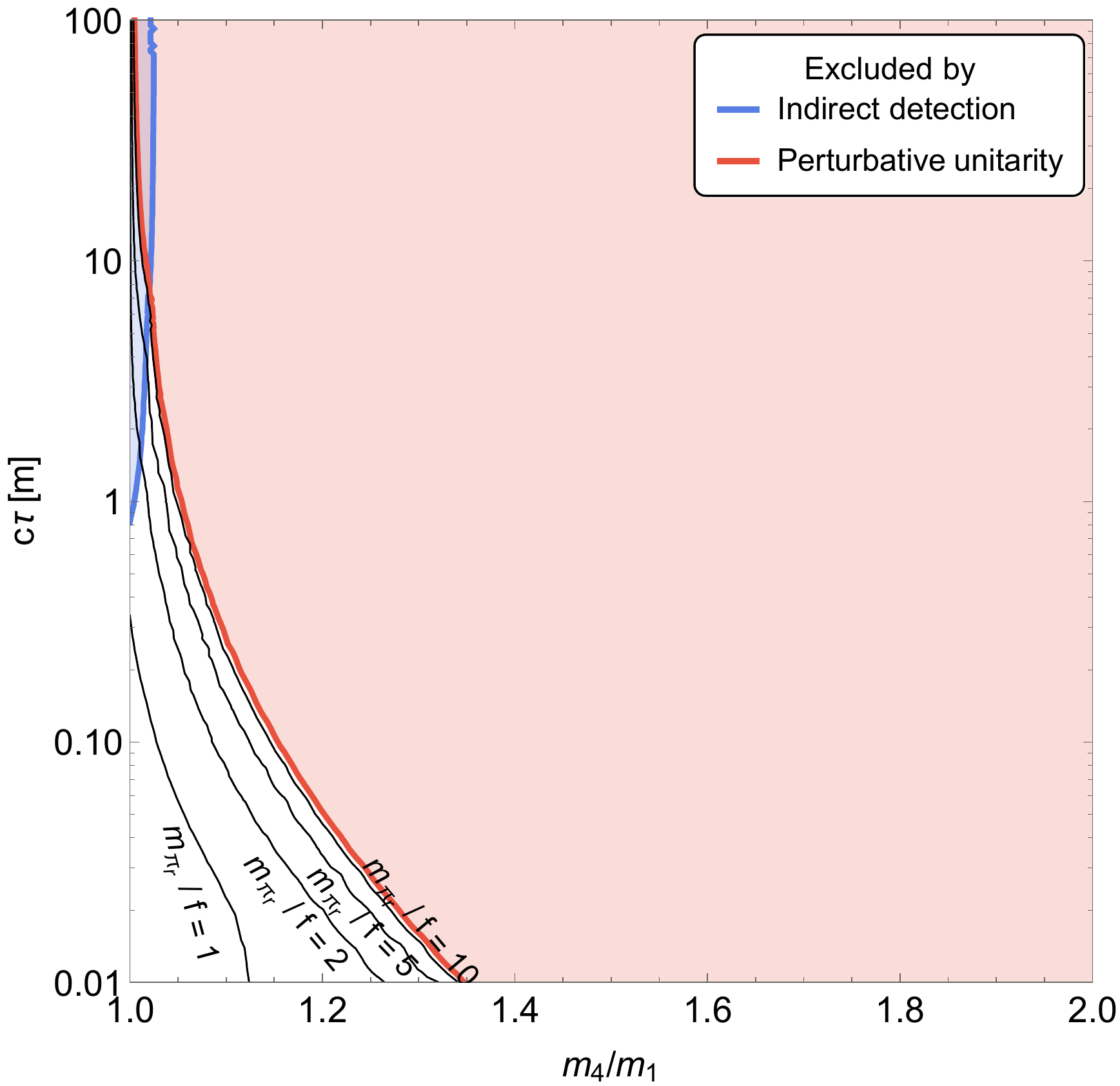}
    \label{fig:CatIIa}
  \end{subfigure}
  ~
  \begin{subfigure}{0.48\textwidth}
    \centering
    \caption{$SU(6)\to Sp(6) \to Sp(4) \times U(1)$}
    \includegraphics[width=\textwidth]{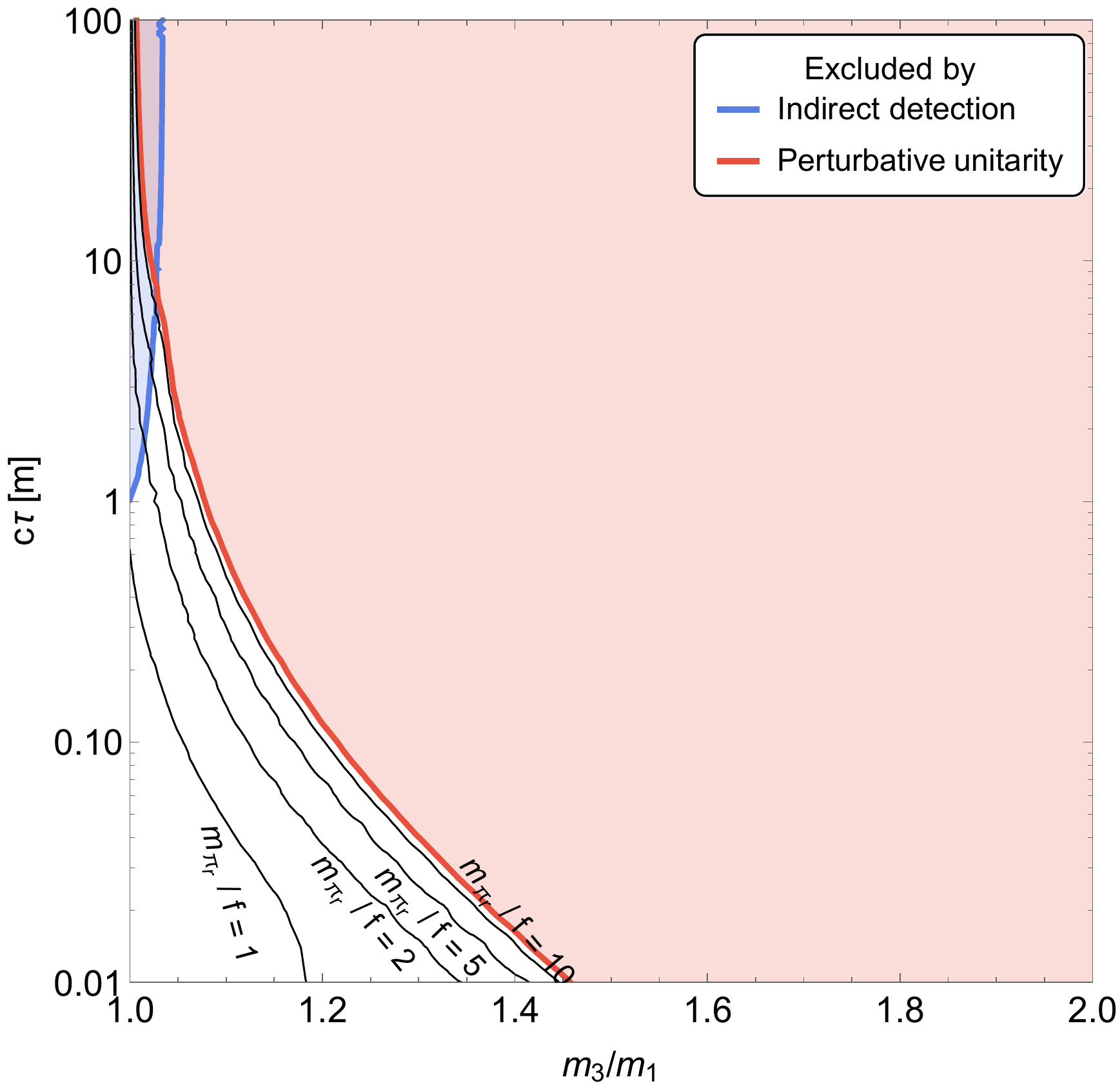}
    \label{fig:CatIIb}
  \end{subfigure}
  ~
  \begin{subfigure}{0.48\textwidth}
    \centering
    \caption{$SU(4) \to SO(4) \to U(1) \times U(1)$ \vspace{0.5cm}}
    \includegraphics[width=\textwidth]{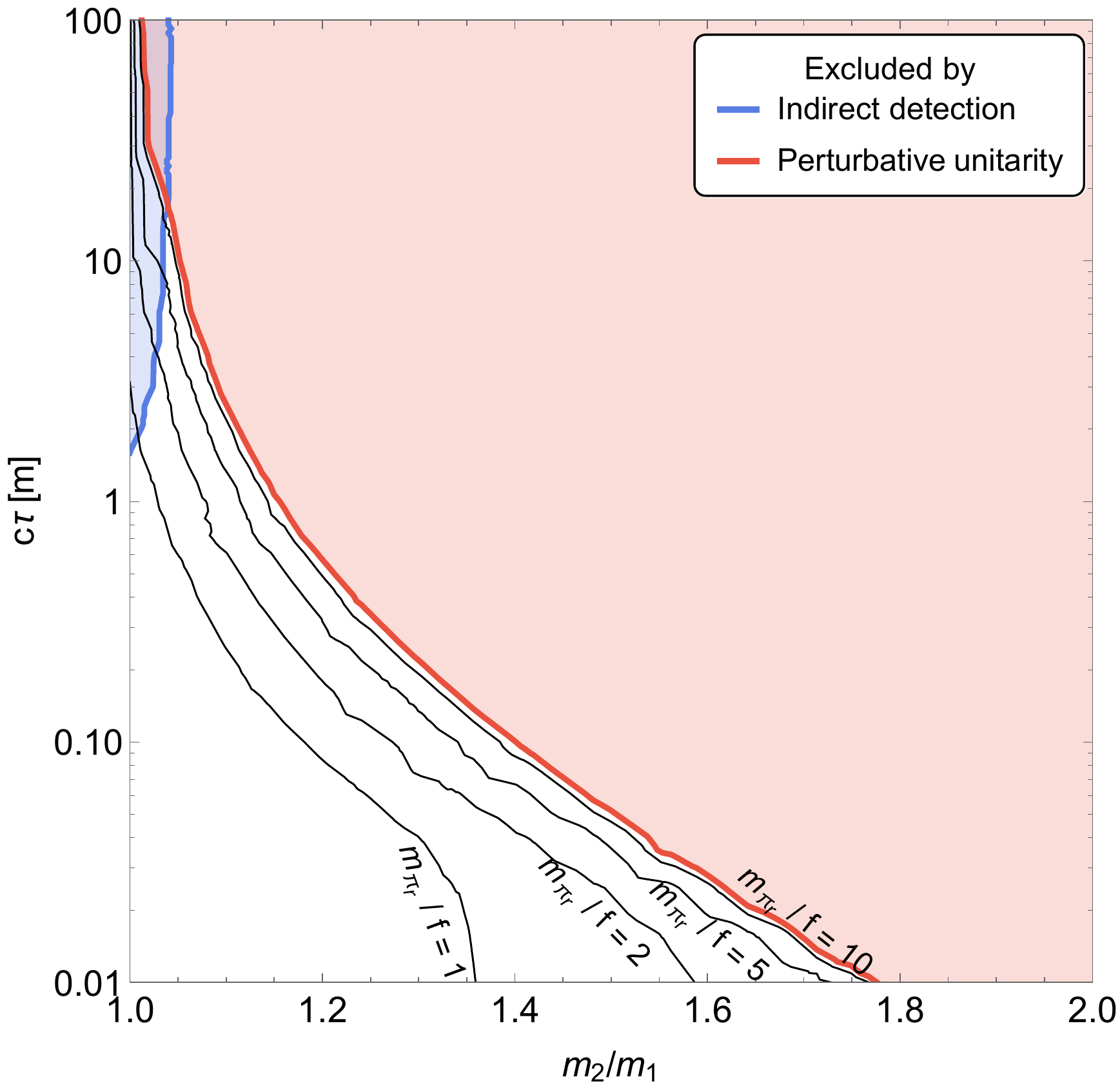}
    \label{fig:CatIIc}
  \end{subfigure}
  \begin{subfigure}{0.48\textwidth}
    \centering
    \caption{$SU(5) \times SU(5) \to SU(5) \to SU(3) \times SU(2) \times U(1)$}
    \includegraphics[width=\textwidth]{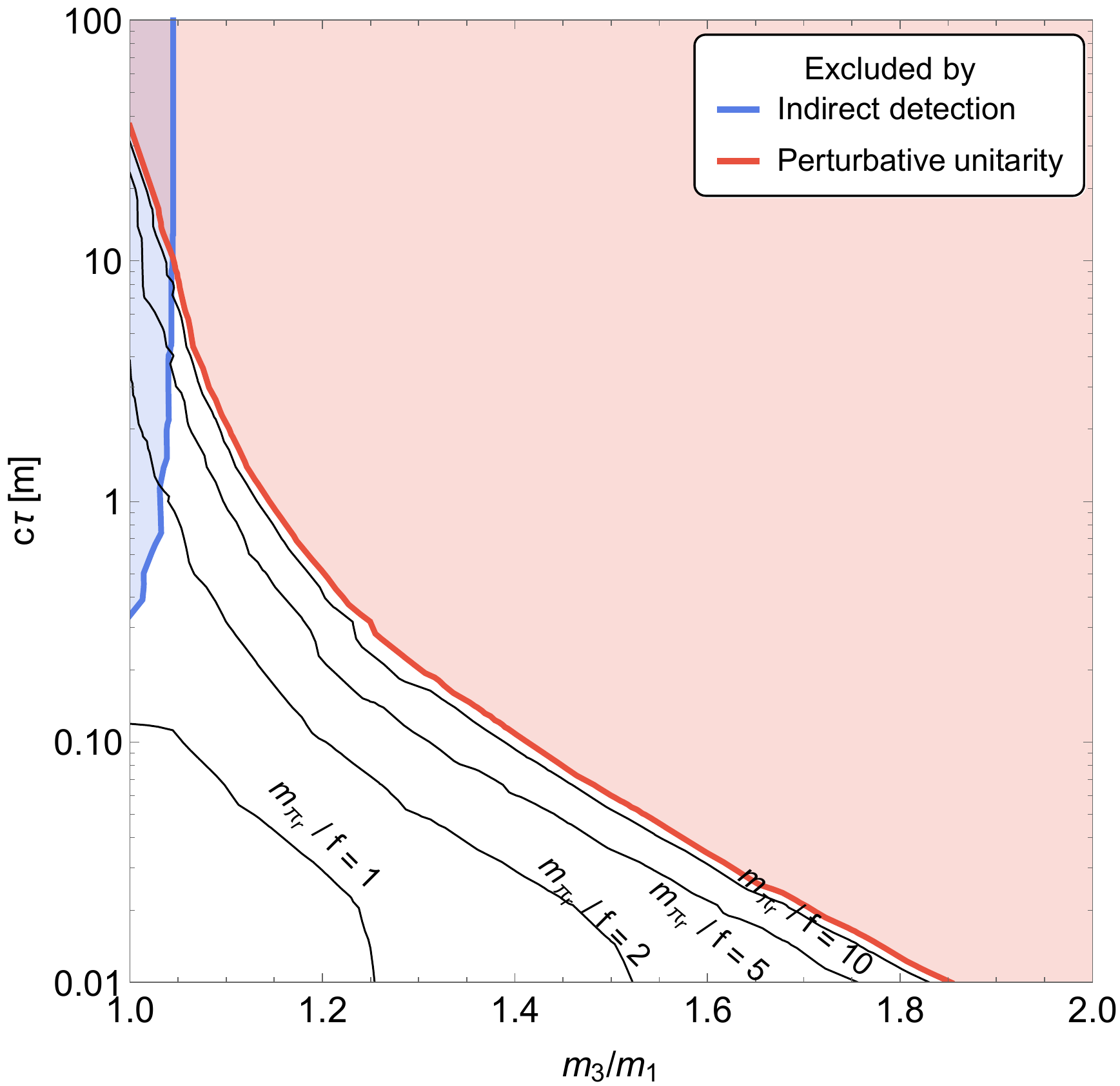}
    \label{fig:CatIId}
  \end{subfigure}
  \captionsetup{justification=justified}
\caption{Allowed region of parameter space for different benchmarks of category II. For (a), (b), (c) and (d), the masses of the $(\mathbf{8}, 0)$ of $SU(3)\times U(1)$, the $(\mathbf{5}, 0)$ of $Sp(4) \times U(1)$, the $(\pm 1, 0)$ of $U(1) \times U(1)$ and the $(\mathbf{8}, \mathbf{1}, 0)$ of $SU(3) \times SU(2) \times U(1)$ are respectively set to 75 GeV. The black contour lines represent constant values of $m_{\pi_r}/f$, where $m_{\pi_r}$ is simply the representation mass kept constant, i.e. 75 GeV. The blue region is excluded by indirect detection DM searches, while the red one by perturbative unitarity constraints.}\label{fig:CatIIOther}
\end{figure}

We begin this section by considering examples with only one unstable pion. Benchmark CIIb is a good example of this. As this is simply benchmark CIa with the mass hierarchy inverted, the same discussion about the parameter space applies. The constraints are shown in Fig.~\ref{fig:CIIb}. As can be seen, a larger decay width or a larger mass splitting must be compensated by a smaller pion decay constant, which eventually becomes incompatible with unitarity. The excluded corner in the upper left comes from indirect detection and the slowly dropping curve on the right from unitarity. The indirect detection signal is suppressed in most of the parameter space because a sizable mass splitting reduces the abundance of the doublets which in turn reduces the effective indirect detection cross section. There is an upper limit on the decay length of the unstable pion. For the range of pion masses we consider, it is of $\mathcal{O}(10)$ to $\mathcal{O}(100)$ m and decreases as the pion masses increase. However, such an upper limit is only possible close to a very specific value of $m_3/m_1$, which seemingly does not have any special theoretical justification. In contrast to category~I, a much larger range for the pion decay constant is possible. Between our benchmarks of 25 and 100~GeV, the $m_{\pi_{\mathbf{3}}^s}/f$ can range from $\mathcal{O}(1)$ to its unitarity bound of $4\pi$. This will result in larger possible variety for dark jets. Other benchmarks are shown in Fig.~\ref{fig:CatIIOther}, keeping one of the pions at 75 GeV.

\subsubsection*{Two unstable pions}
\begin{figure}[t!]
  \centering
   \captionsetup{justification=centering}
  \begin{subfigure}{0.48\textwidth}
    \centering
    \caption{$\Gamma_{\pi_1^m}=\Gamma_{\pi_2^m}$}
    \includegraphics[width=\textwidth]{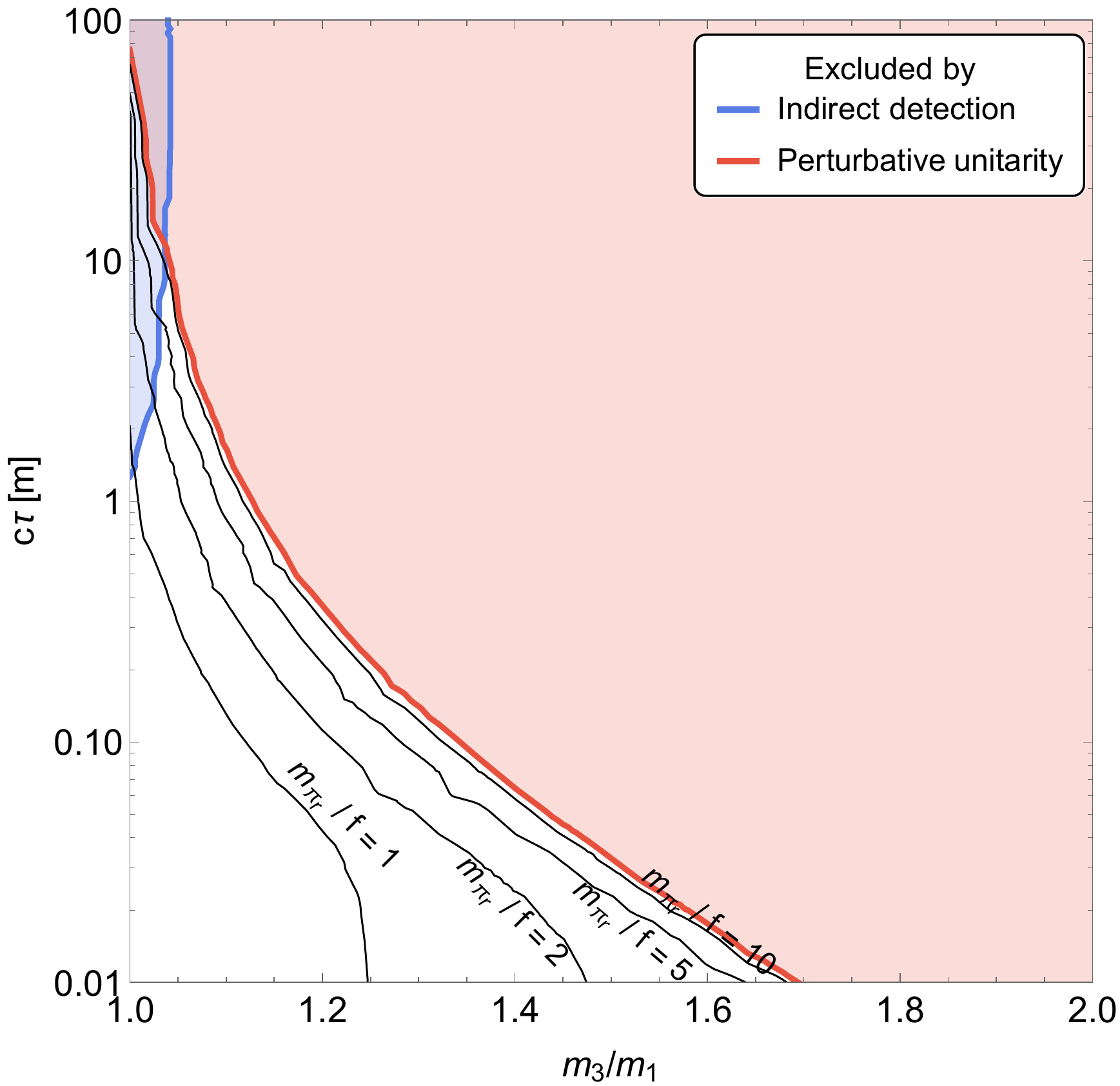}
    \label{fig:CatIIe}
  \end{subfigure}
  ~
  \begin{subfigure}{0.48\textwidth}
    \centering
    \caption{$\Gamma_{\pi_1^m}=2\Gamma_{\pi_2^m}$}
    \includegraphics[width=\textwidth]{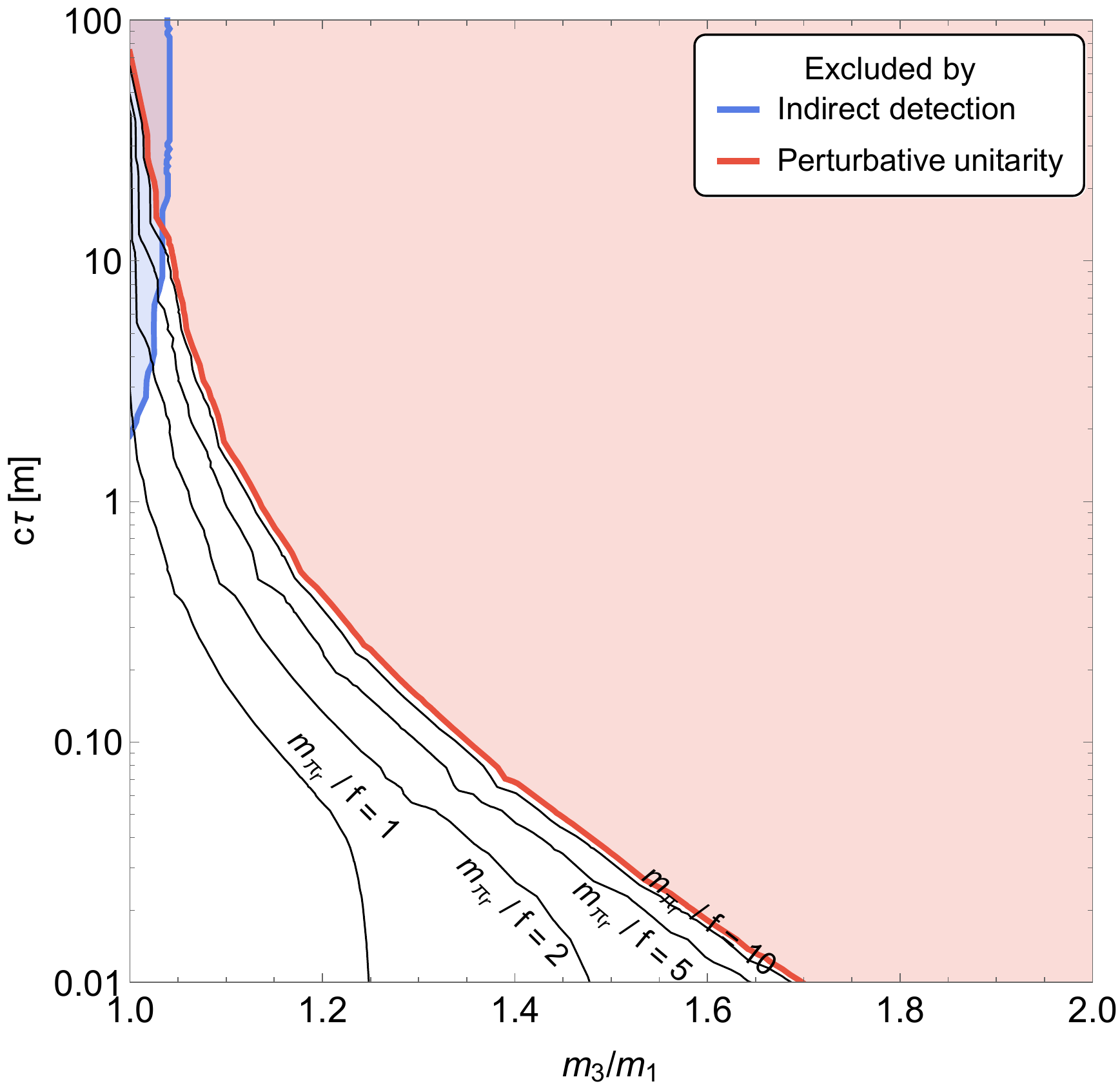}
    \label{fig:CatIIf}
  \end{subfigure}
  \captionsetup{justification=justified}
\caption{Allowed region of parameter space for the benchmark CIIe for different ratios of the decay widths. The mass of triplet of the preserved $SU(2)$ is kept at a constant value of 75~GeV. The black contour lines represent constant values of $m_{\pi_{\mathbf{3}}}/f$. The blue region is excluded by indirect detection DM searches, while the red one by perturbative unitarity constraints. The decay length is that of the lightest unstable pion.}\label{fig:CatII2Unstable}
\end{figure}

A good example of a pion structure of category II with two unstable pions is the benchmark CIIe. This structure is characterized by 7 parameters:
\begin{equation}\label{eq:ParametersCIIe}
  \frac{m_3}{m_1}, \frac{m_4}{m_1}, m_{\pi_1^m}, \Gamma_{\pi_1^m}, \Gamma_{\pi_2^m}, f, N_c.
\end{equation}
The parameters $f$ and $N_c$ are set as before. For the sake of the presentation, we assume $\frac{m_3}{m_1}$ and $\frac{m_4}{m_1}$ to be equal and $\Gamma_{\pi_1^m}$ and $\Gamma_{\pi_2^m}$ to be proportional. The results are shown in Fig.~\ref{fig:CatII2Unstable}. As can be seen, the presence of an extra unstable pion makes a small difference at small $\frac{m_3}{m_1}$, but this effect quickly disappears because of Boltzmann suppression.

\subsection{Limits on decay lengths for category III}\label{sSec:DLcatIII}
\begin{figure}[t!]
  \centering
   \captionsetup{justification=centering}
  \begin{subfigure}{0.48\textwidth}
    \centering
    \caption{$\Gamma_{\pi_d^1}=10\Gamma_{\pi_{\text{others}}}$}
    \includegraphics[width=\textwidth]{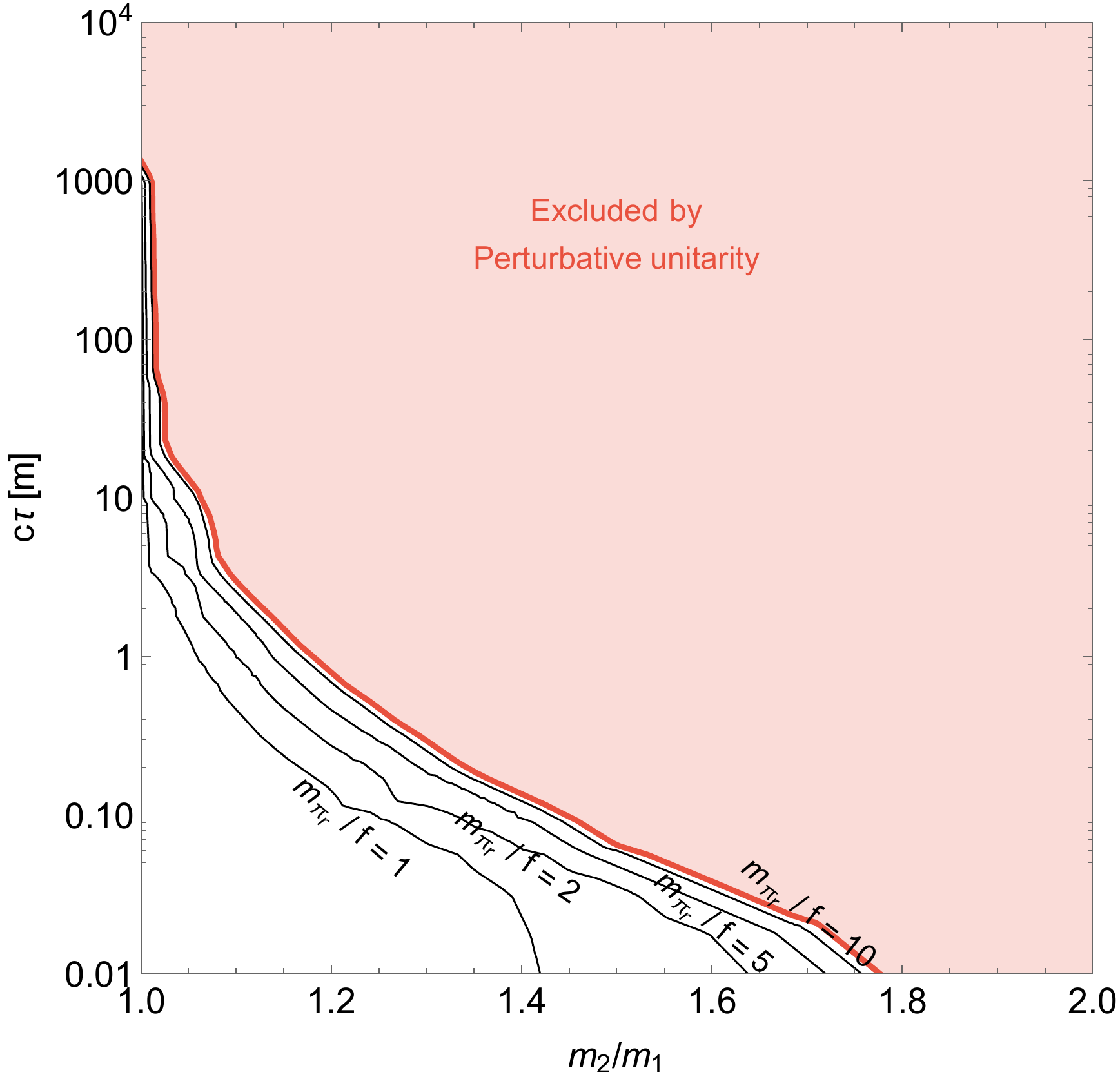}
    \label{fig:CIIIa}
  \end{subfigure}
  ~
  \begin{subfigure}{0.48\textwidth}
    \centering
    \caption{$\Gamma_{\pi_d^1}=5\Gamma_{\pi_{\text{others}}}$}
    \includegraphics[width=\textwidth]{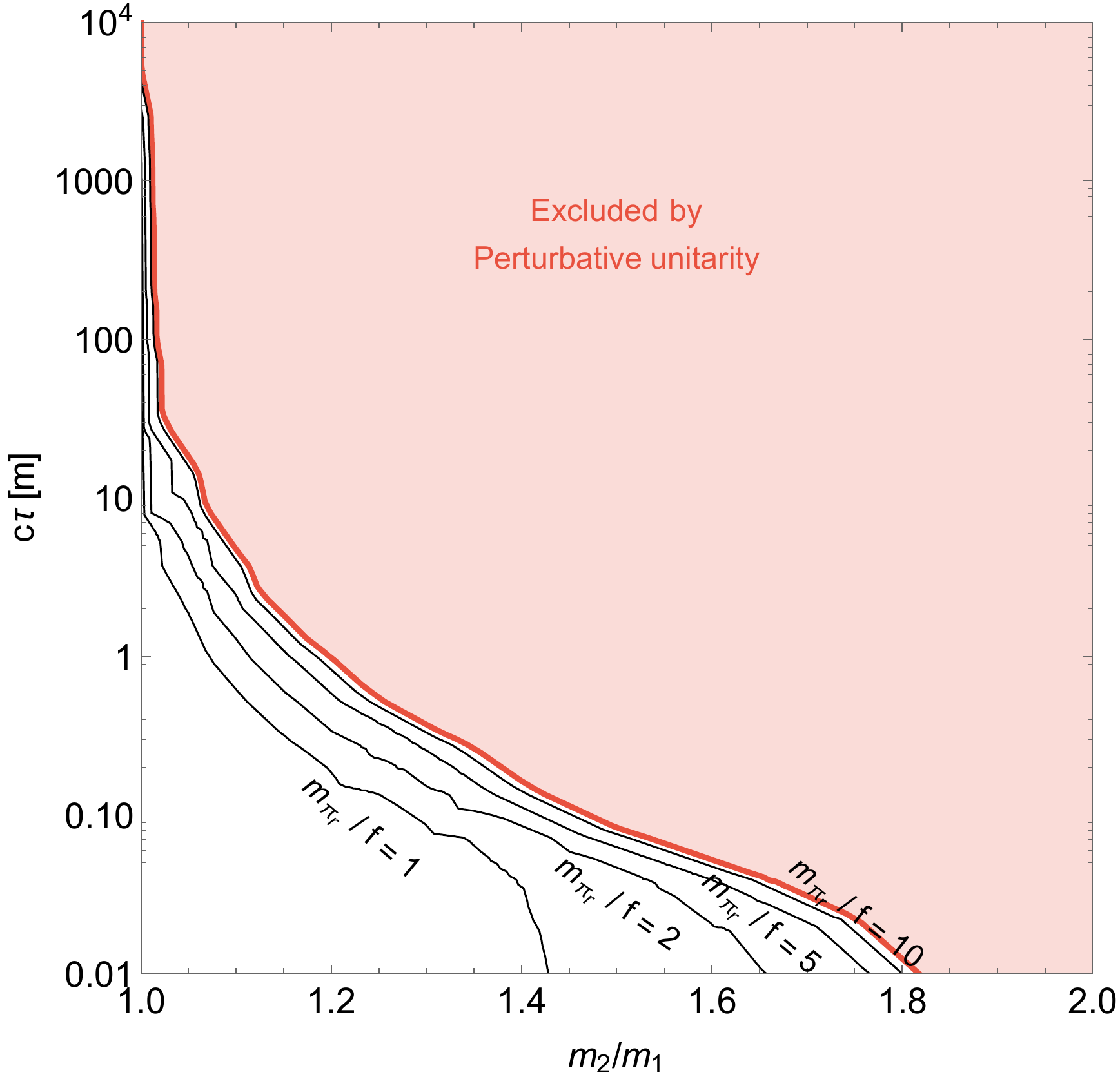}
    \label{fig:CaIIIb}
  \end{subfigure}
  \captionsetup{justification=justified}
\caption{Allowed region of parameter space for the benchmark CIIIa for different ratios of the decay widths. The mass of the lightest pion is kept at a constant value of 75~GeV. The contour lines represent constant values of $m_{\pi^a_1}/f$. The red region is excluded by perturbative unitarity constraints. The decay length is that of the $\pi^d_1$.}\label{fig:CatIII}
\end{figure}

Finally, structures of category III are those for which all annihilations of two stable pions to at least an unstable one are kinematically forbidden. These are very similar to category II, except for the fact that the bounds from indirect detection essentially do not exist anymore. As such, we will present only one example, which we take to be CIIIa. It is characterized by 11 parameters:
\begin{equation}\label{eq:ParametersCIIIa}
  \frac{m_2}{m_1}, m_{\pi_1^a}, f, N_c
\end{equation}
and seven decay widths. We set $f$ and $N_c$ as usual. For the sake of presentation, we assume the decay widths to be proportional to each others, with the decay width of $\pi^d_1$ to be the largest. The results are shown in Fig.~\ref{fig:CatIII}. As expected, they are similar to those of category II without the bounds from indirect detection. This allows for considerably longer decay lengths.

\section{Dark jet properties}\label{Sec:TAJS}
In this section, we investigate the properties of dark jets. In particular, we present results for the jet multiplicity, the fraction of unstable particles in a dark jet ($r_\mathrm{inv}$) and the thrust. We qualitatively outline how these variables vary due to the different features of the models within a given category. In the following, we fix the value of the confinement scale as $\Lambda_d = f / \sqrt{N_c}$ \cite{Veneziano:1976wm}. 

\subsection{Multiplicity}\label{sSec:Multiplicity}
The parton shower is followed by a non-perturbative process, the fragmentation process, and there is a relation between the number of the radiated partons and the number of the produced hadrons. The average number of hadrons is just the first Mellin moment of the fragmentation function \cite{Ellis:1991qj}. In the modified leading-log approximation, the hadron multiplicity is calculable up to a normalization constant~\cite{Mueller:1982cq,Ellis:1991qj}
\begin{equation}
\langle N(\hat{s})\rangle = A \, \mathrm{exp}\left[ \frac{1}{b_1} \sqrt{\frac{2 C_A}{\pi \alpha_s(\hat{s})}}+\left(\frac{1}{4}+\frac{N\, T_F}{\pi b_1}\frac{C_A-C_F}{3C_A} \right)\ln\alpha_s(\hat{s})+ \mathcal{O}(\alpha_s) \right] ,
\label{eq:multiplicity}
\end{equation}
where $A$ is the normalization constant, $C_F$ and $C_A$ are the Casimir invariants for the fundamental and adjoint representations, while $T_F$ is the normalization of the generators $\mathrm{Tr}(t^a t^b) = T_F \delta^{ab}$. We normalize the generators of all the groups in the standard way $T_F=1/2$. Therefore, for the different gauge groups, we have
\bea
C_F^{SU(N_c)}&=&\frac{N_c^2-1}{2N_c}\qquad C_F^{Sp(N_c)}=\frac{N_c+1}{4}\qquad C_F^{O(N_c)}=\frac{N_c-1}{4},\\
C_A^{SU(N_c)}&=&N_c \qquad C_A^{Sp(N_c)}=\frac{N_c+2}{2}\qquad C_A^{O(N_c)}=\frac{N_c-2}{2}.
\label{eq:CFandCA}
\eea
Furthermore, $b_1$ is the one loop coefficient of the beta function, and $\alpha_s(\hat{s})=(b_1 \log \hat{s}/\Lambda_d^2)^{-1}$ is the running strong coupling constant for a centre of mass energy squared $\hat{s}$. The one loop coefficients of the beta function are:
\be
b_1^{SU(N_c)}=\frac{11 N_c - 2N}{12\pi},\qquad b_1^{Sp(N_c)}=\frac{11N_c+22-4N}{24\pi},\qquad b_1^{O(N_c)}=\frac{11N_c-22-4N}{24\pi}.
\ee
This prediction for the average multiplicity is valid for sufficiently large values of the coefficient $b_1$ and has been verified experimentally for QCD \cite{Schmelling:1994py}. Notice also that the group $O(N_c)$ stops confining for a smaller number of fermions with respect to $SU(N_c)$ or $Sp(N_c)$. Furthermore, its one loop coefficient of the $\beta$ function is smaller than the one of the other groups and therefore it radiates more.  

The average multiplicity depends on the number of colors $N_c$, the number of flavors $N$ and the ratio between the scale of the centre of mass energy and the confinement scale $\sqrt{\hat{s}}/\Lambda_d$.
\begin{figure}[t!]
  \centering
   \captionsetup{justification=centering}
  \begin{subfigure}{0.48\textwidth}
    \centering
    \caption{$SU(3)$}
    \includegraphics[width=\textwidth]{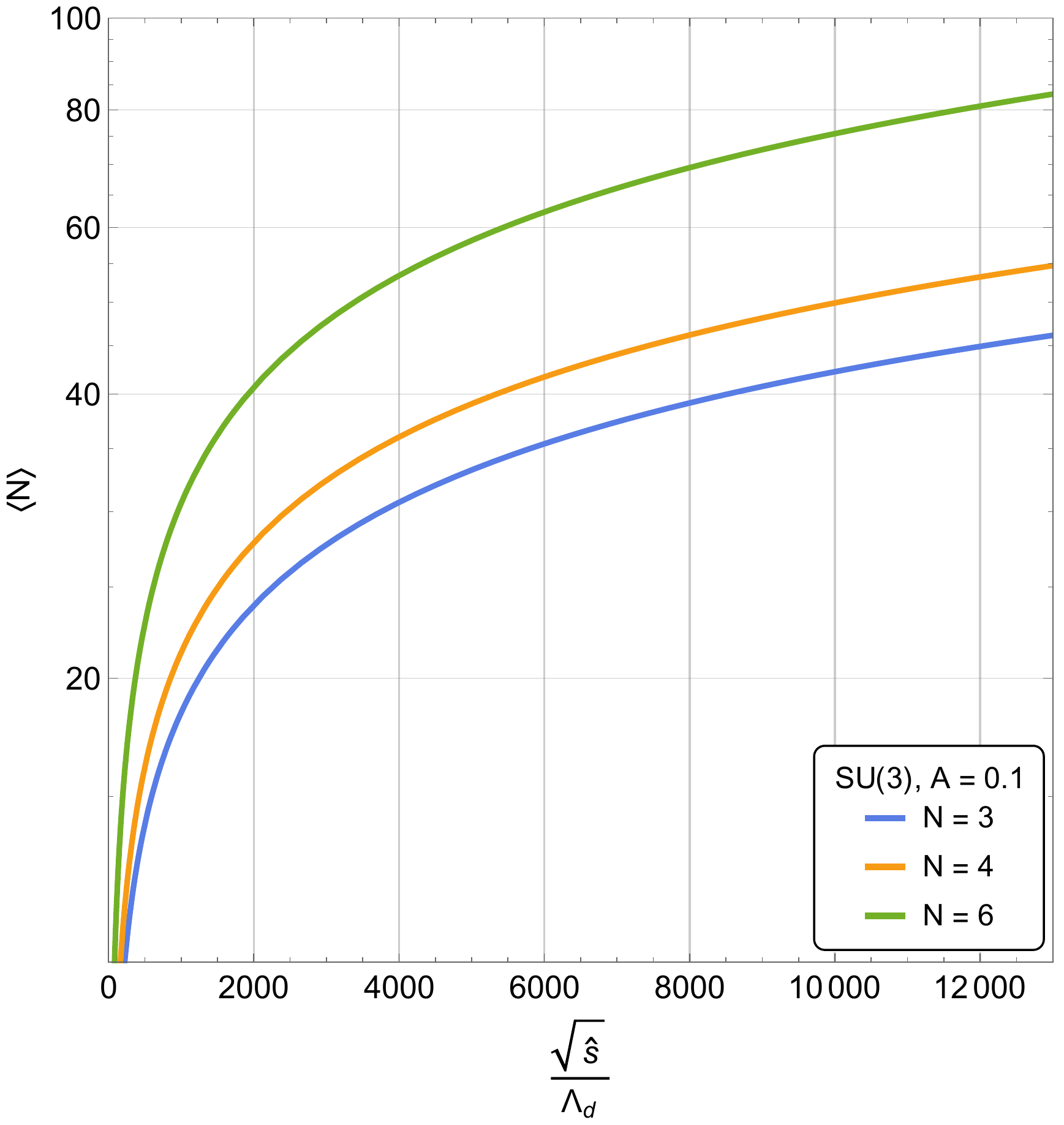}
    \label{fig:multSUN}
  \end{subfigure}
  ~
  \begin{subfigure}{0.48\textwidth}
    \centering
    \caption{$Sp(4)$}
    \includegraphics[width=\textwidth]{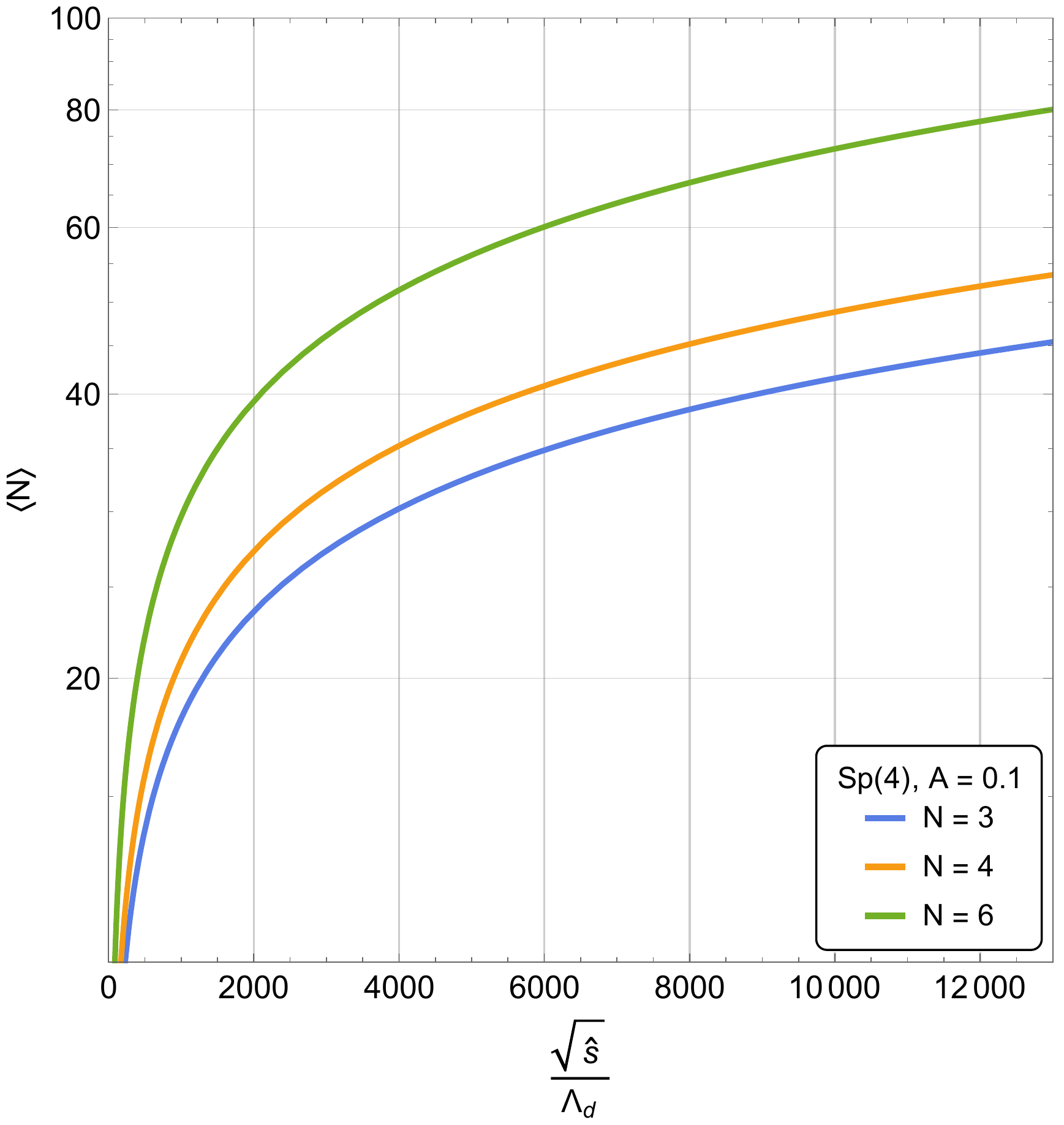}
    \label{fig:multSpN}
  \end{subfigure}
  ~
  \begin{subfigure}{0.48\textwidth}
    \centering
    \caption{$O(4)$}
    \includegraphics[width=\textwidth]{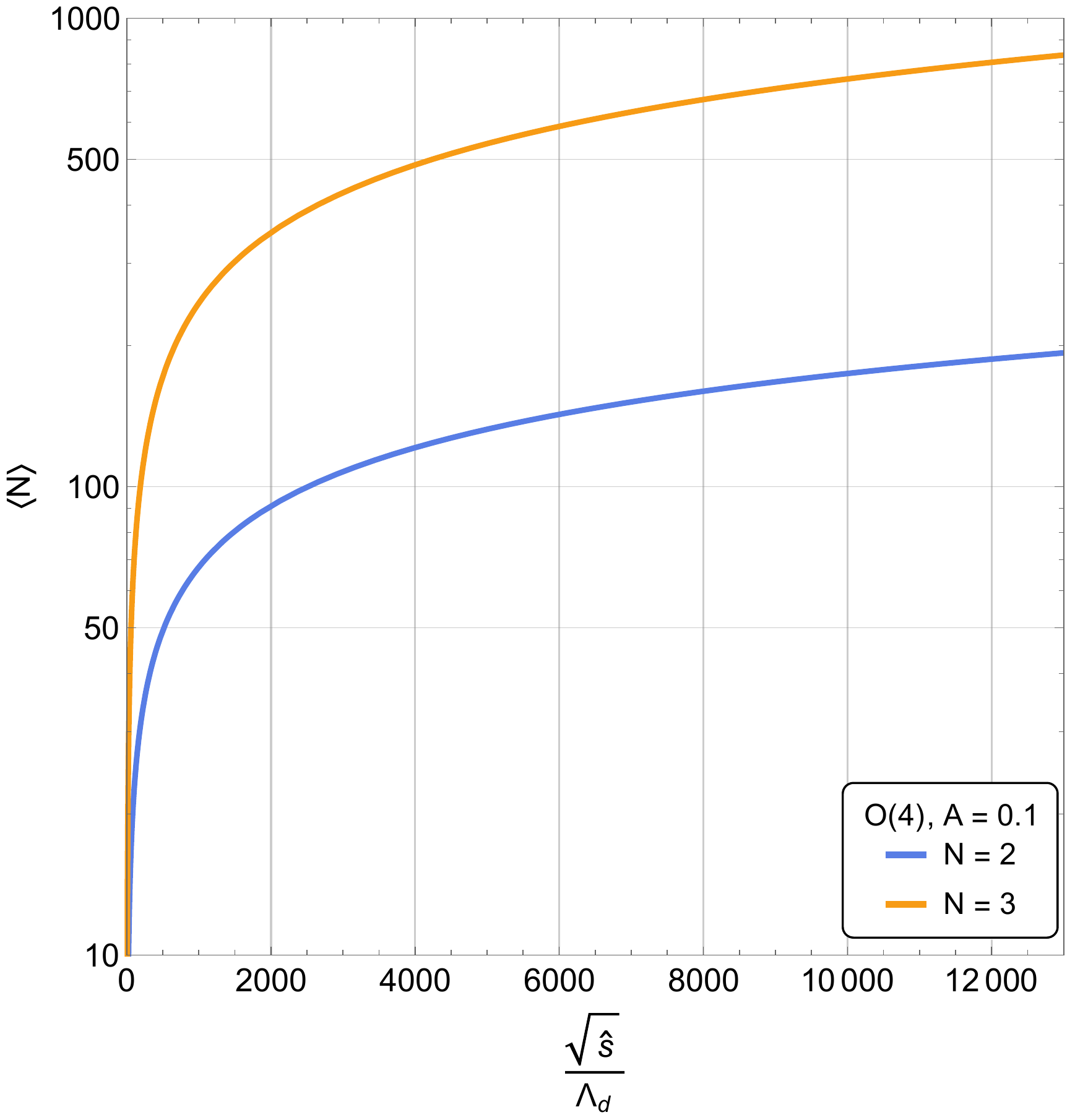}
    \label{fig:multON}
  \end{subfigure}
  \captionsetup{justification=justified}
\caption{Theory prediction of the average dark meson multiplicity as a function of the ratio $\sqrt{\hat{s}}/\Lambda_d$ for (a) $SU(3)$ , (b) $Sp(4)$ and (c) $O(4)$ for different values of $N$. The normalization is fixed as $A=0.1$.}\label{fig:multiplicity}
\end{figure}
For larger $N$, the running of the dark sector coupling to smaller values is slower and more partons are radiated at higher scales. This results in a larger number of dark pions. In Fig.~\ref{fig:multiplicity}, we show the behavior of the average multiplicity as a function of the ratio $\sqrt{\hat{s}}/\Lambda_d$ for $SU(3)$ (\ref{fig:multSUN}), $Sp(4)$ (\ref{fig:multSpN}) and $O(4)$ (\ref{fig:multON}). Notice that we keep the same normalization factor for each curve, $A=0.1$. The normalization factor $A$ cannot be theoretically determined unless we know how gluons fragment into hadrons. Since it can only be obtained by fit to experimental data,\footnote{For example, for QCD with $\Lambda_{QCD}=226$ MeV we would have $A\simeq0.12$ \cite{Schmelling:1994py}.} these curves should be taken as qualitative behaviors rather than explicit values. Increasing the number of dark colors $N_c$ has similar effects on the average multiplicity to lowering $N$, since they both enter the one loop coefficient of the beta function $b_1$. Fig.~\ref{fig:multiplicity} shows that the average multiplicity is also strongly dependent on the value of the dark confinement scale $\Lambda_d$, that enters logarithmically in the strong coupling constant $\alpha_s(\hat{s})$. In particular, the larger is $\Lambda_d$, the smaller is the average multiplicity of dark mesons in a jet produced in a collision with centre of mass energy $\sqrt{\hat{s}}$.

If we assume that $\Lambda_d \sim f/\sqrt{N_c}$,  we expect to have larger multiplicity for smaller $f$. This means that for the models of category I with one unstable pion, we have larger multiplicity for smaller dark quark splitting (compare Figs.~\ref{fig:CIa} and \ref{fig:CatIOther}). Notice that this conclusion changes if we have more unstable pions, depending on the parameters of the model, see Fig.~\ref{fig:CatI2Unstable}. On the other hand, for category II and III the multiplicity increases for larger mass splittings (see Figs.~\ref{fig:CIIb}, \ref{fig:CatIIOther}, \ref{fig:CatII2Unstable} and \ref{fig:CatIII}).

\subsection{Average collinear missing transverse energy fraction, $r_{\mathrm{inv}}$}\label{sSec:rinv}
Generically, both stable and unstable dark pions can be produced in a parton shower. Assuming the unstable pions decay hadronically, this will result in the production of semivisible jets \cite{Cohen:2015toa, Cohen:2017pzm, Beauchesne:2017yhh}. These roughly look like generic jets, but are accompanied by collinear stable particles that escape the detector and contribute to the missing transverse energy (MET). A useful quantity is therefore $r_\mathrm{inv}$, defined as the average fraction of the energy of a given semivisible jet that is transmitted to MET. This variable can take on any value between zero (no MET present) and one (no visible jets present). The purpose of this section is to provide a mapping between the allowed parameter space obtained by cosmological observations and the variable $r_\mathrm{inv}$. 

In the following, we will neglect the contribution of dark baryons. This is justified by the fact that in an $Sp(N_c)$ theory there are no stable baryons, since a baryon would decay into $N_c/2$ mesons \cite{Witten:1983tx}. In theories where stable baryons do exist, their production is suppressed at the 10$\%$ level in QCD and should be further suppressed in the large $N_c$ limit \cite{Witten:1979kh}. On the other hand, vector meson production might not be overly suppressed. Their stability will however not in general be insured by any symmetry. Under our assumptions, they should mostly decay quickly to dark pions, in a similar fashion to how SM vector mesons decay mainly to SM pseudoscalar mesons, which will generally be kinematically allowed.\footnote{Vector mesons are expected to have masses $m_V \sim 4\pi f/\sqrt{N_c}$ \cite{Soldate:1989fh, Chivukula:1992gi, Georgi:1992dw}.} As a consequence, the quantity $r_{\text{inv}}$ is roughly controlled by the fraction of stable pions modified by suppression factors for the production of heavier mesons.

In the Lund string model for fragmentation, the production of heavier mesons during hadronization is exponentially suppressed \cite{Andersson:1983ia}. This is described by the suppression factor for estimating the ratio of $q_i$ to $q_j$ production
\be
T_{ij} = \mathrm{exp}\left[ -\frac{4 \pi (m_i^2-m_j^2)}{\Lambda_d^2}  \right],
\label{eq:tunn}
\ee 
where $m_i>m_j$ are the masses of the dark quarks and $\Lambda_d$ is the dark confinement scale. The larger the mass splitting between $q_i$ and $q_j$ compared to the confinement scale $\Lambda_d$, the more suppressed will be the production of dark mesons containing the heavier quark. As a consequence, this may reduce the number of stable states in the dark pion shower. Notice that the values of the quark masses are related to $m_\pi$ and $B_0$.

\begin{figure}[t!]
  \centering
   \captionsetup{justification=centering}
  \begin{subfigure}{0.48\textwidth}
    \centering
    \caption{CIa: $m_{\pi_3^s} = 75$ GeV, $B_0 = m_1$}
    \includegraphics[width=\textwidth]{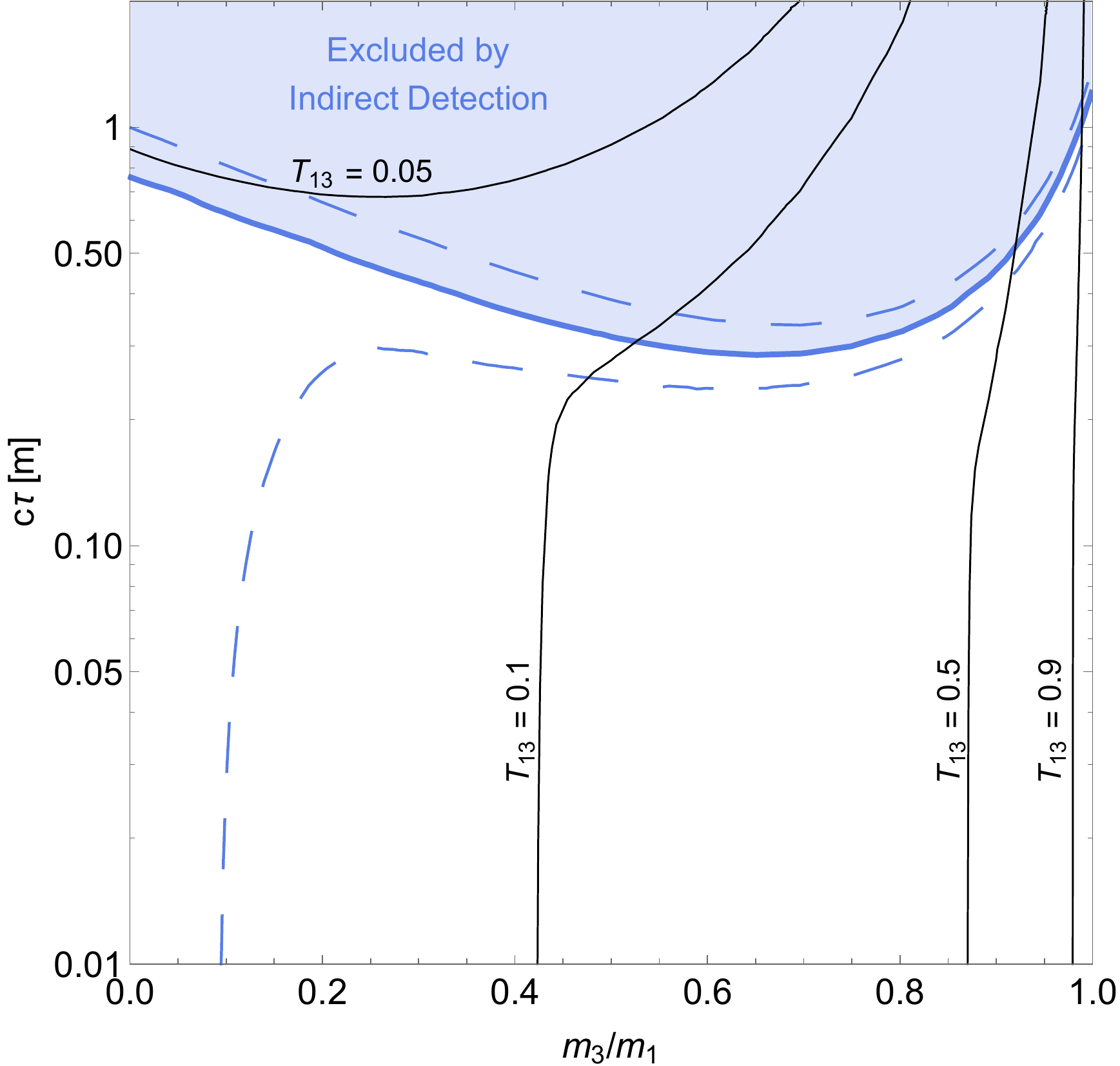}
    \label{fig:CatIa_rinv}
  \end{subfigure}
  ~
  \begin{subfigure}{0.48\textwidth}
    \centering
    \caption{CIa: $m_{\pi_3^s} = 75$ GeV, $B_0 = f$}
    \includegraphics[width=\textwidth]{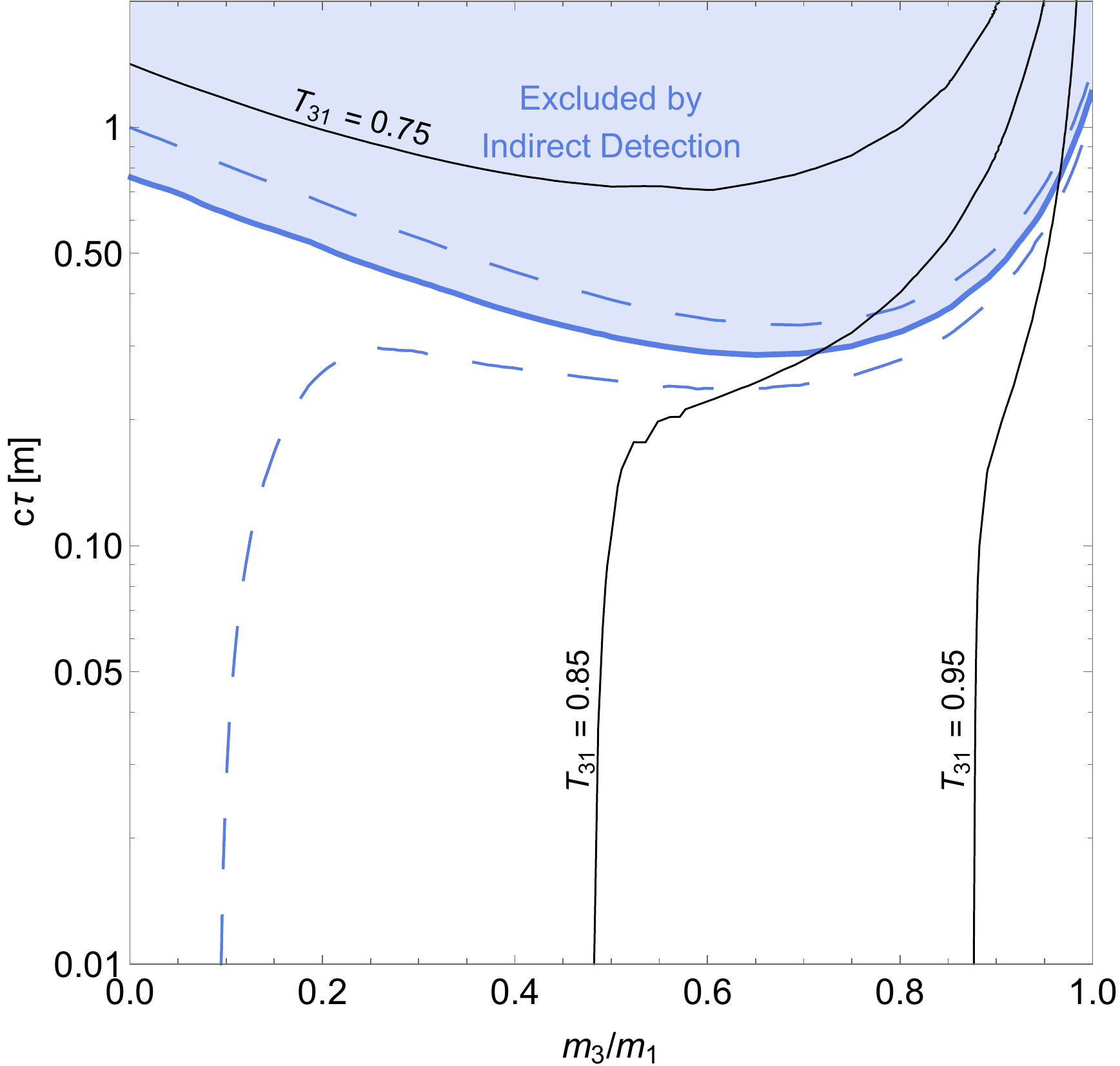}
    \label{fig:CatIa_rinv2}
  \end{subfigure}
  \captionsetup{justification=justified}
\caption{Allowed region of parameter space for the benchmark CIa for (a) $B_0 = m_1$ and (b) $B_0 = f$. The blue region is excluded by indirect detection. The black contour lines represent constant values of $T_{13}$ (a) or $T_{31}$ (b). The dashed blue lines correspond to what the exclusion limit from indirect detection would be if the limit on the effective cross section were to vary by $\pm10\%$.}\label{fig:rinv}
\end{figure}

Consider the spectrum of the case CIa (Fig. \ref{fig:CatIa_rinv}). In this scenario, we have $m_1=m_2=m_{\pi_3^s}^2/(2 B_0)$. Therefore, the suppression factors are 
\be
T_{13}=\mathrm{exp}\left[ -\frac{4 \pi}{\Lambda_d^2}\frac{ m_{\pi_3^s}^4}{4  B_0^2}(1-r^2)  \right],\qquad T_{11}=T_{12}=1,
\label{eq:tunnCIa}
\ee
where $r=m_3/m_1$.

For a small mass difference between the dark quarks $m_1=m_2$ and $m_3$, the factor $T_{13}$ in Eq.~\eqref{eq:tunnCIa} will be close to one and therefore we expect equal production of all three dark quarks. The value of $r_\mathrm{inv}$ will be close to the ratio between the number of stable and unstable states, given roughly by $r_{\mathrm{inv}}\sim7/8$ (see tables in Appendix \ref{sapp:Benchmark} for details on the stability structure). As the mass splitting increases, the production of the mesons containing only $q_1$ or $q_2$ is suppressed with respect to the mesons containing $q_3$. As a consequence, we expect that $T_{13}$ decreases for larger mass splitting. How fast $T_{13}$ decreases depends on the value of the quark masses, or equivalently the ratio between $m_\pi$ and $B_0$.  Fig.~\ref{fig:CatIa_rinv} shows that for large splitting and $B_0=m_1$ we obtain $T_{13} < 0.1$. On the other hand, a different value of $B_0$ would produce slightly different results. Choosing the more sensible $B_0=f$, the suppression factor $T_{13}$ decreases slowly reaching a value $\lesssim0.8$ for large mass splitting. This happens because the increasing value in the mass splitting ($r\ll1$) is almost compensated by the increasing value of $f$. This is shown in Fig. \ref{fig:CatIa_rinv2}. In either cases, a larger mass splitting implies more visible jets.

Interestingly, case CIIb (with the same breaking pattern, but with a different mass hierarchy $m_1=m_2<m_3$) leads to a different behavior. For small mass splitting, $T_{31}\sim1$ and $r_\mathrm{inv}\sim7/8$ as for case Ia. However, as the mass splitting increases, $T_{31}$ tends to $0$ very quickly, leading to very invisible dark jets. The behavior is similar for both $B_0=m_1$ and $B_0=f$. 

All models of category I are expected to have a similar behavior as that of benchmark CIa. Conversely, all models of category II and III have similar behaviors to those of benchmark CIIb. The exact range over which $r_{\mathrm{inv}}$ can vary however depends on the details of the model.

\subsection{Thrust}\label{sSec:Thrust}
Shape variables \cite{Ellis:1991qj} are a common approach in order to study the jet-like properties of hadronic final states. These quantities can characterize whether the distribution of the hadrons in a jet is pencil-like, planar or spherical. This is a strong way to obtain information on a jet. A widely used quantity is the thrust \cite{Kunszt:1989km,Ellis:1991qj}, defined as 
\be
T=\mathrm{max}_{\vec{n}}\frac{\sum_i |\vec{p}_i\cdot \vec{n}|}{\sum_i|\vec{p}_i|},
\ee
where $\vec{p}_i$ are the final-state hadron momenta and $\vec{n}$ is a unit vector. This variable is bounded between $1/2\leq T \leq 1$, where a pencil-like event has $T=1$, while a spherical one has $T=1/2$. The thrust is infrared and collinear safe: an additional parton that is soft or radiated collinear to the thrust axis will not change the thrust. Its distribution is well described by \cite{DeRujula:1978vmq}
\be
\langle 1-T(\hat{s}) \rangle = C_F \frac{\alpha_s}{2\pi}\left[-\frac{3}{4}\ln 3 -\frac{1}{18}+4\int_{2/3}^1 \frac{dT}{T} \ln \frac{2 T -1}{1-T}  \right] + \mathcal{O}(\alpha_s^2),
\ee
where $\sqrt{\hat{s}}$ is a high-energy scale associated with the total system being probed by the thrust and $C_F$ is the Casimir invariant. We listed the Casimir invariant for the different gauge groups in Eq.~(\ref{eq:CFandCA}).
\begin{figure}[t!]
  \centering
   \captionsetup{justification=centering}
  \begin{subfigure}{0.48\textwidth}
    \centering
    \caption{$SU(3)$}
    \includegraphics[width=\textwidth]{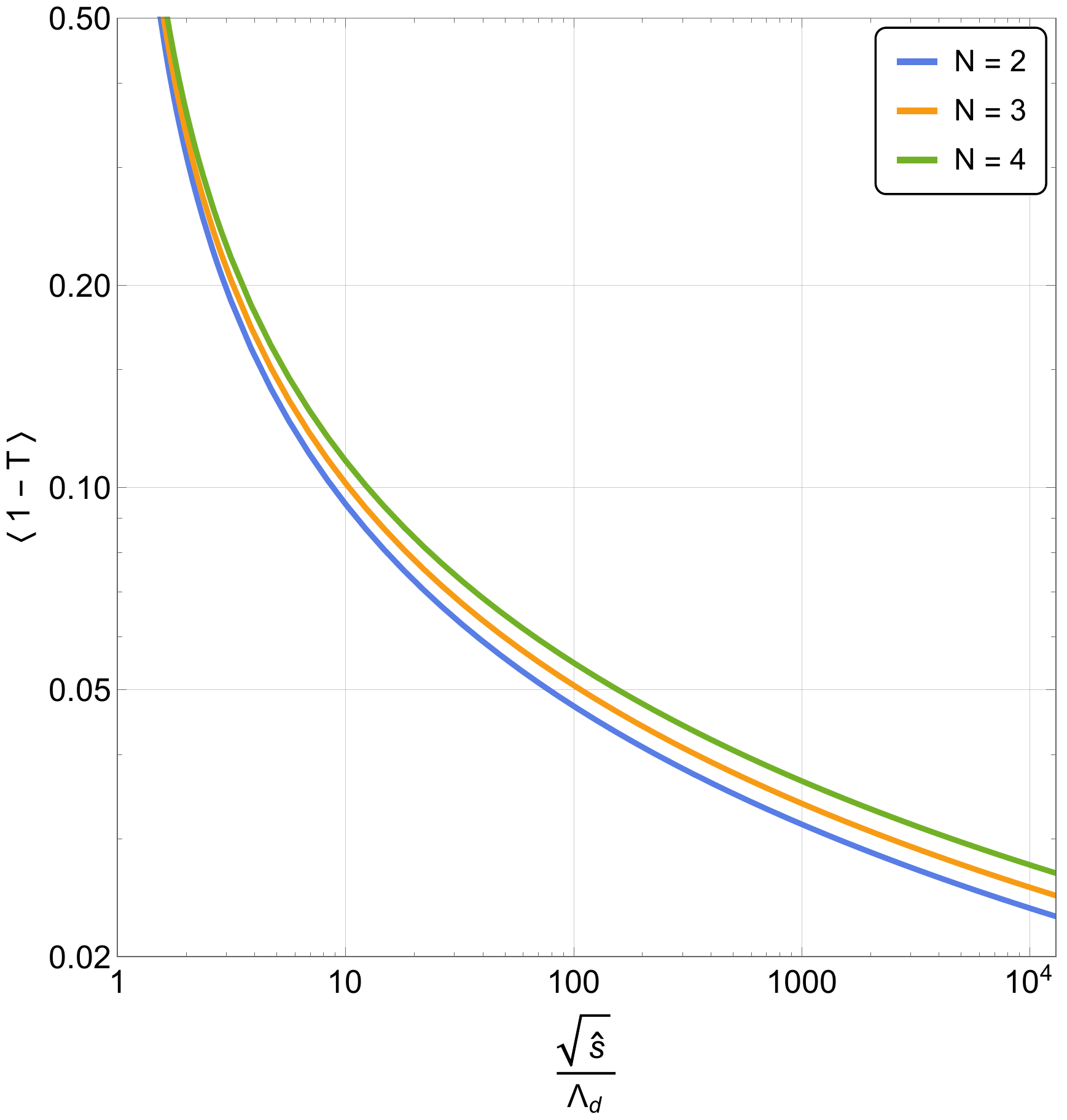}
    \label{fig:thrustSUN}
  \end{subfigure}
  ~
  \begin{subfigure}{0.48\textwidth}
    \centering
    \caption{$Sp(4)$}
    \includegraphics[width=\textwidth]{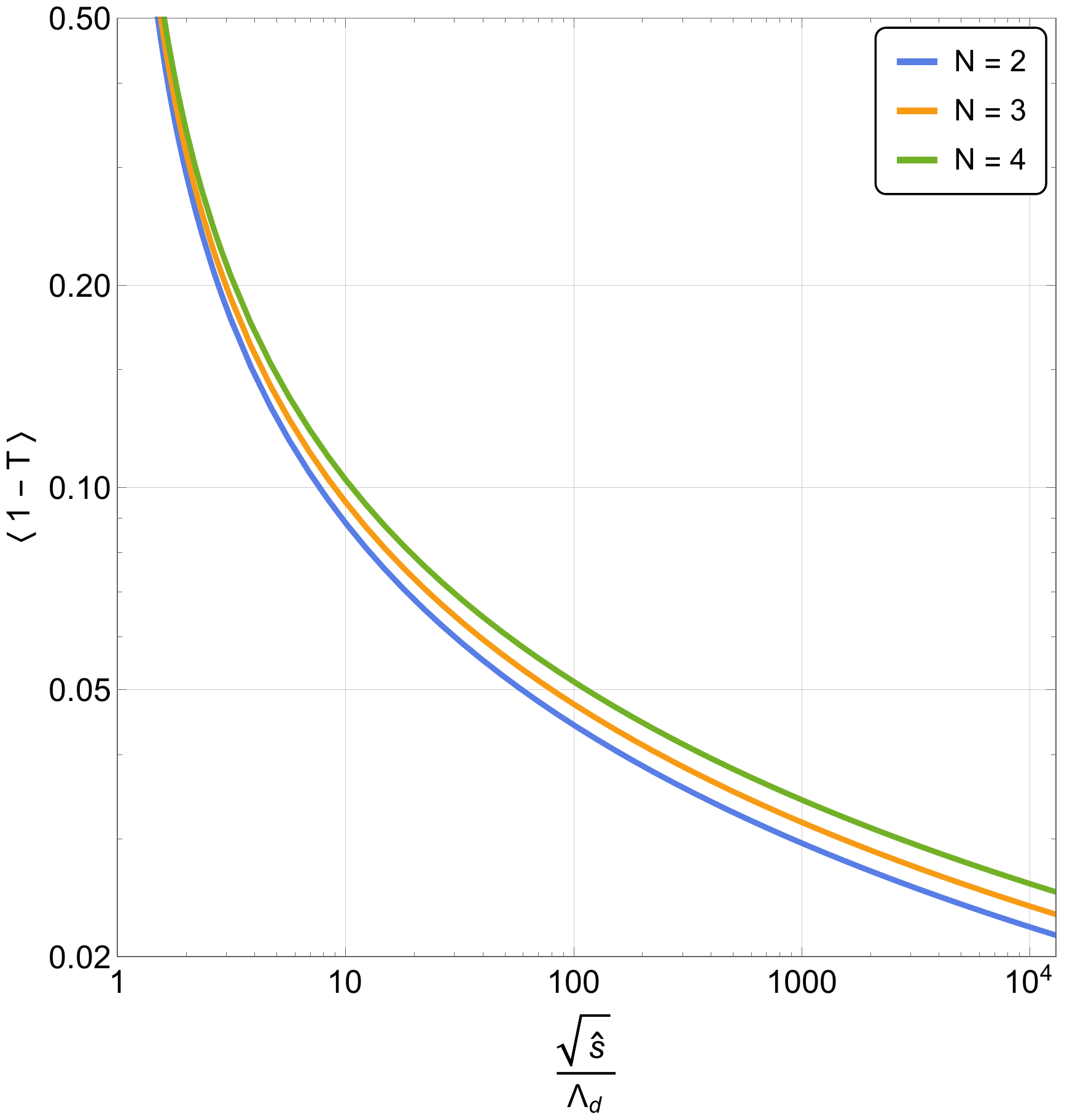}
    \label{fig:thrustSpN}
  \end{subfigure}
  ~
  \begin{subfigure}{0.48\textwidth}
    \centering
    \caption{$O(4)$}
    \includegraphics[width=\textwidth]{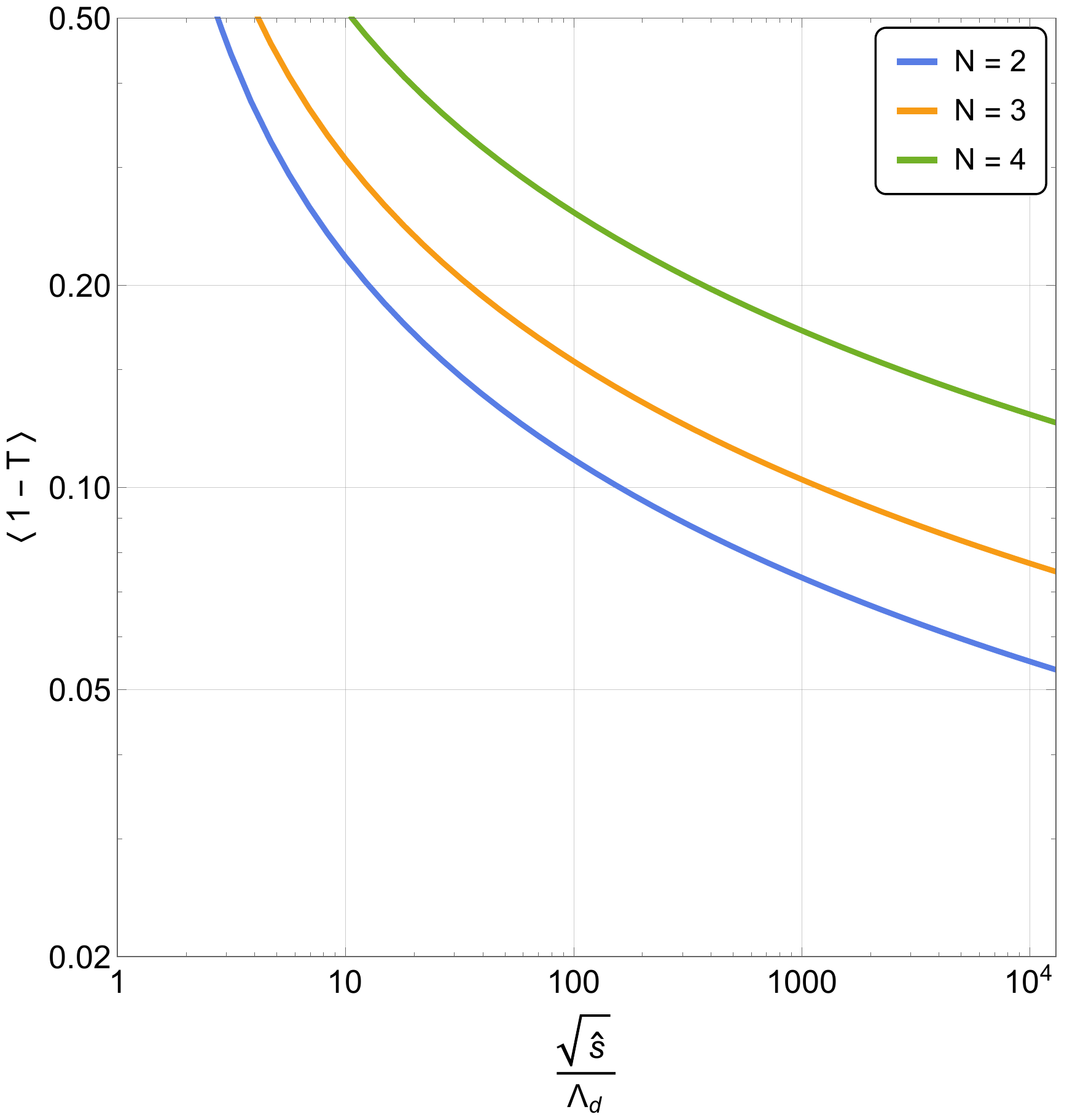}
    \label{fig:thrustON}
  \end{subfigure}
  \captionsetup{justification=justified}
\caption{Theory prediction of the thrust as a function of the ratio $\sqrt{\hat{s}}/\Lambda_d$ for gauge group $SU(3)$ (a), $Sp(4)$ (b) and $O(4)$ (c) for different values of $N=2$ (blue), $N=3$ (orange) and $N=4$ (green).}\label{fig:thrust}
\end{figure}

The thrust depends on the number of colors $N_c$, the number of flavors $N$, the confining scale $\Lambda_d$ and the centre of mass energy $\sqrt{\hat{s}}$. In Fig. \ref{fig:thrust}, we show $\langle 1-T\rangle$ as a function of the ratio $\sqrt{\hat{s}}/\Lambda_d$, for different values of $N$ and for the different groups $SU(3)$ (a), $Sp(4)$ (b) and $O(4)$ (c). The thrust for the groups $SU(3)$ and $Sp(4)$ gives similar results. This is due to the fact that the running of their gauge coupling $\alpha_s$ is similar. In particular, $SU(3)$ with $N=3$ and $Sp(4)$ with $N=3$ have the same coefficient $b_1$. On the other hand, models with a gauge group $O(N_c)$ have a larger strong coupling. This results in generally more spherical jets for $O(N_c)$ models with respect to $SU(N_c)$ or $Sp(N_c)$ models. Furthermore, the figure shows that a larger confining scale $\Lambda_d$ would lead to smaller thrust and therefore more spherical objects. 

\section{Conclusion}\label{Sec:Conclusion}
In this work, we investigated models where the dark matter consists of the stable pseudo-Goldstone mesons of a new confining sector. The combination of spontaneous and explicit symmetry breaking results in a complicated structure of pions in terms of stability and masses. This structure determines which processes are kinematically forbidden or not. In turn, this affects the cosmological evolution, the current constraints and the range of possible collider signatures. We found that all pseudo-Goldstone meson structures with both stable and unstable pions can be classified into three categories.

The first category includes models where all stable dark pions can annihilate with other stable pions to produce unstable ones. As a consequence, all stable pions have similar cross section for the production of unstable ones and therefore decouple from the unstable pions at a similar temperature. As a result, they all have similar densities and there are strong bounds from indirect detection dark matter searches. The jet properties of dark jets cannot vary over a large range and are not that different from QCD jets with a higher confinement scale. At least one unstable pion is required to have a decay length of less than $\mathcal{O}(1)$~m for the masses considered in our benchmarks. These results lead to many different possible signatures at colliders, such as emerging or semivisible jets.

In the second category, not all the stable pions can annihilate with other stable pions to produce unstable ones. Those that are not allowed to do so will have Boltzmann suppressed cross sections for the annihilation to unstable pions, resulting in an earlier decoupling. The dark matter abundance will then be dominated by those pions. Since they do not contribute to the indirect detection signal, the bounds from indirect detection searches are considerably suppressed. The shape of the dark jets, on the other hand, can take many forms, depending on the dark quark mass splitting. Larger splittings will in general require smaller values of $f$ in order to obtain the correct dark matter abundance (and therefore a smaller confining scale $\Lambda_d$). As a result, dark jets will be increasingly narrow, with more stable mesons, larger multiplicity and shorter decay lengths. For smaller splittings, considerably longer decay lengths are allowed for the unstable pions, with some benchmarks allowing for decay lengths of up to $\mathcal{O}(100)$~m.

Finally, the third category includes models where all the annihilations of two stable pions to unstable ones are forbidden, resulting in the relic density being dominated by the lightest stable dark pions. As a consequence, there are essentially no indirect detection bounds. An upper bound on the decay length of the unstable pions comes only from requiring the theory to be unitary. Even longer decay lengths are allowed for the unstable pions, possibly extending all the way to $\mathcal{O}(1)$~km. The dark jets of this category are similar to the ones of category~II. 

\acknowledgments
We would like to thank Yevgeny Kats for useful discussions. HB is grateful to the Azrieli Foundation for the award of an Azrieli Fellowship. GGdC is supported by the National Science Centre, Poland, under research grant no. 2017/26/D/ST2/00225 and would like to express a special thanks to the GGI Institute for Theoretical Physics for its hospitality and support.

\appendix

\section{Relic dark matter abundance calculation}\label{app:RDMAC}
We present in this appendix all the information relevant for computing the dark matter relic abundance in the examples considered in this paper.

\subsection{Lagrangian}\label{sapp:Lagrangian}
The leading order chiral Lagrangian is given by:
\begin{equation}\label{eq:ChiralLagrangian0}
  \mathcal{L}_0 = \frac{f^2}{4}\text{Tr}\left[\partial_\mu U \partial^\mu U^\dagger\right] + \frac{B_0f^2}{2}\text{Tr}\left[(U + U^\dagger)M\right],
\end{equation}
where
\begin{equation}\label{eq:DefU}
    U = \text{exp}\left(\frac{i\sqrt{2}}{f}\Pi\right),
\end{equation}
with $\Pi= \pi^a t^a$. The matrices $t^a$ correspond to the $2N\times 2N$ broken generators (see Sec.~\ref{sapp:Notation} for their expressions and normalization). The $B_0$ parameter is simply a constant that can be traded in practice for the mass of one of the pions. In all cases we consider in this paper, $M$ can be written as:
\begin{equation}\label{eq:Mdefinition}
  M =
  \begin{pmatrix}
    \mathcal{M} & 0           \\
    0           & \mathcal{M}
  \end{pmatrix},
\end{equation}
where $\mathcal{M}$ is a diagonal matrix whose entries satisfy $\mathcal{M}_{ii} = m_{q_i}$. In addition to $\mathcal{L}_0$, the Wess-Zumino-Witten term can possibly be present:
\begin{equation}\label{eq:DefWZW}
  \mathcal{L}_{\text{WZW}}=\frac{N_c}{30\sqrt{2}\pi^2f^5}\epsilon^{\mu\nu\alpha\beta}\text{Tr}\left[\Pi \partial_\mu \Pi\partial_\nu \Pi\partial_\alpha \Pi\partial_\beta \Pi\right],
\end{equation}
where $N_c$ is the number of colors. This expression holds for $SU(N)\times SU(N)$ if the quarks are in the fundamental representations of an $SU(N_c)$ gauge group, for $Sp(2N)$ if the quarks are in the fundamental representation of a $Sp(N_c)$ gauge group with $N_c$ even and for $SO(2N)$ if the quarks are in the vectorial representation of an $O(N_c)$ gauge group. We always assume we are in one such representation.

\subsection{$2\to 2$ scattering}\label{sapp:2to2}
Consider a $2\to 2$ pion scattering process $p$ given by
\begin{equation}\label{eq:2to2scat}
  p:\pi_a \pi_b \to \pi_c \pi_d.
\end{equation}
The amplitude of $p$ is
\begin{equation}\label{eq:Amplitude1}
  \begin{aligned}
    M_p = \frac{1}{12f^2} \bigg[ 2B_0C_1^p
        & + ( - C_2^p + C_3^p + C_4^p - C_5^p + C_6^p - C_7^p)s \\
        & + ( + C_2^p - C_3^p - C_4^p + C_5^p + C_6^p - C_7^p)t \\
        & + ( + C_2^p - C_3^p + C_4^p - C_5^p - C_6^p + C_7^p)u \bigg],
  \end{aligned}
\end{equation}
with the coefficients
\begin{equation}\label{eq:Coeff2to2}
  \begin{aligned}
    C_1^p & = \sum_{\{a,b,c,d\}}                    \text{Tr}[t^a t^b t^c t^d M]               \\
    C_2^p & = \sum_{\substack{\{a, b\} \\\{c, d\}}} \text{Tr}[t^a t^c t^b t^d]                 &
    C_3^p & = \sum_{\substack{\{a, b\} \\\{c, d\}}} \text{Tr}[t^a t^b t^c t^d] \hspace{0.5cm}  &
    C_4^p & = \sum_{\substack{\{a, c\} \\\{b, d\}}} \text{Tr}[t^a t^b t^c t^d]                 \\
    C_5^p & = \sum_{\substack{\{a, c\} \\\{b, d\}}} \text{Tr}[t^a t^c t^b t^d]                 &
    C_6^p & = \sum_{\substack{\{a, d\} \\\{b, c\}}} \text{Tr}[t^a t^b t^d t^c] \hspace{0.5cm}  &
    C_7^p & = \sum_{\substack{\{a, d\} \\\{b, c\}}} \text{Tr}[t^a t^d t^b t^c],
  \end{aligned}
\end{equation}
where the sums are over all permutations of the indices inside the brackets, even repeated ones. For example,
\begin{equation}
  \sum_{\{a, b\}} \text{Tr}[t^a t^b] = \text{Tr}[t^a t^b] + \text{Tr}[t^b t^a],
\end{equation}
even if $a=b$. Using the relation between the Mandelstam variables
\begin{equation}\label{eq:Mandelstam}
  s + t + u = 4(\hat{m}^p)^2,
\end{equation}
where
\begin{equation}\label{eq:Defmhat}
  (\hat{m}^p)^2 = \frac{1}{4}\left(m_{\pi_a}^2 + m_{\pi_b}^2 + m_{\pi_c}^2 + m_{\pi_d}^2\right),
\end{equation} 
Eq.~(\ref{eq:Amplitude1}) can be rewritten as:
\begin{equation}\label{eq:Ampltitude2}
   M_p = \frac{1}{6f^2}\left[a_s^p s + a_t^p t + a_0^p\right],
\end{equation}
where
\begin{equation}\label{eq:Defasi}
  \begin{aligned}
    a_s^p & = -C_2^p + C_3^p + C_6^p - C_7^p \\
    a_t^p & = -C_4^p + C_5^p + C_6^p - C_7^p \\
    a_0^p & = B_0 C_1 +2\left(C_2^p - C_3^p + C_4^p - C_5^p - C_6^p + C_7^p\right)(\hat{m}^p)^2.
  \end{aligned}
\end{equation}
One can then define an effective amplitude square as
\begin{equation}\label{eq:EffAmplitudeSquare}
  |\hat{M}|^2_p = E_{20}^ps^2 +E_{11}^pst + E_{02}^pt^2 + E_{10}^ps + E_{01}^pt + E_{00}^p,
\end{equation}
where
\begin{equation}\label{eq:Defeij}
  E_{ij}^p = \frac{1}{1 + \delta_{ab}}\frac{1}{1 + \delta_{cd}}e_{ij}^p
\end{equation}
with
\begin{equation}\label{eq:DefEij}
  \begin{aligned}
    & e_{20}^p = \frac{(a_s^p)^2  }{36f^4} & \qquad &
      e_{11}^p = \frac{a_s^p a_t^p}{18f^4} & \qquad &
      e_{02}^p = \frac{(a_t^p)^2  }{36f^4} \\
    & e_{10}^p = \frac{a_s^p a_0^p}{18f^4} & \qquad &
      e_{01}^p = \frac{a_t^p a_0^p}{18f^4} & \qquad &
      e_{00}^p = \frac{(a_0^p)^2  }{36f^4}.
  \end{aligned}
\end{equation}
The first possible factor of $1/2$ in Eq.~(\ref{eq:Defeij}) is to avoid double counting collisions between identical particles. The second factor is to take into account identical particles in the final state such that all integrals over the polar angle can be performed from 0 to $\pi$.

A great way to simplify the numerical computations is to use the fact that pions inside an unbroken multiplet maintain the same density at all times. Consider four non-necessarily distinct multiplets $A$, $B$, $C$ and $D$. Define $P:\pi_A \pi_B \to \pi_C \pi_D$ as the set of distinct $p:\pi_a \pi_b \to \pi_c \pi_d$ such that $\pi_a\in A$, $\pi_b\in B$ and so forth. By distinct, we mean for example that only one of the equivalent processes $\pi_1 \pi_2 \to \pi_3 \pi_4$ and $\pi_2 \pi_1 \to \pi_3 \pi_4$ appears in $P$. It is then possible to define an effective amplitude square as:
\begin{equation}\label{eq:EffAmplitudeSquareRep}
  \begin{aligned}
    |\hat{M}|^2_P & = \sum_{p \in P} |\hat{M}|^2_p\\
                  & = E_{20}^Ps^2 +E_{11}^Pst + E_{02}^Pt^2 + E_{10}^Ps + E_{01}^Pt + E_{00}^P,
  \end{aligned}
\end{equation}
where
\begin{equation}\label{eq:Eff}
  E_{ij}^P = \sum_{p \in P} E_{ij}^p.
\end{equation}
This amplitude can then be used to compute an effective cross section $\sigma_P(s)$ using the standard procedure. Of course, factors for identical incoming or outgoing particles should not be reintroduced as they are already taken into account. No average on incoming particles is taken either. One can then compute the thermally-averaged cross section
\begin{equation}\label{eq:ThermallyAveragedCS}
 \langle\sigma v\rangle_P = \frac{\int_{s_{\text{min}}}^\infty \frac{1}{\sqrt{s}} (s - (m_A + m_B)^2) (s - (m_A - m_B)^2) \sigma_P(s) K_1\left(\frac{\sqrt{s}}{T}\right) ds}{8 T m_A^2 m_B^2 K_2\left(\frac{m_A}{T}\right) K_2\left(\frac{m_B}{T}\right)},
\end{equation}
where $s_{\text{min}}=\text{max}((m_A + m_B)^2, (m_C + m_D)^2)$.

\subsection{$3\to 2$ scattering}\label{sapp:3to2}
Consider a $3\to 2$ scattering process $p$ given by
\begin{equation}\label{eq:3to2scat}
  p:\pi_a \pi_b \pi_c \to \pi_d \pi_e.
\end{equation}
In the limit that all pions are degenerate in mass, the corresponding cross section is \cite{Hochberg:2014kqa}
\begin{equation}\label{eq:3to2CS}
  \langle\sigma v^2\rangle_p = \frac{m_\pi^5 N_c^2 T^2_{abcde}}{2^{10} 3 \sqrt{5} \pi^5 f^{10} x^2},
\end{equation}
where
\begin{equation}\label{eq:DefT}
  T_{abcde} = \sum_{\{a, b, c, d, e\}} \text{Tr}[t^a t^b t^c t^d t^e] \text{sign}(\{a, b, c, d, e\}),
\end{equation}
where $\text{sign}(\{a, b, c, d, e\})$ is the sign of the permutation of $\{a, b, c, d, e\}$, e.g the sign of $\{1, 2, 3\}$ is $+1$ and that of $\{1, 3, 2\}$ is $-1$. Note that Eq.~(\ref{eq:DefWZW}) implies that five different pions must be involved in the Lagrangian term. This means that there is no subtlety concerning identical particles in the ingoing or outgoing states. A complex pion can however appear as both itself and its conjugate, meaning that a pion can appear on one side of a $3\to 2$ process and the other at the same time.

We neglect the mass splitting between the pions for two reasons. First, the $3\to 2$ processes that we consider are always between particles that are close in mass and have far more mass incoming than outgoing. The mass splitting is therefore a small effect, specially considering that these processes take place at high temperature. Second, in the end, the dark matter abundance is mainly determined by when the stable pions decouple from the unstable ones. The $3\to 2$ processes do not affect this much.

As was done for $2\to 2$ processes, an effective cross section can be defined by merging processes involving pions from the same multiplets. Define $P:\pi_A \pi_B \pi_C \to \pi_D \pi_E$ as the set of distinct $p:\pi_a \pi_b \pi_c \to \pi_d \pi_e$ such that $\pi_a\in A$, $\pi_b\in B$ and so forth. The effective cross section is
\begin{equation}\label{eq:3to2CSEff}
  \langle\sigma v^2\rangle_P = \frac{m_\pi^5 N_c^2 T^2_{ABCDE}}{2^{10} 3 \sqrt{5} \pi^5 f^{10} x^2},
\end{equation}
where
\begin{equation}\label{eq:DefTEff}
  T_{ABCDE} = \sum_{p \in P} T_{abcde}.
\end{equation}

\subsection{Boltzmann equation}\label{sapp:Boltzmann}
The results of the previous sections can then be combined to write down Boltzmann equation. Let's begin by introducing additional notation:
\begin{itemize}
  \item Define $\mathcal{P}^{2\to 2}$ as the set of all distinct $2\to 2$ scatterings between representations.
  \item Define $\mathcal{P}^{3\to 2}$ as the set of all distinct $3\to 2$ scatterings between representations.
  \item Define $N_A$ as the size of the representation $A$.
  \item Define the entropy density as $s$.
  \item Define $H$ as the Hubble constant.
  \item Define $\hat{Y}_A$ as the number density of pions from $A$ per entropy density divided by $N_A$.
  \item Define $\hat{Y}_A^{\text{eq}}$ as the equilibrium value of $\hat{Y}_A$.
  \item Define $\#A_{\text{in}}^P$ as the number of pions from $A$ in the in state of a process $P$.
  \item Define $\#A_{\text{out}}^P$ as the number of pions from $A$ in the out state of a process $P$.
  \item Define $\Gamma_A=\sum\limits_{a\in A} \Gamma_a$, where $\Gamma_a$ is the decay width of pion $a$.
  \item Define $\langle\Gamma_A\rangle$ as $\Gamma_A \langle 1/\gamma_A\rangle$, where $\langle 1/\gamma_A\rangle$ is the average of the inverse of the gamma factor for a pion from multiplet~$A$.
\end{itemize}
With all this notation established, the Boltzmann equation can be written as:
\begin{equation}\label{eq:BoltzmannEq}
  \begin{aligned}
    & N_A  \frac{d\hat{Y}_A}{dx} = -\frac{s}{xH}\bigg[ \frac{\langle\Gamma_A\rangle}{s}\left(\hat{Y}_A - \hat{Y}_A^{\text{eq}} \right)\\
    & + \sum_{\substack{P\in \mathcal{P}^{2\to 2} \\ \#A_{\text{in}}^P > \#A_{\text{out}}^P}}(\#A_{\text{in}}^P - \#A_{\text{out}}^P)\left(\hat{Y}_{P_1}\hat{Y}_{P_2} - \frac{\hat{Y}_{P_1}^{\text{eq}}\hat{Y}_{P_2}^{\text{eq}}}{\hat{Y}_{P_3}^{\text{eq}}\hat{Y}_{P_4}^{\text{eq}}}\hat{Y}_{P_3}\hat{Y}_{P_4}\right) \langle \sigma v\rangle_{P:\pi_{P_1}\pi_{P_2}\to \pi_{P_3} \pi_{P_4}} \\
    & + s\sum_{P\in \mathcal{P}^{3\to 2}}(\#A_{\text{in}}^P - \#A_{\text{out}}^P)\left(\hat{Y}_{P_1}\hat{Y}_{P_2}\hat{Y}_{P_3} - \frac{\hat{Y}_{P_1}^{\text{eq}}\hat{Y}_{P_2}^{\text{eq}}\hat{Y}_{P_3}^{\text{eq}}}{\hat{Y}_{P_4}^{\text{eq}}\hat{Y}_{P_5}^{\text{eq}}}\hat{Y}_{P_4}\hat{Y}_{P_5}\right) \langle \sigma v^2\rangle_{P:\pi_{P_1}\pi_{P_2} \pi_{P_3} \to \pi_{P_4} \pi_{P_5}}\bigg]
  \end{aligned}
\end{equation}
where the first sum is over processes such that $\#A_{\text{in}}^P > \#A_{\text{out}}^P$ to avoid double counting.

\subsection{Numerical procedure}\label{sapp:NumericalProcedure}
We now proceed to describe how we solve the Boltzmann equation in practice. At first, the pions are kept at their thermal equilibrium value by $3\to 2$ processes. The validity of this assumption is tested periodically by comparing the rate of $3\to 2$ processes to the Hubble constant. When this approximation becomes remotely close to failing, the pion densities are evolved as a whole by assuming that they are in chemical equilibrium with each other. The validity of this assumption is tested periodically by computing the rates of $2\to 2$ scattering if one of the pion densities were to vary a little and comparing this rate to the Hubble constant. When one particle comes remotely close to failing this assumption, we begin to evolve it individually. The densities of the other pions are computed as before but also by keeping track of how many were converted to pions that are not in chemical equilibrium anymore. Once none of the pions are in chemical equilibrium anymore, all pions are evolved using the full Boltzmann equation.

\section{Symmetry breaking benchmarks}\label{app:SBB}
In this appendix, we present relevant information for each benchmark scenario of symmetry breaking of Table~\ref{tab:Benchmarks}.

\subsection{Notation}\label{sapp:Notation}
First, we introduce some notation for the pions. The pion matrix can be written in all cases we consider as \cite{Peskin:1980gc}
\begin{equation}\label{eq:PionMatrixGeneral}
  \Pi = \frac{1}{\sqrt{2}}
  \begin{pmatrix}
    C         & D  \\
    D^\dagger & C^T
  \end{pmatrix}.
\end{equation}
The matrix $C$ is Hermitian, traceless and can be written for all patterns of chiral symmetry breaking with $N$ flavors as:
\begin{equation}\label{eq:MatC}
  \begin{tabular}{ccc}
  $N = 2$ & $N = 3$ & ...\\
  $C =
  \begin{pmatrix}
    \frac{\pi^d_1}{\sqrt{2}} & \pi^o_1                   \\
    \bar{\pi}^o_1            & -\frac{\pi^d_1}{\sqrt{2}}
  \end{pmatrix}$,
  &
  $C =
  \begin{pmatrix}
    \frac{\pi^d_1}{\sqrt{2}} + \frac{\pi^d_2}{\sqrt{6}} & \pi^o_1 & \pi^o_2 \\
    \bar{\pi}^o_1                                       & -\frac{\pi^d_1}{\sqrt{2}} + \frac{\pi^d_2}{\sqrt{6}} & \pi^o_3 \\
    \bar{\pi}^o_2                                       & \bar{\pi}^o_3                                        & -\sqrt{\frac{2}{3}}\pi^d_2
  \end{pmatrix}$,
  &
  ...
  \end{tabular}
\end{equation}
The matrix $D$ can be written for each pattern of chiral symmetry breaking as:
\begin{equation}\label{eq:MatD}
  \begin{tabular}{ccccc}
    & $N = 2$ & $N = 3$ & ... & \\
    \begin{tabular}{c}
      $SU(N)\times SU(N) \to SU(N)$:\\
      $(D=0)$
    \end{tabular}
    & 
    $D =
    \begin{pmatrix}
      0 & 0 \\
      0 & 0 
    \end{pmatrix}$,
    & 
    $D =
    \begin{pmatrix}
      0 & 0 & 0\\
      0 & 0 & 0\\
      0 & 0 & 0
    \end{pmatrix}$,
    & 
    ...\\\\
    \begin{tabular}{c}
      $SU(2N) \to Sp(2N)$:\\
      $(D=-D^T)$
    \end{tabular}
    & 
    $D =
    \begin{pmatrix}
      0         & \pi^b_1 \\
      - \pi^b_1 & 0 
    \end{pmatrix}$,
    & 
    $D =
    \begin{pmatrix}
      0         & \pi^b_1   & \pi^b_2 \\
      - \pi^b_1 & 0         & \pi^b_3 \\
      - \pi^b_2 & - \pi^b_3 & 0
    \end{pmatrix}$,
    & 
    ...\\\\
    \begin{tabular}{c}
      $SU(2N) \to SO(2N)$:\\
      $(D=D^T)$
    \end{tabular}
    & 
    $D =
    \begin{pmatrix}
      \pi^a_1 & \pi^b_1 \\
      \pi^b_1 & \pi^a_2 
    \end{pmatrix}$,
    & 
    $D =
    \begin{pmatrix}
      \pi^a_1 & \pi^b_1 & \pi^b_2 \\
      \pi^b_1 & \pi^a_2 & \pi^b_3 \\
      \pi^b_2 & \pi^b_3 & \pi^a_3
    \end{pmatrix}$,
    & 
    ...
  \end{tabular}
\end{equation}
The generalization to larger $N$ is trivial in both cases. The different types of pions are:
\begin{itemize}
  \item Pions $\pi_i^d$ consist of a linear superposition of a quark and the corresponding antiquark. They are present for all patterns a chiral symmetry breaking.
  \item Pions $\pi_i^o$ consist of a quark and a different antiquark. They are present for all patterns a chiral symmetry breaking.
  \item Pions $\pi_i^b$ consist of two different quarks. They are present for $SU(2N)\to Sp(2N)$ and $SU(2N) \to SO(2N)$.
  \item Pions $\pi_i^a$ consist of two identical quarks. They are present for $SU(2N) \to SO(2N)$ only.
\end{itemize}
Obviously, pions $\bar{\pi}_i^j$ are the conjugate of $\pi_i^j$. The generators $t^a$ can then be read by simply decomposing $\Pi = t^a \pi^a$. They are normalized such that $\text{Tr}[t^a t^b]=\delta^{a\bar{b}}$, i.e. 1 if $\pi^a$ is the conjugate of $\pi^b$ and 0 otherwise.

\subsection{Benchmarks}\label{sapp:Benchmark}
We then proceed to discuss the benchmarks. For each benchmark, we will provide the list of representations of $h$, the pions that constitute them, their masses and their stability. They are ordered from heaviest to lightest. In some cases, the mass eigenstates are linear combinations of several pions of Eq.~(\ref{eq:PionMatrixGeneral}). We express such states as the combination of these pions within brackets, eg. $(\pi^d_1, \pi^d_2)$. A pion will be considered stable if there does not exist a set of pions which can be combined to reproduce the same charges and whose total mass is inferior. We will also indicate in the column ID whether that representation contributes to indirect detection or not. If it does, we provide an example. Assumed relations between the quark masses are included. Relevant comments are provided. When multiple $U(1)$ symmetries are present, they can be combined in different ways and we provide one example only. Their charges are normalized to the smallest integers. For representations of $h_1\times h_2$ with $h_1$ non-Abelian and $h_2$ Abelian, the notation $(\mathbf{n}, \pm m)$ stand for the combination of $(\mathbf{n}, +m)$ and $(\bar{\mathbf{n}}, -m)$. Generalizations of this notation are straightforward.

\subsubsection*{CIa: $SU(3) \times SU(3)\to SU(3) \to SU(2)\times U(1)$}
\begin{center}
  \begin{tabular}{cccccc}
    $h$                   & Pions                                                  & $m^2/B_0$                 & Stability & ID  & Example                                     \\\hline
    $(\mathbf{3}, 0)$     & $\pi^d_1$, $\pi^o_1$, $\bar{\pi}^o_1$                  & $2m_1$                    & Yes       & Yes & $\pi^o_1 \bar{\pi}^o_1 \to \pi^d_2 \pi^d_2$ \\
    $(\mathbf{2}, \pm 1)$ & $\pi^o_2$, $\bar{\pi}^o_2$, $\pi^o_3$, $\bar{\pi}^o_3$ & $m_1 + m_3$               & Yes       & Yes & $\pi^o_2 \bar{\pi}^o_2 \to \pi^d_2 \pi^d_2$ \\
    $(\mathbf{1}, 0)$     & $\pi^d_2$                                              & $\frac{2}{3}(m_1 + 2m_3)$ & No        &     &
  \end{tabular}
\end{center}
Condition(s):
\begin{itemize}
  \item $m_1 = m_2 > m_3$
\end{itemize}
Comment(s):
\begin{itemize}
  \item Generalization to a larger $SU(N) \times SU(N)\to SU(N) \to SU(N - 1)\times U(1)$ is trivial.
\end{itemize}

\subsubsection*{CIb: $SU(6) \to Sp(6) \to Sp(4)\times U(1)$}
\begin{center}
  \begin{tabular}{cccccc}
    $h$                   & Pions                                                                                                          & $m^2/B_0$   & Stability & ID  & Example                                     \\\hline
    $(\mathbf{5}, 0)$     & $\pi^d_1$, $\pi^o_1$, $\bar{\pi}^o_1$, $\pi^b_1$, $\bar{\pi}^b_1$                                              & $2m_1$      & Yes       & Yes & $\pi^o_1 \bar{\pi}^o_1 \to \pi^d_2 \pi^d_2$ \\
    $(\mathbf{4}, \pm 1)$ & $\pi^o_2$, $\bar{\pi}^o_2$, $\pi^o_3$, $\bar{\pi}^o_3$, $\pi^b_2$, $\bar{\pi}^b_2$, $\pi^b_3$, $\bar{\pi}^b_3$ & $m_1 + m_3$ & Yes       & Yes & $\pi^o_2 \bar{\pi}^o_2 \to \pi^d_2 \pi^d_2$ \\
    $(\mathbf{1}, 0)$     & $\pi^d_2$                                                                                                      & $\frac{2}{3}(m_1 + 2m_3)$ & No        &     &
  \end{tabular}
\end{center}
Condition(s):
\begin{itemize}
  \item $m_1 = m_2 > m_3$
\end{itemize}
Comment(s):
\begin{itemize}
  \item Generalization to a larger $SU(2N) \to Sp(2N) \to Sp(2N - 2)\times U(1)$ is trivial.
\end{itemize}

\subsubsection*{CIc: $SU(3) \times SU(3)\to SU(3) \to U(1)\times U(1)$}
\begin{center}
  \begin{tabular}{cccccc}
    $h$          & Pions                          & $m^2/B_0$   & Stability & ID  & Example                                     \\\hline
    $(0, 0)$     & $\pi^m_1 = (\pi^d_1, \pi^d_2)$ & $m_1^m$     & No        &     &                                             \\
    $\pm(1,  0)$ & $\pi^o_1$, $\bar{\pi}^o_1$     & $m_1 + m_2$ & Yes       & Yes & $\pi^o_1 \bar{\pi}^o_1 \to \pi^m_2 \pi^m_2$ \\
    $\pm(1, -1)$ & $\pi^o_2$, $\bar{\pi}^o_2$     & $m_1 + m_3$ & Yes       & Yes & $\pi^o_2 \bar{\pi}^o_2 \to \pi^m_2 \pi^m_2$ \\
    $\pm(0,  1)$ & $\pi^o_3$, $\bar{\pi}^o_3$     & $m_2 + m_3$ & Yes       & Yes & $\pi^o_3 \bar{\pi}^o_3 \to \pi^m_2 \pi^m_2$ \\
    $(0, 0)$     & $\pi^m_2 = (\pi^d_1, \pi^d_2)$ & $m_2^m$     & No        &     &    
  \end{tabular}
\end{center}
where $m_1^m$ and $m_2^m$ are respectively the largest and smallest eigenvalues of
\begin{equation}\label{eq:DetCIb}
  \begin{pmatrix}
    m_1 + m_2                  & \frac{m_1 - m_2}{\sqrt{3}} \\
    \frac{m_1 - m_2}{\sqrt{3}} & \frac{m_1 + m_2 + 4m_3}{3}
  \end{pmatrix}.
\end{equation}
Condition(s):
\begin{itemize}
  \item $m_1 > m_2 > m_3$
\end{itemize}
Comment(s):
\begin{itemize}
  \item Any other hierarchy of dark quark masses would lead to this symmetry breaking pattern being part of category I as long as no two dark quark masses are equal.
  \item Two relations are relevant:
  \begin{equation}\label{eq:RelationCIb}
     m_1^m > m_1 + m_2 \hspace{1cm} m_2^m < m_2 + m_3.
  \end{equation}
\end{itemize}

\subsubsection*{CIIa: $SU(4) \to SO(4) \to U(1) \times U(1)$}
\begin{center}
  \begin{tabular}{cccccc}
    $h$          & Pions                      & $m^2/B_0$   & Stability & ID  & Example                                     \\\hline
    $\pm(2,  0)$ & $\pi^a_2, \bar{\pi}^a_2$   & $2m_2$      & Yes       & Yes & $\pi^a_2 \bar{\pi}^a_2 \to \pi^d_1 \pi^d_1$ \\
    $\pm(1, -1)$ & $\pi^o_1$, $\bar{\pi}^o_1$ & $m_1 + m_2$ & Yes       & Yes & $\pi^o_1 \pi^b_1 \to \pi^d_1 \pi^a_1$       \\
    $\pm(1,  1)$ & $\pi^b_1$, $\bar{\pi}^b_1$ & $m_1 + m_2$ & Yes       & Yes & $\pi^o_1 \pi^b_1 \to \pi^d_1 \pi^a_1$       \\
    $(0, 0)$     & $\pi^d_1$                  & $m_1 + m_2$ & No        &     &                                             \\
    $\pm(0,  2)$ & $\pi^a_1, \bar{\pi}^a_1$   & $2m_1$      & Yes       & No  &
  \end{tabular}
\end{center}
Condition(s):
\begin{itemize}
  \item $m_1 < m_2$
\end{itemize}
Comment(s):
\begin{itemize}
  \item Any other hierarchy of dark quark masses would lead to this symmetry breaking pattern being part of category II as long as no two dark quark masses are equal.
\end{itemize}

\subsubsection*{CIIb: $SU(3) \times SU(3)\to SU(3) \to SU(2)\times U(1)$}
\begin{center}
  \begin{tabular}{cccccc}
    $h$                   & Pions                                                  & $m^2/B_0$                 & Stability & ID  & Example                                     \\\hline
    $(\mathbf{1}, 0)$     & $\pi^d_2$                                              & $\frac{2}{3}(m_1 + 2m_3)$ & No        &     &                                             \\
    $(\mathbf{2}, \pm 1)$ & $\pi^o_2$, $\bar{\pi}^o_2$, $\pi^o_3$, $\bar{\pi}^o_3$ & $m_1 + m_3$               & Yes       & Yes & $\pi^o_2 \bar{\pi}^o_2 \to \pi^d_2 \pi^d_1$ \\
    $(\mathbf{3}, 0)$     & $\pi^d_1$, $\pi^o_1$, $\bar{\pi}^o_1$                  & $2m_1$                    & Yes       & No  &
  \end{tabular}
\end{center}
Condition(s):
\begin{itemize}
  \item $m_1 = m_2 < m_3$
\end{itemize}
Comment(s):
\begin{itemize}
  \item Generalization to a larger $SU(N) \times SU(N)\to SU(N) \to SU(N - 1)\times U(1)$ is trivial.
\end{itemize}

\subsubsection*{CIIc: $SU(6) \to Sp(6)\to Sp(4)\times U(1)$}
\begin{center}
  \begin{tabular}{cccccc}
    $h$                   & Pions                                                                                                          & $m^2/B_0$                 & Stability & ID  & Example                               \\\hline
    $(\mathbf{1}, 0)$     & $\pi^d_2$                                                                                                      & $\frac{2}{3}(m_1 + 2m_3)$ & No        &     &                                             \\
    $(\mathbf{4}, \pm 1)$ & $\pi^o_2$, $\bar{\pi}^o_2$, $\pi^o_3$, $\bar{\pi}^o_3$, $\pi^b_2$, $\bar{\pi}^b_2$, $\pi^b_3$, $\bar{\pi}^b_3$ & $m_1 + m_3$               & Yes       & Yes & $\pi^o_2 \bar{\pi}^o_2 \to \pi^d_2 \pi^d_1$ \\
    $(\mathbf{5}, 0)$     & $\pi^d_1$, $\pi^o_1$, $\bar{\pi}^o_1$, $\pi^b_1$, $\bar{\pi}^b_1$                                              & $2m_1$                    & Yes       & No  &  
  \end{tabular}
\end{center}
Condition(s):
\begin{itemize}
  \item $m_1 = m_2 < m_3$
\end{itemize}
Comment(s):
\begin{itemize}
  \item Generalization to a larger $SU(2N) \to Sp(2N) \to Sp(2N - 2)\times U(1)$ is trivial.
\end{itemize}

\subsubsection*{CIId: $SU(5) \times SU(5)\to SU(5) \to SU(3) \times SU(2) \times U(1)$}
\begin{center}
  \begin{tabular}{cccccc}
    $h$                               & Pions                                                                                    & $m^2/B_0$                  & Stability & ID  & Example                                     \\\hline
    $(\mathbf{8}, \mathbf{1}, 0)$     & $\pi^m_1=(\pi^d_2, \pi^d_3 ,\pi^d_4)$,                                                   & $2m_3$                     & Yes       & Yes & $\pi^o_8 \bar{\pi}^o_8 \to \pi^m_3 \pi^m_3$ \\
                                      & $\pi^m_2=(\pi^d_2, \pi^d_3 ,\pi^d_4)$,                                                   &                            &           &     &                                             \\
                                      & $\pi^o_8$, $\bar{\pi}^o_8$, $\pi^o_9$, $\bar{\pi}^o_9$, $\pi^o_{10}$, $\bar{\pi}^o_{10}$ &                            &           &     &                                             \\
    $(\mathbf{3}, \mathbf{2}, \pm 1)$ & $\pi^o_2$, $\bar{\pi}^o_2$, $\pi^o_3$, $\bar{\pi}^o_3$, $\pi^o_4$, $\bar{\pi}^o_4$,      & $m_1 + m_3$                & Yes       & Yes & $\pi^o_2 \bar{\pi}^o_2 \to \pi^m_3 \pi^m_3$ \\
                                      & $\pi^o_5$, $\bar{\pi}^o_5$, $\pi^o_6$, $\bar{\pi}^o_6$, $\pi^o_7$, $\bar{\pi}^o_7$       &                            &           &     &                                             \\
    $(\mathbf{1}, \mathbf{1},     0)$ & $\pi^m_3=(\pi^d_2, \pi^d_3 ,\pi^d_4)$                                                    & $\frac{2}{5}(3m_1 + 2m_3)$ & No        &     &                                             \\
    $(\mathbf{1}, \mathbf{3}, 0)$     & $\pi^d_1$, $\pi^o_1$, $\bar{\pi}^o_1$                                                    & $2m_1$                     & Yes       & No  &
  \end{tabular}
\end{center}
Condition(s):
\begin{itemize}
  \item $m_1 = m_2 < m_3 = m_4 = m_5$
\end{itemize}
Comment(s):
\begin{itemize}
  \item Inverting the inequality would still lead to a benchmark that is part of category II.
\end{itemize}

\subsubsection*{CIIe: $SU(4) \times SU(4)\to SU(4) \to SU(2) \times U(1) \times U(1)$}
\begin{center}
  \begin{tabular}{cccccc}
    $h$                      & Pions                                                  & $m^2/B_0$   & Stability & ID  & Example                                     \\\hline
    $(\mathbf{1}, 0, 0)$     & $\pi^m_2=(\pi^d_2, \pi^d_3)$                           & $m_1^m$     & No        &     &                                             \\
    $\pm(\mathbf{1}, 1, -1)$ & $\pi^o_6$, $\bar{\pi}^o_6$                             & $m_3 + m_4$ & Yes       & Yes & $\pi^o_6 \bar{\pi}^o_6 \to \pi^m_1 \pi^m_1$ \\
    $(\mathbf{2}, 0, \pm 1)$ & $\pi^o_3$, $\bar{\pi}^o_3$, $\pi^o_5$, $\bar{\pi}^o_5$ & $m_1 + m_4$ & Yes       & Yes & $\pi^o_3 \bar{\pi}^o_3 \to \pi^m_1 \pi^m_1$ \\
    $(\mathbf{2}, \pm 1, 0)$ & $\pi^o_2$, $\bar{\pi}^o_2$, $\pi^o_4$, $\bar{\pi}^o_4$ & $m_1 + m_3$ & Yes       & Yes & $\pi^o_2 \bar{\pi}^o_2 \to \pi^m_1 \pi^m_1$ \\
    $(\mathbf{1}, 0, 0)$     & $\pi^m_1=(\pi^d_2, \pi^d_3)$                           & $m_2^m$     & No        &     &                                             \\
    $(\mathbf{3}, 0, 0)$     & $\pi^d_1$, $\pi^o_1$, $\bar{\pi}^o_1$                  & $2m_1$      & Yes       & No  &
  \end{tabular}
\end{center}
where $m_1^m$ and $m_2^m$ are respectively the largest and smallest eigenvalues of
\begin{equation}\label{eq:DetCIIe}
  \begin{pmatrix}
    \frac{2m_1 + 4m_3}{3}         & \frac{\sqrt{2}}{3}(m_1 - m_3) \\
    \frac{\sqrt{2}}{3}(m_1 - m_3) & \frac{2m_1 + m_3 + 9m_4}{6}
  \end{pmatrix}.
\end{equation}
Condition(s):
\begin{itemize}
  \item $m_1 = m_2 < m_3 \leq m_4$
\end{itemize}
Comment(s):
\begin{itemize}
  \item Some alternative hierarchies lead to this symmetry not being part of category II.
  \item Two relations are relevant:
  \begin{equation}\label{eq:RelationCIe}
     m_1^m > m_3 + m_4 \hspace{1cm} m_2^m < m_1 + m_3.
  \end{equation}
\end{itemize}

\subsubsection*{CIIIa: $SU(4) \to SO(4) \to U(1)$}
\begin{center}
  \begin{tabular}{ccccc}
    $h$     & Pions                      & $m^2/B_0$   & Stability & ID  \\\hline
    $\pm 1$ & $\pi^a_2$, $\bar{\pi}^a_2$ & $2m_2$      & No        &     \\
    $\pm 1$ & $\pi^b_1$, $\bar{\pi}^b_1$ & $m_1 + m_2$ & No        &     \\
    $0$     & $\pi^o_1$                  & $m_1 + m_2$ & No        &     \\
    $0$     & $\bar{\pi}^o_1$            & $m_1 + m_2$ & No        &     \\
    $0$     & $\pi^d_1$                  & $m_1 + m_2$ & No        &     \\
    $\pm 1$ & $\pi^a_1, \bar{\pi}^a_1$   & $2m_1$      & Yes       & No  
  \end{tabular}
\end{center}
Condition(s):
\begin{itemize}
  \item $m_1 < m_2$
\end{itemize}
Comment(s):
\begin{itemize}
  \item Any other hierarchy of dark quark masses would lead to this symmetry breaking pattern being part of category III as long as no two dark quark masses are equal.
\end{itemize}

\subsubsection*{CIIIb: $SU(6) \to SO(6) \to U(1)$}
\begin{center}
  \begin{tabular}{ccccc}
    $h$     & Pions                          & $m^2/B_0$   & Stability & ID  \\\hline
    $\pm 1$ & $\pi^a_1$, $\bar{\pi}^a_1$     & $2m_1$      & No        &     \\
    $0$     & $\pi^m_1 = (\pi^d_1, \pi^d_2)$ & $m_1^m$     & No        &     \\
    $\pm 1$ & $\pi^b_1$, $\bar{\pi}^b_1$     & $m_1 + m_2$ & No        &     \\
    $0$     & $\pi^o_1$                      & $m_1 + m_2$ & No        &     \\
    $0$     & $\bar{\pi}^o_1$                & $m_1 + m_2$ & No        &     \\
    $\pm 1$ & $\pi^a_2$, $\bar{\pi}^a_2$     & $2m_2$      & No        &     \\
    $\pm 1$ & $\pi^b_2$, $\bar{\pi}^b_2$     & $m_1 + m_3$ & No        &     \\
    $0$     & $\pi^o_2$                      & $m_1 + m_3$ & No        &     \\
    $0$     & $\bar{\pi}^o_2$                & $m_1 + m_3$ & No        &     \\
    $\pm 1$ & $\pi^b_3$, $\bar{\pi}^b_3$     & $m_2 + m_3$ & No        &     \\
    $0$     & $\pi^o_3$                      & $m_2 + m_3$ & No        &     \\
    $0$     & $\bar{\pi}^o_3$                & $m_2 + m_3$ & No        &     \\
    $0$     & $\pi^m_2 = (\pi^d_1, \pi^d_2)$ & $m_2^m$     & No        &     \\
    $\pm 1$ & $\pi^a_3, \bar{\pi}^a_3$       & $2m_3$      & Yes       & No  
  \end{tabular}
\end{center}
Condition(s):
\begin{itemize}
  \item $m_1 > m_2 > m_3$
  \item $2m_2 > m_1 + m_3$
\end{itemize}
Comment(s):
\begin{itemize}
  \item Any other hierarchy of dark quark masses would lead to this symmetry breaking pattern being part of category III as long as no two dark quark masses are equal.
  \item The masses $m_1^m$ and $m_2^m$ are given by Eq.~(\ref{eq:DetCIb}).
\end{itemize}

\subsubsection*{CIIIc: $SU(4) \times SU(4)\to SU(4) \to SU(2)$}
\begin{center}
  \begin{tabular}{ccccc}
    $h$          & Pions                                                                             & $m^2/B_0$   & Stability & ID  \\\hline
    $\mathbf{3}$ & $\pi^d_1$, $\pi^o_1$, $\bar{\pi}^o_1$                                             & $2m_1$      & No        &     \\
    $\mathbf{3}$ & $\pi^m_1 =(\pi^o_2, \pi^o_5)$, $\pi^o_3$, $\pi^o_4$                               & $m_1 + m_3$ & No        &     \\
    $\mathbf{3}$ & $\bar{\pi}^m_1 =(\bar{\pi}^o_2, \bar{\pi}^o_5)$, $\bar{\pi}^o_3$, $\bar{\pi}^o_4$ & $m_1 + m_3$ & No        &     \\
    $\mathbf{1}$ & $\pi^m_2 =(\pi^o_2, \pi^o_5)$                                                     & $m_1 + m_3$ & No        &     \\
    $\mathbf{1}$ & $\bar{\pi}^m_2 =(\bar{\pi}^o_2, \bar{\pi}^o_5)$                                   & $m_1 + m_3$ & No        &     \\
    $\mathbf{1}$ & $\pi^d_2$                                                                         & $m_1 + m_3$ & No        &     \\
    $\mathbf{3}$ & $\pi^d_3$, $\pi^o_6$, $\bar{\pi}^o_6$                                             & $2m_3$      & Yes       & No
  \end{tabular}
\end{center}
Condition(s):
\begin{itemize}
  \item $m_1 = m_2 > m_3 = m_4$
\end{itemize}
\begin{itemize}
  \item Inverting the inequality would still lead to this symmetry breaking pattern being part of category III.
\end{itemize}

\bibliography{biblio}
\bibliographystyle{JHEP}

\end{document}